\pdfoutput=1
\documentclass[11pt,a4paper,oneside,openany]{book}
\usepackage[top=2.5cm,bottom=2.5cm,left=2.5cm,right=2.5cm]{geometry}

\usepackage[T1]{fontenc}
\usepackage[utf8]{inputenc}
\usepackage{lmodern}
\usepackage[english]{babel}
\usepackage{microtype}

\usepackage{amsmath}
\usepackage{amssymb}
\usepackage{graphicx}
\usepackage{array}
\usepackage{booktabs}
\usepackage{caption}
\usepackage{xcolor}
\usepackage{lastpage}

\usepackage[numbers,sort&compress]{natbib}

\usepackage{fancyhdr}
\fancypagestyle{plain}{\fancyhf{}\cfoot{\thepage}}

\usepackage[nottoc,notlot,notlof]{tocbibind}

\usepackage{xurl}
\usepackage[strings]{underscore}
\usepackage{hyperref}
\hypersetup{colorlinks=true,linkcolor=black,citecolor=black,urlcolor=black,
  pdftitle={Parallel Computing Systems for the Optimal Selection of Process Equipment},
  pdfauthor={Andrey B. Borisenko},
  pdfkeywords={branch-and-bound, metaheuristics, ant colony optimization, simulated annealing, parameter tuning, GPU, CUDA, OpenMP, MPI, multiproduct batch plants}}
\makeatletter
\renewcommand{\toc@chapter}[1]{\chapter*{#1}\prw@mkboth{#1}\phantomsection\addcontentsline{toc}{chapter}{#1}}
\makeatother

\definecolor{keyword}{HTML}{1A00FF}
\definecolor{string}{HTML}{B40000}
\definecolor{comment}{HTML}{008000}
\definecolor{brackets}{HTML}{B40000}
\usepackage{listings}
\title{Parallel Computing Systems for the Optimal Selection of Process Equipment}
\author{Andrey B. Borisenko}
\date{2026}

\begin{document}
%%%% Title page
\begin{titlepage}
\thispagestyle{empty}
\newcommand{\tprule}[1]{\noindent\rule{\textwidth}{#1}\par}
\begin{center}
\vspace*{0.5cm}
{\Large A.\,B.~BORISENKO\par}

\vspace{0.9cm}
\tprule{1.6pt}\vspace{-0.62\baselineskip}
\tprule{0.4pt}

\vspace{4.2cm}
{\fontsize{24.88}{30}\selectfont\bfseries
PARALLEL COMPUTING\\
SYSTEMS FOR THE OPTIMAL\\
SELECTION OF PROCESS\\
EQUIPMENT\par}

\vspace{1.3cm}
{\centering\rule{3cm}{0.6pt}\par}
\vspace{0.6cm}
{\Large Monograph\par}
\vspace{0.4cm}
{\normalsize\itshape English edition of the author's manuscript\par}

\vfill
\tprule{0.4pt}\vspace{-0.62\baselineskip}
\tprule{1.6pt}
\vspace{0.5cm}
{\large Tambov\\[3pt] 2026\par}
\end{center}
\end{titlepage}
\setcounter{page}{2}
%%%% Reverse of the title page
\thispagestyle{empty}
\begingroup
\setlength{\parindent}{0pt}%
\setlength{\parskip}{0pt}%
\emergencystretch=2em%

UDC 004.272.2:519.854:66.011\par

\vspace{0.9cm}
\begin{center}
\begin{minipage}{0.86\textwidth}
\centering\small\itshape
The Russian edition of this monograph was recommended for publication by the Scientific and Technical Council of Tambov State Technical University
\end{minipage}
\end{center}

\vspace{0.5cm}
\textbf{Reviewers of the Russian edition:}\par\vspace{3pt}
{\small
\leftskip=1.25cm
\textit{E.\,N.~Tugolukov}, Doctor of Engineering Sciences, Professor, Department of Engineering and Technology of Nanoproduct Manufacturing, Tambov State Technical University;\par\vspace{2pt}
\textit{A.\,D.~Obukhov}, Doctor of Engineering Sciences, Associate Professor, Department of Automated Decision Support Systems, Tambov State Technical University;\par\vspace{2pt}
\textit{D.\,S.~Solovjev}, Doctor of Engineering Sciences, Associate Professor, Department of Mathematical Modeling and Information Technologies, Derzhavin Tambov State University\par}

\vspace{0.7cm}
\textbf{Borisenko, A.\,B.}\par
{\leftskip=1.25cm
Parallel Computing Systems for the Optimal Selection of Process Equipment\,: monograph / A.\,B.~Borisenko.~---~Tambov, 2026.~---~\pageref{LastPage}~p.\par}

\vspace{0.5cm}
{\small
\setlength{\parindent}{1.25cm}%
The monograph is devoted to the development and study of parallel algorithms for the optimal selection of equipment for chemical-engineering systems of multiproduct batch plants, which is a large-scale NP-hard discrete optimization problem. It presents a mathematical model of the operation of a chemical-engineering system; sequential and parallel variants of the branch-and-bound method for shared-memory and distributed-memory systems (OpenMP, MPI) and for graphics processing units (CUDA); a hybrid metaheuristic algorithm combining Ant Colony Optimization and Simulated Annealing; and methods for tuning its parameters on one or several GPUs. Results of computational experiments on multi-core processors, computing clusters, and GPUs of various architectures are reported.\par

The book is intended for researchers, engineers, and software developers working in parallel and high-performance computing, as well as for doctoral and senior undergraduate students in related fields.\par}

\vspace{0.5cm}
{\small\itshape
This English edition is the author's translation of the Russian-language monograph \emph{Primenenie parallel'nykh vychislitel'nykh sistem dlya optimal'nogo vybora tekhnologicheskogo oborudovaniya} (Tambov, 2026), with minor corrections.\par}

\vfill
\rule{\textwidth}{0.4pt}\par\vspace{4pt}
\hfill\copyright~A.\,B.~Borisenko, 2026\par
\endgroup
\clearpage

\tableofcontents

%%%% List of abbreviations
\chapter*{Abbreviations}
\phantomsection
\addcontentsline{toc}{chapter}{Abbreviations}
\label{list_of_acronyms}
\begingroup
\setlength{\parindent}{0pt}
\setlength{\parskip}{2pt}
\begin{tabular}{@{}>{\raggedright\arraybackslash}p{0.17\textwidth}>{\raggedright\arraybackslash}p{0.78\textwidth}@{}}
ACO & Ant Colony Optimization \\
API & Application Programming Interface \\
B\&B & Branch-and-Bound \\
BFS & Breadth-First Search \\
CC & Compute Capability \\
CES & Chemical-Engineering System \\
CO & Combinatorial Optimization \\
CPU & Central Processing Unit \\
CSP & Constraint Satisfaction Problem \\
CUDA & Compute Unified Device Architecture \\
DFS & Depth-First Search \\
DOE & Design of Experiments \\
DRAM & Dynamic Random-Access Memory \\
DSP & Digital Signal Processor \\
FP32 & 32-bit (single-precision) Floating Point \\
FPGA & Field-Programmable Gate Array \\
GPGPU & General-Purpose computing on Graphics Processing Units \\
GPU & Graphics Processing Unit \\
HLSL & High-Level Shading Language \\
LIFO & Last-In, First-Out \\
MAC & Multiple Ant Colonies \\
MILP & Mixed-Integer Linear Programming \\
MINLP & Mixed-Integer Nonlinear Programming \\
MIR & Multiple Independent Runs \\
MPI & Message Passing Interface \\
NP & Nondeterministic Polynomial (time) \\
NVCC & NVIDIA CUDA Compiler \\
OFAT & One-Factor-At-a-Time \\
OpenCL & Open Computing Language \\
OpenMP & Open Multi-Processing \\
RAM & Random-Access Memory \\
SA & Simulated Annealing \\
SIMD & Single Instruction, Multiple Data \\
SIMT & Single Instruction, Multiple Threads \\
SM & Streaming Multiprocessor \\
SMP & Symmetric Multiprocessing \\
STL & Standard Template Library \\
\end{tabular}
\endgroup

%%%% Introduction
\chapter*{Introduction}
\phantomsection
\addcontentsline{toc}{chapter}{Introduction}
\label{intro}

NP-hard combinatorial optimization (CO) problems are among the most difficult computational problems to solve in practice.
A CO problem is an optimization problem in which a finite set of objects is given together with an objective function that assigns a cost to each object.
The goal is to find an object with the minimum (or maximum) value of the objective function.
The objects are typically integers, subsets of a set of elements, permutations of a set of elements, and so on.
Well-known examples of CO problems include the Travelling Salesman Problem~\cite{Fujimoto2011highly}, the Assignment Problem~\cite{Czapinski2013effective}, the Knapsack Problem~\cite{Boyer2012solving}, and various scheduling problems~\cite{Pellerin2020}.

A characteristic feature of NP-hard problems is that, when exact (deterministic) algorithms such as exhaustive enumeration are used, the time required to find the optimal solution grows exponentially with the size of the search space; this phenomenon is known as \emph{combinatorial explosion}~\cite{gavanelli2009}.

The main ways of increasing computational speed are the following: using more advanced algorithms; tuning the parameters of the algorithms employed; compiler optimization of the programs; optimization with the help of profilers; and using more powerful hardware, including the development of parallel computing technologies.
Of these, the strongest influence on computational speed comes from the choice of the solution algorithm together with an optimal setting of its parameters, and from the hardware used.

All known algorithms for NP-hard problems, including CO problems, can be roughly divided into three categories: exact~\cite{Woeginger2003}, heuristic~\cite{Zerovnik2015}, and metaheuristic~\cite{Talbi2009} algorithms.

Exact algorithms include, for example, exhaustive search~\cite{Nievergelt2000}, the branch-and-bound method~\cite{Tomazella2020}, the Hungarian algorithm~\cite{Date2016}, and dynamic programming~\cite{Bouman2018}.

Heuristic approaches rely on plausible but not mathematically justified assumptions about the properties of the optimal solution~\cite{Marti2011}.
Such algorithms usually yield practically good solutions, but they do not guarantee that the best solution will be found.
Moreover, heuristics are built on ``common sense'' and therefore depend strongly on the particular problem.

Metaheuristics combine basic heuristic methods within higher-level algorithmic frameworks aimed at an efficient exploration of the search space~\cite{Panteleev2009,yang2010,Ershov2015}.
Among metaheuristic algorithms are Tabu Search~\cite{Glover2013}, Simulated Annealing~\cite{Aarts2014}, Ant Colony Optimization~\cite{Dorigo2005}, Genetic Algorithms~\cite{Man2012}, Particle Swarm Optimization~\cite{Clerc2013}, Artificial Bee Colony Optimization~\cite{Karaboga2014}, and others.
Their advantage over classical heuristics based on local search is that they can explore a larger part of the search space in order to find a near-optimal solution, whereas local search methods stop as soon as a local optimum in the chosen neighborhood has been reached.
It should also be noted that hybridization of metaheuristics is currently one of the most promising ways of designing new, highly efficient algorithms~\cite{Anghinolfi2008,blum2010brief}.

A distinctive feature of metaheuristic algorithms is that they contain a large number of free parameters.
The efficiency of an algorithm depends strongly on the values of these parameters; however, as a rule, there are no universal formal guidelines for choosing them on the basis of the properties of the optimization problem at hand~\cite{Birattari2009}.
The development of methods for tuning metaheuristics is therefore a topical research problem~\cite{Joshi2020,Huang2020}.

Over the past decades, parallelization of algorithms and the use of high-performance computing systems have become one of the most promising ways of improving algorithm performance~\cite{voevodin2002,Grama2003,gebali2011algorithms,Strongin2013}.
Such systems include shared-memory systems, distributed-memory systems~\cite{Padua2011}, and heterogeneous systems.
Shared-memory systems include computers with one or more multi-core processors.
Distributed-memory systems consist of several compute nodes, each with its own main memory, connected into a computing cluster by communication links (e.g., Ethernet or InfiniBand).
Most modern high-performance computing systems have a heterogeneous architecture comprising a general-purpose multi-core Central Processing Unit (CPU) and a set of various accelerators~\cite{Chakroun2013combining,Kolpakov2019,Semenov2020}.
The performance of heterogeneous systems is increased not only by increasing the number of processors and cores, but also by incorporating (either integrated or discrete) specialized accelerators for parallel computing, such as Graphics Processing Units (GPUs), Digital Signal Processors (DSPs), and Field-Programmable Gate Arrays (FPGAs)~\cite{Sluga2014heterogeneous}.
Traditional approaches to developing parallel programs for CPU-based systems are not sufficient here.
Efficient implementation of algorithms on heterogeneous systems is possible only if the specific features and limitations of their architecture are taken into account, including the interaction between the CPU and the accelerator, the memory hierarchy and the rules for using it efficiently, and the orientation towards massively parallel data processing.

Thus, because of the high computational complexity of practically important optimization problems and the rapidly growing availability of high-performance computing systems of various kinds, the development of parallel metaheuristic algorithms is becoming increasingly important.

This book is devoted to the development and study of parallel discrete optimization algorithms, the tuning of their parameters, and their efficient implementation on various parallel computing systems, including distributed-memory systems, shared-memory systems, and heterogeneous systems equipped with GPUs.

As a test problem, we use the optimal selection of equipment for chemical-engineering systems (CES)~\cite{Sandoval2017}: among all available alternatives, one has to find the combination of equipment units for which the objective function (the total capital cost of the equipment) attains its minimum.

The book is organized as follows.
Chapter~\ref{chapter1} states the problem.
Chapter~\ref{chapter2} presents a sequential algorithm for solving it based on the branch-and-bound (B\&B) method.
Chapter~\ref{chapter3} presents parallel implementations of the algorithm for shared-memory and distributed-memory systems.
Chapter~\ref{chapter4} is devoted to the implementation of branch-and-bound on heterogeneous systems equipped with GPUs.
Chapter~\ref{chapter5} develops a parallel hybrid metaheuristic algorithm for the problem and its implementation on heterogeneous systems.
Chapter~\ref{chapter6} deals with the selection and optimization of the parameters of the hybrid metaheuristic algorithm.

%%%% Chapter 1
\chapter{Optimal Equipment Selection for Chemical-Engineering Systems: Problem Statement}
\label{chapter1}

Selecting the equipment of a chemical-engineering system (CES) is one of the main problems that arise both in designing new~\cite{Borisenko2014hierarchy} and in retrofitting existing~\cite{karpushkin2012} multiproduct batch plants.

Such plants are characterized by relatively small production volumes, a wide range of products, high product prices, equipment operating in batch or semi-continuous mode and, accordingly, production in batches, the ability to switch quickly from one product to another using the same set of equipment (high flexibility), and a large number of processing stages.

Typical examples of multiproduct batch plants are the production of chemical reagents, pharmaceuticals, cosmetics, dyes, and intermediates~\cite{kafarov1990,fletcher1991flexible}.

At each processing stage, the materials are processed for a certain time and then transferred to the next stage.

Batches of each product must pass successively through the equipment of all production stages, and the units of each stage must be able to receive and process a batch completely.
For example, a batch of material from a unit with a volume of $10$~m$^3$ cannot be transferred to a unit with a working volume of $0.1$~m$^3$.
If such a situation arises, the CES is not operable.

The operating cycle of a unit is an ordered sequence of main and auxiliary operations of finite duration.
The most common of them are loading, the main processing operations, unloading, and cleaning with preparation for the next cycle.
In addition, a unit may be idle during operation.

The processing time at a stage depends on the main size (working volume or working surface area) of the unit in which the processing takes place, because a change in the unit size changes the amount of material processed in it.
Therefore, whenever the main sizes of the equipment change, the production time $T$ has to be recalculated.
The plant being designed must produce the required amount of products $Q$ within a given time horizon $T_{max}$.

A solution of the multiproduct plant design problem is a CES variant, i.e., the number of main units at each stage of the CES and the main size (working volume or working surface area) of each unit, together with the operating parameters (product batch sizes and production periods) that ensure the required production rate for all products.
The main size is usually chosen from a set of standard values.

One has to find the optimal combination of the number and main sizes of units at the stages of the CES among the set of possible variants, together with the operating parameters of a system with the given production rate; the optimality criterion may be, for example, the total capital cost of the equipment~\cite{malygin2002,malygin2003,karpushkin2011}.
The solution of this problem largely determines the quality of design decisions on equipment layout and production scheduling.
In this book, we consider the selection of the main equipment of the CES stages.
The selection of auxiliary equipment (measuring tanks, receivers, pumps, etc.) is a separate problem, which is solved using the results of the main equipment selection~\cite{MalyginKarpushkinMikhailova2011,Borisenko2012auxiliary}.

The problem under consideration belongs to the class of Mixed-Integer Nonlinear Programming (MINLP) problems~\cite{Hamzaoui_2010,Ponsich_2007,Chu2014}.
In a number of works, it is reduced to a Mixed-Integer Linear Programming (MILP) problem~\cite{Sandoval20161,Fumero2013}.

For each operable CES variant, the operating parameters of the system are determined uniquely.
Therefore, the search for the optimal CES variant can be regarded as a discrete optimization problem.
Finding the global optimum by exhaustive enumeration of all variants is impractical because of the large size of the problem.

As the size of the search space grows (for this problem, as the number of available standard equipment sizes increases), the number of combinations to be examined (all possible system variants $E$) grows exponentially because of the combinatorial explosion, so that even the most powerful computers may require an unacceptable amount of time to find a solution.

Various approaches to this problem have been proposed in the literature, for example, Monte Carlo methods, genetic algorithms, heuristic methods, and the reduction of the original problem to a nonlinear or linear programming problem~\cite{Ponsich_2007,Borisenko2014hierarchy,Hamzaoui_2010,lin2001,Seid2013design}.
Heuristic approaches aim at finding acceptable (suboptimal) solutions, whereas mathematical programming methods guarantee exact results.
However, as the problem size grows, exact methods face the combinatorial explosion, which makes them inefficient.
In this book, we propose parallel algorithms for the equipment selection problem for CES of multiproduct batch plants and study their applicability to finding the optimal solution.

The chapter is organized as follows.
Section~\ref{chap1_sect1} gives the mathematical formulation of the optimal CES equipment selection problem and describes the representation of the search space as a tree of variants.
Section~\ref{chap1_sect_model} considers the mathematical model of CES operation, which determines the form of the constraints on the operability and on the production time of the system as well as of the optimality criterion, and analyzes the computational properties of the model that are essential for designing sequential and parallel algorithms.
Section~\ref{chap1_sect2} describes the test case used in the computational experiments throughout the subsequent chapters.

\section{Mathematical Formulation of the Problem}
\label{chap1_sect1}

A CES of a multiproduct batch plant is a set of processing stages equipped with reactors, tanks, filters, dryers, and other process equipment, and intended to carry out the production steps for a given range of products.
Assuming that one main unit can be installed at each processing stage of the CES, the problem can be formulated as follows~\cite{Borisenko2011,borisenko2012}.

A CES is a sequence of $I$ processing stages.
Each stage $i$ of the system can be equipped with a main unit chosen from a finite set $X_i$ containing $J_i$ elements.
The set $X_i$ comprises all possible main sizes of the units for stage $i$: $X_i=\{x_{i,j}\}, i=\overline{1,I}, j=\overline{1,J_i}$, where $x_{i,j}$ is the main size (working volume or working surface area) of the $j$th unit that can be installed at stage $i$ of the CES.

Each CES variant $\Omega_e, e=\overline{1,E}$, where $E=\prod_{i=1}^I J_i$, is an ordered set of main sizes of the units chosen from the corresponding sets.

\begin{figure}[htb]
	\centering
	\includegraphics[width=0.97\textwidth]{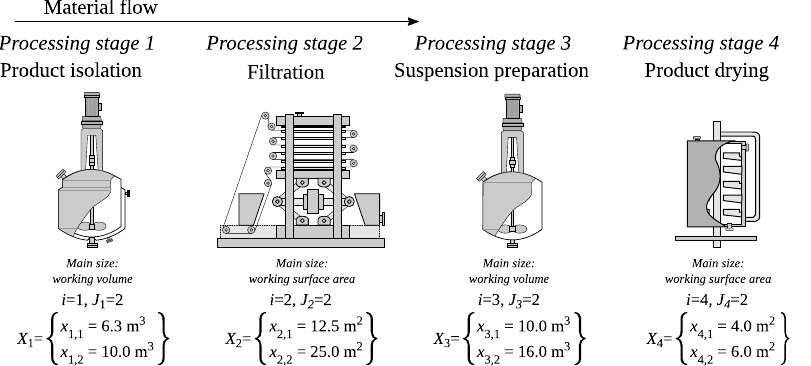}
	\caption{Example of a simple chemical-engineering system}
	\label{fig:simple_CES}
\end{figure}

Figure~\ref{fig:simple_CES} shows an example of a simple CES consisting of $4$ stages ($I=4$).

For each processing stage of the CES, two standard unit sizes are available, i.e., $J_1=J_2=J_3=J_4=2$.
Here $X_1=\{x_{1,1},x_{1,2}\}$, $X_2=\{ x_{2,1},x_{2,2}\}$, $X_3=\{ x_{3,1},x_{3,2}\}$, $X_4=\{ x_{4,1},x_{4,2}\}$.
The total number of possible CES variants is $E=J_1 \cdot J_2 \cdot J_3 \cdot J_4=2^4=16$.

The order in which product batches are processed by the main units of the CES stages is fixed (it is determined in advance when the structure of the material flows is designed); for example, the variants $\{x_{1,1},x_{2,1},x_{3,2},x_{4,2}\}$ and $\{x_{1,2},x_{2,1},x_{3,1},x_{4,1}\}$ are admissible, whereas the variants $\{x_{3,1},x_{2,1},x_{4,1}, x_{1,2}\}$ and $\{x_{2,2},x_{4,2},x_{3,1},x_{1,1}\}$ are obviously inadmissible.

Every operable CES variant must satisfy the conditions for the joint operation of the equipment of its stages, $S(\Omega_e)=0$ (the \emph{compatibility constraint}).
A CES variant is operable if the units of every one of its processing stages, without exception, are able to process the batch volume of the product, taking into account the maximum and minimum filling degrees of the units.

In addition, an operable CES variant must ensure the production of the given product range and amounts within a given period of time, i.e., $T(\Omega_e)\leq T_{max}$ (the \emph{processing time constraint}), where $T_{max}$ is the given time horizon.

Thus, the CES equipment selection problem can be formulated as follows: on the set of available process equipment (\ref{eq_2})--(\ref{eq_3}), find a CES variant $\Omega^*\in \Omega_e, e=\overline{1,E}$, for which the optimality criterion, the capital cost $Cost(\Omega_e)$ (\ref{eq_1}), attains its minimum and the compatibility constraint (\ref{eq_4}) and the processing time constraint (\ref{eq_5}) are satisfied:

\begin{equation}
\label{eq_1}
\Omega^*=\textit{argmin}\ \textit{Cost}(\Omega_e), e=\overline{1,E}
\end{equation}
\begin{equation}
\label{eq_2}
\Omega_e = \{x_{1,j_1},x_{2,j_2},\ldots, x_{I,j_I}|j_i=\overline{1,J_i},i=\overline{1,I}\}, e=\overline{1,E}
\end{equation}
\begin{equation}
\label{eq_3}
x_{i,j}\in {X_i},\,i=\overline{1,I}, j=\overline{1,J_i}
\end{equation}
\begin{equation}
\label{eq_4}
S(\Omega_e)=0, e=\overline{1,E}
\end{equation}
\begin{equation}
\label{eq_5}
T(\Omega_e) \leq T_{max}, e=\overline{1,E}
\end{equation}

\begin{figure}[htb]
	\centering
	\includegraphics[width=0.97\textwidth]{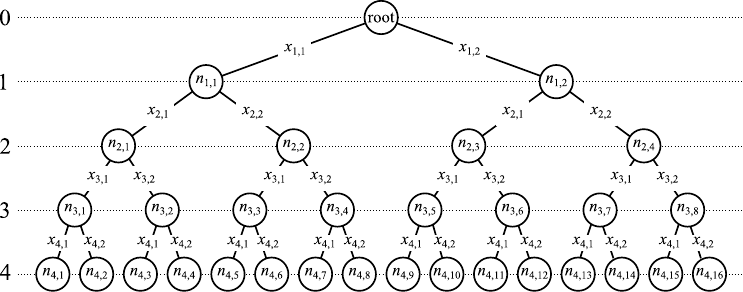}
	\caption{Search space}
	\label{fig:search_tree}
\end{figure}

The search space for the CES shown in Fig.~\ref{fig:simple_CES} is a tree consisting of $I$ levels (see Fig.~\ref{fig:search_tree}).

Each level $i$ of the tree corresponds to a stage of the CES.
Each edge $x_{i,j}$ of the tree corresponds to the selected unit variant $j$ for stage $i$ (one selected element of the set $X_i=\{x_{i,j}\},\,i=\overline{1,I}, j=\overline{1,J_i}$).
Each internal node $n_{i,k},\,i=\overline{1,I}, k=\overline{1,K_i}, K_i=\prod_{l=1}^i(J_l)$ of the tree corresponds to a fragment of the CES consisting of stages $1$ to $i$.

Note that the problem under consideration is NP-hard~\cite{Ponsich_2007,DeMiranda2019,Borisenko2019}.

The form of the functions $S(\Omega_e)$ and $T(\Omega_e)$, as well as of the optimality criterion, is determined by the mathematical model of CES operation discussed in Section~\ref{chap1_sect_model}.

\section{Mathematical Model of CES Operation}
\label{chap1_sect_model}

In the problem formulation (\ref{eq_1})--(\ref{eq_5}), the compatibility and processing time constraints are written in terms of the functions $S(\Omega_e)$ and $T(\Omega_e)$, whose specific form is determined by the mathematical model of system operation~\cite{Malygin2005mathematical}.
The model is presented here to the extent needed to understand the computational side of the problem: which quantities are computed when a single CES variant is evaluated, which of these computations can be performed for incomplete variants (internal nodes of the search tree), and which data remain unchanged throughout the search.
The process engineering justification of the model relations and the procedure for preparing the input data from process specifications are beyond the scope of this book; they are described in detail in~\cite{Malygin2005mathematical}.
Models of the operation of multiproduct batch plants are widely represented in the literature~\cite{Rippin1983,BarbosaPovoa2007}.
In the most common formulation, each stage is characterized by a size factor (a counterpart of the material index used below) and a batch processing time, the production rate of the system is determined by the limiting cycle time, and the time constraint is written as the sum, over all products, of the number of batches multiplied by the corresponding limiting cycle time~\cite{Kocis1988,RavemarkRippin1996}.
Within this formulation, the literature considers parallel units operating in phase or out of phase~\cite{Montagna2000}, intermediate storage tanks and discrete sets of standard equipment sizes~\cite{RavemarkRippin1996}, as well as units duplicated in series~\cite{Moreno2009}.
The model used in this book~\cite{Malygin2005mathematical} is detailed down to the individual operations of the unit cycles: it explicitly accounts for the overlap of loading and unloading with the processing operations, for the splitting and merging of batches at particular stages, and for the resulting ``filled idle times'' of the equipment.
The most complete description of the model and of the methods for solving its relations is given in~\cite{MalyginKarpushkin1999,Karpushkin2006,Karpushkin2007diss}.

The notation is consistent with Section~\ref{chap1_sect1}: $i=\overline{1,I}$ is the index of a processing stage of the CES, $j=\overline{1,J_i}$ is the index of a standard unit size, and $\Omega_e$ is a CES variant.
The main size of the unit selected for stage $i$ in the variant $\Omega_e=\{x_{1,j_1},\ldots,x_{I,j_I}\}$ will be denoted by $\hat{x}_i=x_{i,j_i}$.

\subsection{Input Data and Assumptions}
\label{chap1_model_data}

The CES manufactures a range of $P$ products, $p=\overline{1,P}$.
For each product, the production amount $Q_p$~(t) is given; the total time available for producing the whole product range is $T_{max}$.

Product $p$ is processed at a subset of processing stages $I_p\subseteq\{1,\ldots,I\}$.
The processing order is defined by the \emph{process route}, a directed graph $\Gamma_p$ whose vertices are the stages in $I_p$ and whose arcs correspond to the transfer of materials from stage to stage.
Let $\mathcal{P}_{p,i}$ denote the set of stages from which materials are fed to stage $i$ when product $p$ is produced; stages with $\mathcal{P}_{p,i}=\varnothing$ will be called \emph{initial} stages.
Dye production typically has routes in the form of a converging tree: several independent branches in which raw materials and intermediates are prepared merge at the stages of chemical transformation, and the route ends with a single final stage.
For example, in the test case (Section~\ref{chap1_sect2}), when the first product is manufactured, stage~110 (azo coupling) receives materials from stages~103 and~106, and stage~145 (drying) is the final stage for all three products.

According to the type of the main unit, the set of stages is partitioned into four disjoint classes:
$\mathcal{A}$, batch vessels (reactors and tanks equipped with agitators);
$\mathcal{B}$, filters and dryers, in which the processing time is determined by the amount of material processed and by the working surface area (or volume) of the unit;
$\mathcal{C}$, chamber and plate-and-frame filter presses whose target product is the solid phase of the suspension;
$\mathcal{D}$, batch vacuum dryers.

Stage $i$ may be equipped with $\nu_i\geq 1$ identical units of the same standard size $\hat{x}_i$; the number $\nu_i$ is part of the input data and is not an optimization variable.
For $\nu_i>1$, the units operate either \emph{out of phase} (each unit processes a whole batch in turn) or \emph{in phase} (a batch is divided equally among the units).

The model is based on the following assumptions.
\begin{enumerate}
	\item At any moment, the CES manufactures only one product; the products of the range are manufactured one after another.
	\item The operating cycle of a unit at any stage is an ordered sequence of operations of six types: loading, ``filled idle time'' during loading, physicochemical transformations (hereafter referred to as \emph{processing}), unloading, ``filled idle time'' during unloading, and cleaning (preparation for the next cycle).
	A ``filled idle time'' is a period during which the unit stays partially loaded or partially unloaded, waiting for the next portion of material or for the receiving unit to become ready.
	\item Loading and unloading are either performed as separate operations or overlap in time with processing (for example, when a suspension is fed continuously to a filter).
	\item The duration of each operation is either constant or a known function of the amount of material processed and of the main size of the unit; it does not depend on the batch number or on the cycle number.
	\item At particular stages, a batch may be divided into equal portions processed in succession, or several batches may be merged and processed together.
\end{enumerate}

Note that the FPAKM automatic chamber filter presses installed at the filtration stages of the test CES (stages~$108$, $111$, and~$134$) belong to class~$\mathcal{B}$ despite their name: they operate cyclically like ordinary filters (suspension feeding, filtration, cake discharge, filter cloth regeneration), and the filtration time is determined by their working surface area and the batch size.
In the model, loading and unloading of these units overlap with filtration, while the time for cake discharge and cloth regeneration is accounted for by the fraction $1-h_{p,i}$ of auxiliary operations (see~(\ref{eq:m_dc})).
The test case contains no stages of classes~$\mathcal{C}$ and~$\mathcal{D}$.

The way in which a batch of product $p$ is processed at stage $i$ is specified by the \emph{batch size ratio} $r_{p,i}$: for $r_{p,i}=1$, the batch is processed as a whole; for $r_{p,i}=\gamma>1$, it is divided into $\gamma$ equal portions; for $r_{p,i}=1/\gamma$, $\gamma>1$, $\gamma$ batches are merged at the stage.
The fraction of a batch that resides in one unit of the stage at a time is given by the coefficient
\begin{equation}
	\label{eq:m_u}
u_{p,i}=
\begin{cases}
	1/r_{p,i}, & \nu_i=1 \text{ or out-of-phase operation},\\
	1/(\nu_i\, r_{p,i}), & \text{in-phase operation}.
\end{cases}
\end{equation}

The process specification for product $p$ provides the following characteristics for each stage $i\in I_p$:
$v_{p,i}$ and $\mu_{p,i}$, the volumetric (m$^3$/t) and mass (kg/t) material indices, i.e., the volume and mass of materials processed at the stage per tonne of final product;
$\varphi^{\min}_{p,i}$ and $\varphi^{\max}_{p,i}$, the lower and upper bounds of the admissible filling degree of the unit;
$\tau_{p,i}$, the processing time in a vessel;
$\sigma_{p,i}$, the specific capacity of a filter or dryer;
$h_{p,i}$, the fraction of main operations in the total time during which a filter or dryer is occupied by a batch;
$\delta_{p,i}$, the admissible cake thickness in a filter press;
and the durations of loading, unloading, and cleaning of a vessel.

\subsection{Batch Size and the Compatibility Constraint}
\label{chap1_model_batch}

Each stage $i\in I_p$ with the selected unit $\hat{x}_i$ can process a batch of product $p$ only within a certain range of sizes $w^{\min}_{p,i}\leq w_p\leq w^{\max}_{p,i}$.
For vessels, the bounds of this range are determined by the admissible filling degrees; for filter presses, by the admissible cake thickness:
\begin{equation}
	\label{eq:m_wmin}
	w^{\min}_{p,i}(\hat{x}_i)=
	\begin{cases}
		\dfrac{\varphi^{\min}_{p,i}\,\hat{x}_i}{u_{p,i}\,v_{p,i}}, & i\in\mathcal{A},\\[2ex]
		0, & i\in\mathcal{B}\cup\mathcal{C}\cup\mathcal{D},
	\end{cases}
\end{equation}
\begin{equation}
	\label{eq:m_wmax}
	w^{\max}_{p,i}(\hat{x}_i)=
	\begin{cases}
		\dfrac{\varphi^{\max}_{p,i}\,\hat{x}_i}{u_{p,i}\,v_{p,i}}, & i\in\mathcal{A}\cup\mathcal{D},\\[2ex]
		\dfrac{\delta_{p,i}\,\hat{x}_i}{u_{p,i}\,v_{p,i}}, & i\in\mathcal{C},\\[2ex]
		+\infty, & i\in\mathcal{B}.
	\end{cases}
\end{equation}
For stages of class $\mathcal{B}$, the batch size is not limited directly: the main size of the unit determines not the admissible amount of material but the time needed to process it (see Section~\ref{chap1_model_durations}).

For example, in the CES shown in Fig.~\ref{fig:simple_CES}, the product isolation and suspension preparation stages are equipped with vessels (class~$\mathcal{A}$), whose main size is the working volume, while the filtration and drying stages (class~$\mathcal{B}$) are equipped with units whose main size is the working surface area.
This sequence of stages (isolation, filtration, slurrying, drying) forms the final section of the routes of all three products in the test case (stages~$122$, $134$, $138$, and~$145$; Fig.~\ref{fig:test_CES}).

A batch of product $p$ must be processed by the units of all stages of the route; therefore, the admissible batch sizes form the intersection of the intervals defined by the individual stages.
Consider a fragment of a CES variant comprising stages $1$ to $i$ (which corresponds to a node at level $i$ of the search tree, Fig.~\ref{fig:search_tree}) and introduce the quantities
\begin{equation}
	\label{eq:m_wfragment}
	w^{\min}_p(\Omega_e,i)=\max_{l\in I_p,\ l\leq i} w^{\min}_{p,l}(\hat{x}_l), \qquad
	w^{\max}_p(\Omega_e,i)=\min_{l\in I_p,\ l\leq i} w^{\max}_{p,l}(\hat{x}_l).
\end{equation}
The fragment is operable if the corresponding interval is nonempty for every product.
We define the function
\begin{equation}
	\label{eq:m_S}
	S(\Omega_e,i)=\sum_{p=1}^{P}\bigl[\,w^{\min}_p(\Omega_e,i)>w^{\max}_p(\Omega_e,i)\,\bigr],
\end{equation}
where $[\,\cdot\,]$ is the indicator function, equal to one if the condition in brackets holds and to zero otherwise.
The value $S(\Omega_e,i)$ equals the number of products for which there is no batch size admissible simultaneously at all stages $1$ to $i$ of the fragment.
For $i=I$, the fragment coincides with the complete CES variant; hence the function $S(\Omega_e)$ in~(\ref{eq_4}) is a special case of the function just introduced: $S(\Omega_e)=S(\Omega_e,I)$.

If $S(\Omega_e)=0$, then constraint~(\ref{eq_4}) is satisfied and the CES variant is operable (for every product, there exists a batch size admissible at all stages of its process route), so it makes sense to proceed with checking the processing time constraint (\ref{eq_5}).
A nonzero value of $S(\Omega_e)$ means that the variant is not operable, and the value itself is the number of products whose batches cannot be processed.

Function (\ref{eq:m_S}) has a property that is fundamental for the branch-and-bound method: when another stage is added to the fragment, the maximum in (\ref{eq:m_wfragment}) can only increase and the minimum can only decrease, i.e., the interval of admissible batch sizes can only shrink.
Consequently,
\begin{equation}
	\label{eq:m_Smonotone}
	S(\Omega_e,i)\leq S(\Omega_e,i+1),\quad i=\overline{1,I-1},
\end{equation}
and if the fragment corresponding to a node of the search tree is not operable, then all CES variants in the subtree of this node are not operable either.
It is precisely this property that justifies checking \texttt{S(W,~i)} at the internal nodes of the tree in the algorithms considered in the subsequent chapters.

If a CES variant is operable, the batch size of each product is set to the largest admissible value:
\begin{equation}
	\label{eq:m_w}
	w_p=w^{\max}_p(\Omega_e,I),\quad p=\overline{1,P}.
\end{equation}
The upper bound is chosen because the number of system cycles needed to produce the required amount of product is inversely proportional to the batch size (see (\ref{eq:m_Z})), while the processing times in vessels do not depend on it.
This makes precise the statement made at the beginning of the chapter that, for an operable CES variant, the operating parameters of the system are determined uniquely.
Note that in design problems where the unit sizes are not known in advance, batch sizes are chosen differently, namely, so as to load the equipment of the coupled stages as evenly as possible~\cite{Malygin2005mathematical}; in the formulation considered here, where CES variants are composed of standard sizes, no such adjustment is needed.

For example, for the solution given in Table~\ref{tab:CESSolution}, when the first product is manufactured, the lower bound of the batch size range is determined by stage~138 (slurrying) and the upper bound by stage~115 (diazotization); the range is $[2.459,\ 4.437]$~t, and according to (\ref{eq:m_w}) we take $w_1=4.437$~t.

\subsection{Operation Durations}
\label{chap1_model_durations}

Let $d^{\mathrm{w}}_{p,i}$ denote the processing time of a batch of product $p$ in a unit of stage $i$.
For vessels, it is given by the process specification; for filters and dryers, it is determined by the amount of material in the unit and by the unit capacity:
\begin{equation}
	\label{eq:m_dw}
	d^{\mathrm{w}}_{p,i}=
	\begin{cases}
		\tau_{p,i}, & i\in\mathcal{A},\\[1ex]
		\dfrac{h_{p,i}\,g_{p,i}\,u_{p,i}\,w_p}{\sigma_{p,i}\,\hat{x}_i}, & i\in\mathcal{B},\\[2ex]
		\dfrac{h_{p,i}\,\mu_{p,i}\,\delta_{p,i}}{v_{p,i}\,\sigma_{p,i}}, & i\in\mathcal{C},\\[2ex]
		\dfrac{h_{p,i}\,\mu_{p,i}\,u_{p,i}\,w_p}{\sigma_{p,i}\,F(\hat{x}_i)}, & i\in\mathcal{D},
	\end{cases}
\end{equation}
where $g_{p,i}$ is either the mass material index $\mu_{p,i}$ or the volumetric material index $v_{p,i}$ (depending on whether the specific capacity $\sigma_{p,i}$ refers to the mass or to the volume of material), and $F(\hat{x}_i)$ is the heat-transfer area of a vacuum dryer with working volume $\hat{x}_i$, taken from the equipment catalogue.
For filter presses, the processing time depends neither on the batch size nor on the unit size, since filtration continues until a cake of the given thickness has formed.

The cleaning (regeneration) time of filters and dryers is proportional to the duration of the main operations:
\begin{equation}
	\label{eq:m_dc}
	d^{\mathrm{c}}_{p,i}=d^{\mathrm{w}}_{p,i}\,\frac{1-h_{p,i}}{h_{p,i}},\quad i\in\mathcal{B}\cup\mathcal{C}\cup\mathcal{D};
\end{equation}
for vessels, the durations of loading, unloading, and cleaning are given by the process specification.
In the test case (Section~\ref{chap1_sect2}), only the processing times are specified for vessels, and the initial durations of the other operations are set to zero.
The actual durations of loading and unloading are then determined when the schedule is computed, as a result of their alignment with the coupled operations of the neighboring stages.

The transfer of a portion of material from one stage to another is a pair of simultaneous operations: unloading from the unit of the source stage and loading into the unit of the receiving stage.
Both operations of the pair must start and end at the same moments; therefore, their durations are aligned to the larger of the two.
If loading at the receiving stage overlaps with processing (for example, a suspension is fed continuously to a filter), unloading from the unit of the source stage continues throughout the filtration.

Thus, in the CES of Fig.~\ref{fig:simple_CES}, the suspension is unloaded from the unit of the product isolation stage throughout the filtration; hence the duration of this unloading is determined by the filter capacity, i.e., by its working surface area $\hat{x}_2$ and by the batch size.

Expressions (\ref{eq:m_wmax}), (\ref{eq:m_w}), and (\ref{eq:m_dw}) show that a CES variant affects the production time in two ways: through the batch sizes (and hence the number of system cycles), which depend on the unit sizes at all stages, and through the processing times at the stages of classes $\mathcal{B}$ and $\mathcal{D}$, which depend both on the sizes of the corresponding units and on the batch size.

\subsection{Schedule of a System Cycle}
\label{chap1_model_schedule}

A \emph{system cycle} in the production of product $p$ is the time interval from the start of the first operation at the initial stages of the route to the end of the last operation at the final stage.
During one system cycle,
\begin{equation}
	\label{eq:m_b}
	b_p=\frac{1}{\min\limits_{i\in I_p} r_{p,i}}
\end{equation}
batches of the product are produced, and the units of stage $i$ perform $\kappa_{p,i}=r_{p,i}\,b_p$ cycles of their own.
If the ratios $r_{p,i}$ of neighboring stages differ, several loadings or unloadings take place during one unit cycle; for a stage $i$ that receives materials from stage $i'\in\mathcal{P}_{p,i}$ and passes them on to stage $i''$, their numbers are
\begin{equation}
	\label{eq:m_L}
	L^{\mathrm{l}}_{p,i}=\max\left\{1,\frac{r_{p,i'}}{r_{p,i}}\right\},\qquad
	L^{\mathrm{u}}_{p,i}=\max\left\{1,\frac{r_{p,i''}}{r_{p,i}}\right\}.
\end{equation}
The number of portions transferred from the source stage during a system cycle must equal the number of portions received by the receiving stage:
\begin{equation}
	\label{eq:m_balance}
	\kappa_{p,i'}\,L^{\mathrm{u}}_{p,i'}=\kappa_{p,i}\,L^{\mathrm{l}}_{p,i},\quad i'\in\mathcal{P}_{p,i}.
\end{equation}

Let $\mathcal{O}_p$ denote the set of all operations of all unit cycles at all stages of the route during one system cycle.
Each operation $o\in\mathcal{O}_p$ has a type, a duration $d_o$, a start time $t^{\mathrm{s}}_o$, and a finish time $t^{\mathrm{f}}_o=t^{\mathrm{s}}_o+d_o$.
Operations are related in two ways: by the process sequence within a unit cycle (for an operation $o$, let $o^{-}$ denote the immediately preceding operation of the same cycle) and by the material link between stages (for a loading operation $o$, let $o^{\ast}$ denote the coupled unloading operation at the source stage).
The start times of the operations are determined recursively:
\begin{equation}
	\label{eq:m_ts}
	t^{\mathrm{s}}_o=
	\begin{cases}
		t^{\mathrm{f}}_{o^{-}}, & \text{if } o^{-} \text{ exists},\\
		t^{\mathrm{s}}_{o^{\ast}}, & \text{if } o \text{ is the first in its cycle and } o^{\ast} \text{ exists},\\
		\theta_c, & \text{if } o \text{ is the first in cycle } c \text{ of an initial stage},
	\end{cases}
\end{equation}
where $\theta_c$ is the start time of cycle $c$ of the units of the initial stage (in the initial schedule, $\theta_c=0$).

From the data-structure point of view, the schedule of a system cycle is a weighted directed acyclic graph whose vertices are the operations, whose arcs are the relations ``$o^{-}\to o$'' and ``$o^{\ast}\to o$'', and whose vertex weights are the operation durations.
The start and finish times of all operations are computed in a single pass over this graph in an order consistent with the direction of the arcs.

\begin{figure}[htb]
	\centering
	\includegraphics[width=\textwidth]{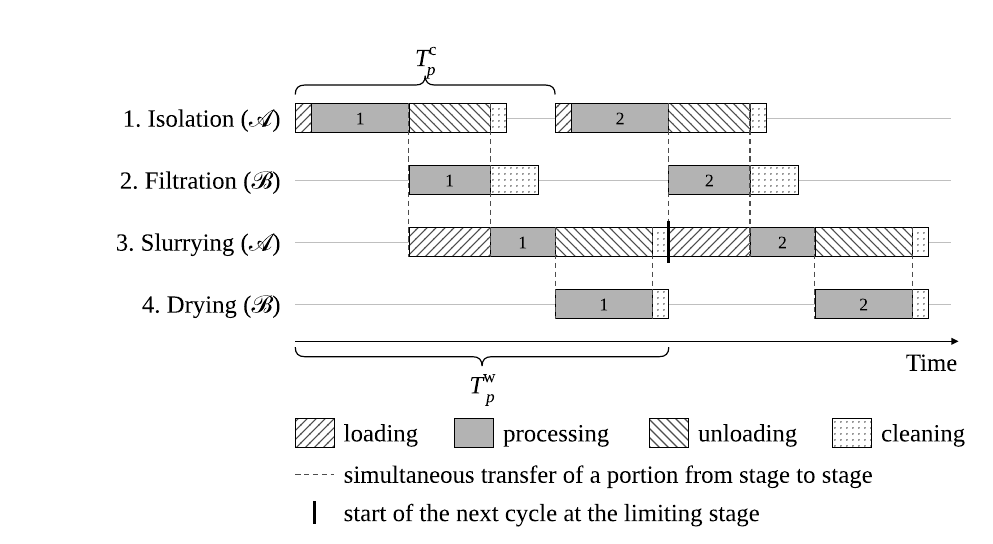}
	\caption{Gantt chart of two CES operating cycles}
	\label{fig:gantt_cycle}
\end{figure}

Figure~\ref{fig:gantt_cycle} schematically shows the schedule of two consecutive operating cycles of the CES from Fig.~\ref{fig:simple_CES} for $r_{p,i}=1$ and $\nu_i=1$ (the numbers on the processing bars are the system cycle numbers).
At the filtration and drying stages (class~$\mathcal{B}$), loading and unloading overlap with processing.
Therefore, unloading from the unit of the product isolation stage and loading of the unit of the suspension preparation stage are performed simultaneously with filtration, and unloading of the suspension is performed simultaneously with drying (dashed lines in the figure).
As a consequence, the unit of the product isolation stage is occupied longer than its processing operation lasts: its occupation time also includes the filtration time.

The durations of ``filled idle times'' are not part of the input data; they are determined by the schedule.
A ``filled idle time'' during loading occurs between two consecutive loadings of a unit if the portions arrive at intervals; its duration equals the gap between the end of one unloading operation and the start of the next one at the source stage and is fully determined by the preceding stages.
A ``filled idle time'' during unloading is more difficult to determine: its duration depends on the operations not only of the immediately following stage but also of all subsequent stages.
It is introduced to satisfy the non-overlap condition for cycles: the next cycle of a unit cannot start before the previous one has finished,
\begin{equation}
	\label{eq:m_nocollision}
	t^{\mathrm{s}}_{\mathrm{first}(c')}\geq t^{\mathrm{f}}_{\mathrm{last}(c)},
\end{equation}
where $\mathrm{first}(c)$ and $\mathrm{last}(c)$ are the first and last operations of cycle $c$, and $c'$ is the index of the next cycle of the same unit: $c'=c+\nu_i$ for $\nu_i>1$ units operating out of phase and $c'=c+1$ otherwise.

The durations of ``filled idle times'' during unloading are determined by an iterative algorithm that minimizes the duration of the system cycle~\cite{Malygin2005mathematical}:
\begin{enumerate}
	\item The initial schedule (\ref{eq:m_ts}) is constructed with zero durations of the ``filled idle times'' during unloading.
	\item The stages of the route are examined in turn, and condition (\ref{eq:m_nocollision}) is checked for each cycle $c$.
	If it is violated, the overlap $\Delta=t^{\mathrm{f}}_{\mathrm{last}(c)}-t^{\mathrm{s}}_{\mathrm{first}(c')}>0$ is computed.
	\item The start of cycle $c'$ must be delayed by $\Delta$.
	To this end, moving from the first operation of cycle $c'$ against the direction of the material flow, one finds the nearest ``filled idle time'' during unloading at one of the preceding stages after which a portion of material is transferred to cycle $c'$, and increases its duration by $\Delta$: the material is held longer in the unit of the source stage.
	If no such idle time exists and the search has reached the initial stage, the start time $\theta$ of the corresponding cycle of the initial stage is increased by $\Delta$.
	\item The durations of the ``filled idle times'' during loading and the start times of the operations are recomputed, after which the check is resumed from step~2.
\end{enumerate}
The algorithm terminates when condition (\ref{eq:m_nocollision}) holds for all cycles at all stages of the route.

If $r_{p,i}=1$ at all stages of the route, then $b_p=1$ and $\kappa_{p,i}=1$, i.e., during a system cycle each unit performs exactly one cycle with one loading and one unloading; in this case, no ``filled idle times'' arise and the iterative part of the algorithm is not executed.
This is exactly the situation in the test case of Section~\ref{chap1_sect2}. Nevertheless, the software implementation supports the model in full.

\subsection{Production Time and the Processing Time Constraint}
\label{chap1_model_time}

The completed schedule yields the main time characteristics of the CES operating mode in the production of product $p$.
The duration of a system cycle is
\begin{equation}
	\label{eq:m_Tw}
	T^{\mathrm{w}}_p=\max_{o\in\mathcal{O}_p} t^{\mathrm{f}}_o-\min_c\theta_c.
\end{equation}

Consecutive system cycles overlap: the next cycle starts as soon as the unit that limits the production rate becomes free.
The time during which the units of stage $i$ are occupied with the materials of one system cycle is
\begin{equation}
	\label{eq:m_tpi}
	t_{p,i}=
	\begin{cases}
	\dfrac{\kappa_{p,i}}{\nu_i}\left(t^{\mathrm{f}}_{\mathrm{last}(1)}-t^{\mathrm{s}}_{\mathrm{first}(1)}\right), & \nu_i>1,\ \text{out-of-phase operation},\\[2ex]
	t^{\mathrm{f}}_{\mathrm{last}(\kappa_{p,i})}-t^{\mathrm{s}}_{\mathrm{first}(1)}, & \text{otherwise},
\end{cases}
\end{equation}
and the \emph{cycle time}, i.e., the minimum possible interval between the starts of consecutive system cycles, is determined by the limiting stage:
\begin{equation}
	\label{eq:m_Tc}
	T^{\mathrm{c}}_p=\max_{i\in I_p} t_{p,i}.
\end{equation}
Installing several units operating out of phase at a stage reduces its occupation time by a factor of $\nu_i$; in the test case, this applies to stage~138 (slurrying), which is equipped with two units.

The number of system cycles required to produce the amount $Q_p$ of product $p$ and the production time of this product are
\begin{equation}
	\label{eq:m_Z}
	Z_p=\left\lceil \frac{Q_p}{b_p\,w_p}\right\rceil,
\end{equation}
\begin{equation}
	\label{eq:m_Tp}
	T_p=T^{\mathrm{w}}_p+(Z_p-1)\,T^{\mathrm{c}}_p.
\end{equation}
Since the products are manufactured one after another, the production time of the whole product range is the sum of the production times of the individual products, and constraint (\ref{eq_5}) takes the form
\begin{equation}
	\label{eq:m_T}
	T(\Omega_e)=\sum_{p=1}^{P} T_p\leq T_{max}.
\end{equation}

Unlike the function $S$, the function $T$ cannot be computed for an incomplete CES variant: the batch sizes (\ref{eq:m_w}) are determined by the units of all stages of the route, so that until units have been selected for all stages, neither the operation durations (\ref{eq:m_dw}) nor the number of cycles (\ref{eq:m_Z}) are defined.
Moreover, $T$ has no monotonicity property similar to (\ref{eq:m_Smonotone}): increasing the unit size at one of the stages may increase the batch size and reduce the number of cycles, but at the same time increase the processing time at the stages of classes $\mathcal{B}$ and $\mathcal{D}$ and thereby the cycle time.
Therefore, constraint (\ref{eq_5}) is checked only for complete CES variants, which are formed at the leaves of the search tree.

\subsection{Optimality Criterion}
\label{chap1_model_cost}

Let $c_{i,j}$ be the cost of equipping stage $i$ with units of standard size $x_{i,j}$.
The optimality criterion (\ref{eq_1}), i.e., the capital cost of the equipment, is given by
\begin{equation}
	\label{eq:m_cost}
	\textit{Cost}(\Omega_e)=\sum_{i=1}^{I} c_{i,j_i}.
\end{equation}
For a fragment comprising stages $1$ to $i$, the partial sum $\textit{Cost}(\Omega_e,i)=\sum_{l=1}^{i} c_{l,j_l}$ is nondecreasing in $i$ when the costs are nonnegative; it therefore serves as a lower bound on the cost of all variants in the subtree of the corresponding node.
This bound is used to prune branches that cannot contain the optimum (the \texttt{DefineCost(W, i)} procedure in the subsequent chapters).
In the test case, the unit costs are taken to be numerically equal to their main sizes; therefore, the value of the criterion for the solution in Table~\ref{tab:CESSolution} equals the sum of the main sizes of the selected units.

\subsection{Computational Aspects of the Model}
\label{chap1_model_comp}

The model determines the computational profile of the problem, which is essential for developing both sequential and parallel algorithms.

First, the evaluation of a CES variant consists of three procedures of very different computational cost.
Computing the criterion (\ref{eq:m_cost}) and the compatibility function (\ref{eq:m_S}) amounts to a single pass over the stages of the fragment and can be updated incrementally when moving from a tree node to its child.
Computing the function $T$, in contrast, requires constructing the schedule of a system cycle for each product, i.e., computing the durations and start times of all operations in the sets $\mathcal{O}_p$ and, in the general case, also the iterative alignment of the cycles.
This explains the ratio of the execution times of the procedures \texttt{S(W, i)} and \texttt{T(W)} reported in Chapter~\ref{chapter2} (Table~\ref{tab:AveragedExecutionTimes}) and justifies the strategy of the sequential algorithm: the expensive check (\ref{eq_5}) is performed only for those complete variants that have passed all inexpensive checks.

Second, the \emph{structure} of the operation graph, i.e., the composition of the sets $\mathcal{O}_p$, the operation types, and the relations ``$o^{-}\to o$'' and ``$o^{\ast}\to o$'', is determined only by the process routes and by the parameters $r_{p,i}$ and $\nu_i$, and does not depend on the CES variant.
Only the vertex weights of the graph (the operation durations) and the cycle start times $\theta_c$ depend on the variant.
Therefore, it suffices to build the schedule structure once before the search starts, representing it as arrays of indices of the preceding and coupled operations; after that, evaluating each variant amounts to filling the arrays of durations and traversing the graph.
This separation of data into an immutable part shared by all variants and a small mutable state specific to a particular variant is of fundamental importance for the GPU implementation: the immutable part can be placed in constant or shared memory and used jointly by all threads, whereas each thread needs its own copy of only the arrays of durations and cycle start times (see Chapters~\ref{chapter4} and~\ref{chapter5}).

Third, the cost of computing $T$ is determined by the number of operations in the sets $\mathcal{O}_p$, which grows with the number of products, the length of the routes, and the batch size ratios, and, when batches are split and merged, also by the number of cycle alignment iterations, which depends on the input data.
The latter makes the computations irregular: the time needed to evaluate different CES variants may vary, which must be taken into account when distributing work among parallel workers.

The main notation of the model is summarized in Table~\ref{tab:model_notation}.

\begin{table}[htbp]
	\centering
	\small
	\caption{Main notation of the mathematical model of CES operation}
	\label{tab:model_notation}
	\begin{tabular}{>{\raggedright\arraybackslash}p{0.22\textwidth}>{\raggedright\arraybackslash}p{0.72\textwidth}}
		\toprule
		Symbol & Meaning \\
		\midrule
		$p=\overline{1,P}$ & product index \\
		$Q_p$ & production amount of product $p$, t \\
		$I_p$, $\mathcal{P}_{p,i}$ & set of stages in the route of product $p$; set of source stages for stage $i$ \\
		$\mathcal{A},\mathcal{B},\mathcal{C},\mathcal{D}$ & stage classes by the type of the main unit \\
		$\hat{x}_i$ & main size of the unit of stage $i$ in variant $\Omega_e$ \\
		$\nu_i$ & number of identical units at stage $i$ \\
		$r_{p,i}$, $u_{p,i}$ & batch size ratio; fraction of a batch in one unit \\
		$v_{p,i}$, $\mu_{p,i}$ & volumetric and mass material indices \\
		$\varphi^{\min}_{p,i}$, $\varphi^{\max}_{p,i}$ & bounds of the admissible filling degree of a unit \\
		$w_p$; $w^{\min}_p$, $w^{\max}_p$ & batch size; bounds of the admissible batch size range, t \\
		$S(\Omega_e,i)$ & number of products whose batches cannot be processed by the fragment of stages $1,\ldots,i$ \\
		$d^{\mathrm{w}}_{p,i}$, $d^{\mathrm{c}}_{p,i}$ & processing and cleaning times, h \\
		$b_p$, $\kappa_{p,i}$ & number of batches per system cycle; number of cycles of the units of stage $i$ per system cycle \\
		$\mathcal{O}_p$; $t^{\mathrm{s}}_o$, $t^{\mathrm{f}}_o$ & set of operations of a system cycle; start and finish times of operation $o$ \\
		$T^{\mathrm{w}}_p$, $T^{\mathrm{c}}_p$ & system cycle duration and cycle time, h \\
		$Z_p$, $T_p$ & number of system cycles and production time of product $p$ \\
		$c_{i,j}$ & cost of equipping stage $i$ with units of standard size $x_{i,j}$ \\
		\bottomrule
	\end{tabular}
\end{table}

\section{Test Case}
\label{chap1_sect2}

\begin{figure}[htb]
	\centering
    \includegraphics[width=0.97\textwidth]{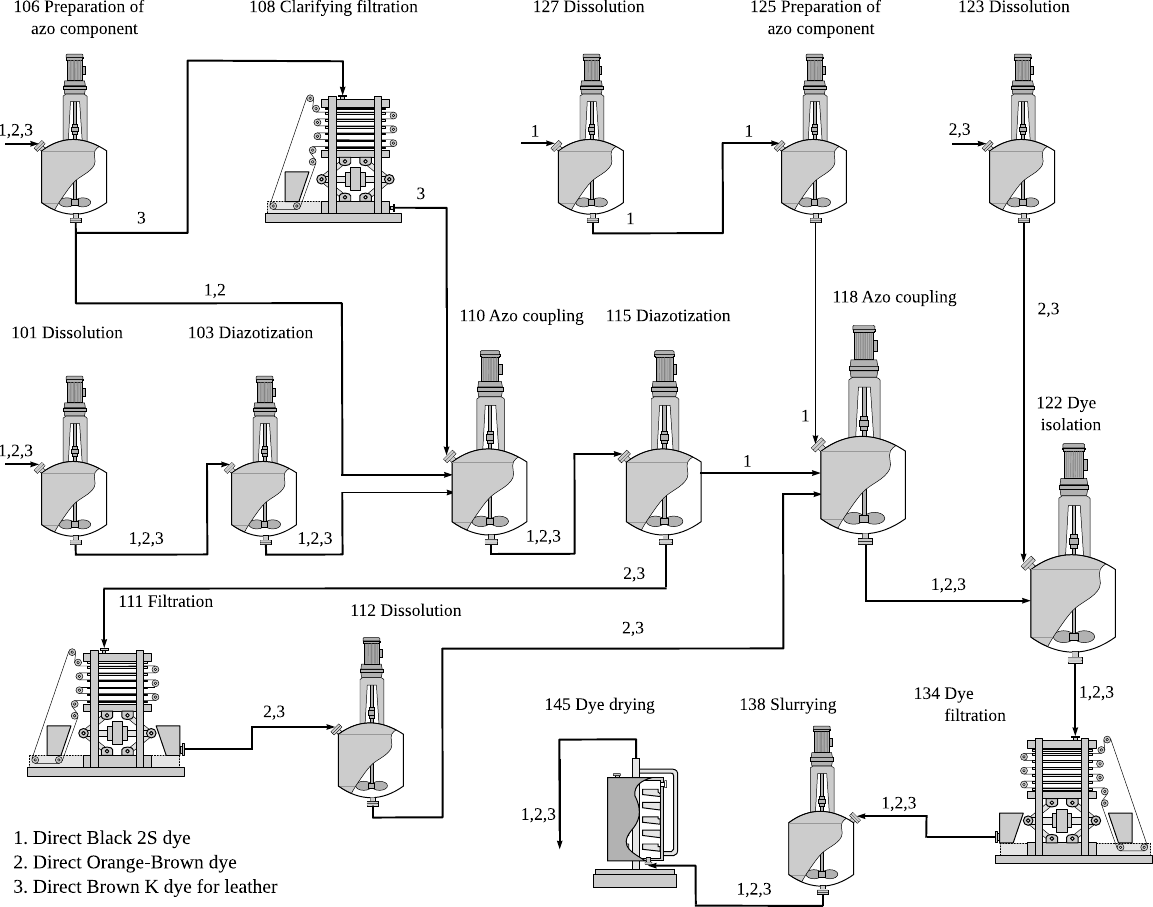}
    \caption{Test chemical-engineering system}
	\label{fig:test_CES}
\end{figure}

The algorithms developed in this book were studied on the equipment selection problem for a CES consisting of $16$ processing stages (see Fig.~\ref{fig:test_CES}).

Within $308$~days ($7392$~h), the CES must produce three dyes in amounts of $2250$, $100$, and $50$~t, respectively.
For each stage, a finite set (from $2$ to $12$) of standard sizes of standard equipment is available for installation.

Units of one standard size, chosen from the available set, are installed at each stage. The number of units is given by the input data (two units are installed at stage~138 and one unit at each of the other stages).

The problem size in the algorithm tests is set by the number of standard equipment sizes available for each stage.
For example, if two standard sizes are available for each stage, the test case shown in Fig.~\ref{fig:test_CES} has $2^{16}$ ($65\,536$) possible CES variants.
If $5$ standard sizes are available, there are $5^{16}$ ($\approx10^{11}$) possible CES variants;
if $12$ standard sizes are available, there are $12^{16}$ ($\approx10^{17}$) possible CES variants.

\begin{table}[htbp]
	\centering
	\small
	\caption{Solution of the equipment selection problem for the test CES}
	\label{tab:CESSolution}
	\begin{tabular}{clc}
		\toprule
		Stage no. & Stage & Main size of equipment \\
		\midrule
		101 & Dissolution & $6.3$~m$^3$ \\
		103 & Diazotization & $10$~m$^3$ \\
		106 & Preparation of the azo component & $5$~m$^3$ \\
		108 & Clarifying filtration & $5$~m$^2$ \\
		110 & Azo coupling & $16$~m$^3$ \\
		115 & Diazotization & $20$~m$^3$ \\
		127 & Dissolution & $0.4$~m$^3$ \\
		125 & Preparation of the azo component & $3.2$~m$^3$ \\
		111 & Filtration & $25$~m$^2$ \\
		112 & Dissolution & $10$~m$^3$ \\
		118 & Azo coupling & $32$~m$^3$ \\
		122 & Dye isolation & $32$~m$^3$ \\
		123 & Dissolution & $3.2$~m$^3$ \\
		134 & Dye filtration & $50$~m$^2$ \\
		138 & Slurrying & $10$~m$^3$ \\
		145 & Dye drying & $20$~m$^3$ \\
		\bottomrule
	\end{tabular}
\end{table}

Since no actual cost data for the equipment were available, the test problems rely on the assumption that the cost of a unit is directly related to its size: in the input data, the cost of each unit is set numerically equal to its main size (in m$^3$ or m$^2$, depending on the type of equipment).
The optimality criterion~(\ref{eq_1}) of a CES variant is thus computed as the sum of the main sizes of the units selected for its stages.
This simplification is sufficient for studying the algorithms, since the test problems are not intended to model a particular operating plant.

An example solution of the equipment selection problem for the test CES is given in Table~\ref{tab:CESSolution}.
The value of the optimality criterion for this solution is $248.1$.
%%%% Chapter 2
\chapter{Sequential Branch-and-Bound Algorithm}
\label{chapter2}
The branch-and-bound method is one of the popular techniques for solving NP-hard optimization problems in various fields, such as combinatorial optimization and artificial intelligence.
The method is based on the idea of successively partitioning the set of feasible solutions into subsets~\cite{Brassard1996}.
A lower bound on the objective function is then determined for each subset.
If the lower bound is not less than the best solution found so far, the subset is discarded, and the process continues until a single solution is obtained.

In effect, branch-and-bound searches the solution space using rules that exclude those regions of the search space in which no solution better than the current one can be found.
A branch-and-bound scheme has three algorithmic components that can be used to fine-tune the behavior of the algorithm: the search strategy, the branching strategy, and the pruning rules.
All three components depend on the specific problem being solved and have a considerable effect on the performance of the algorithm~\cite{Morrison2016}.

The chapter is organized as follows.
Section~\ref{chap2_sect1} presents the basic version of the sequential branch-and-bound algorithm for the problem.
Section~\ref{chap2_sect2} estimates the execution times of individual parts of the algorithm, which makes it possible to identify the most time-consuming operations.
Section~\ref{chap2_sect3} proposes an optimization of the sequential algorithm based on these estimates.

\section{Basic Version of the Algorithm}
\label{chap2_sect1}

\begin{figure}[htb]
	\centering
	\includegraphics[width=0.97\textwidth]{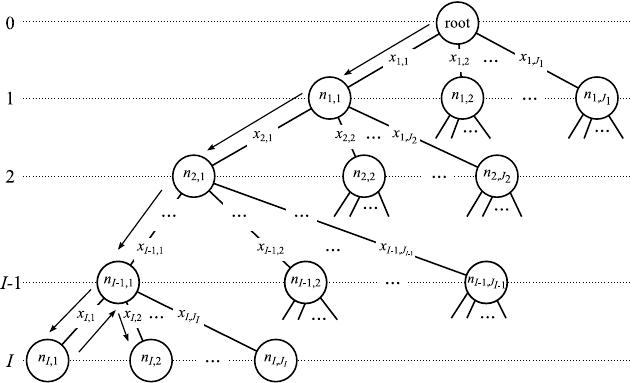}
	\caption{Depth-first traversal of the tree of variants}
	\label{fig:DFS_search}
\end{figure}

The whole set of variants of a CES consisting of $I$ stages can be represented as a tree with $I$ levels (see Fig.~\ref{fig:DFS_search}).

Each level $i$ of the tree corresponds to a stage of the CES.
Each edge $x_{i,j}$ of the tree corresponds to the selected unit variant $j$ for stage $i$ (one selected element of the set $X_i=\{x_{i,j}\},\,i=\overline{1,I}, j=\overline{1,J_i}$).
Each internal node $n_{i,k},\,i=\overline{1,I}, k=\overline{1,K_i}, K_i=\prod_{l=1}^i(J_l)$ of the tree corresponds to a fragment of the CES consisting of stages $1$ to $i$.

The transition from an internal node $n_{i,k}$ to one of its children (i.e., choosing the main size of a unit and appending it to the initial fragment of the CES variant obtained at the previous levels of the tree) will be called \emph{solution construction}.
A complete path from the root of the tree to one of its leaves represents a CES variant.

The possible CES variants are enumerated by a Depth-First Search (DFS) of the tree~\cite{kormen2009algorithm,knut2018}.
For this problem, depth-first search is preferable to Breadth-First Search (BFS), because the depth of the search tree is small, whereas the number of CES variants at each level grows exponentially.
With breadth-first search, the amount of main memory needed to store the tree nodes would also grow exponentially.

At each level of the tree, the main sizes of the units available for this stage of the CES are enumerated.
The selected size is appended to the initial fragment of the CES variant obtained at the previous levels of the tree, and the process continues recursively for all operable initial fragments.
Thus, at the last level of the tree, a complete operable CES variant is obtained; the recursion stops, and the variant obtained is compared with the current optimal one.
If the new variant has a smaller value of the criterion, it replaces the current optimal solution.

A complete traversal of the tree of variants (fixing, for each CES stage, a unit size chosen from the set of standard sizes for that stage and forming a complete CES variant), followed by checking constraints (\ref{eq_4}) and (\ref{eq_5}), would require a very large computational effort.

To reduce the number of enumerated variants, an algorithm based on the Branch-and-Bound (B\&B) method was developed~\cite{Leyffer2009,borisenko2012}.
Branch-and-bound is widely used to solve optimization problems of this kind in various fields (combinatorial optimization, artificial intelligence, etc.)~\cite{Brassard1996}, as well as MINLP and MILP problems~\cite{Bonami2012,Burer2012}.
The method is based on the idea of successively partitioning the set of feasible solutions into subsets~\cite{Brassard1996} and discarding those subsets that certainly do not contain an optimal solution.
To this end, a lower bound on the objective function over a subset is determined.
If the lower bound is not less than the best solution found so far, the subset is discarded, and the process continues until a single solution is obtained.

This approach is deterministic and yields the globally optimal solution.

The pseudocode of the algorithm is shown in Listing~\ref{lst:sequential} (most variable names coincide with those used in the problem statement).

\begin{lstlisting}[%
caption={Sequential version of the algorithm},%
float=htb,%
label=lst:sequential]
FindSolution () { EnumerateVariants (1); }(*@\label{ln101}@*)
/* recursive depth-first traversal of the tree */
EnumerateVariants(i) {
	if (i <= I) {(*@\label{ln104}@*)
		for (j = 1; j <= J[i]; j++) {
			/* build the initial fragment of the CES (stages 1 to i) */
			W[i] = X[i][j];
			/* check the compatibility constraint (1.4) */
			if (S(W, i) == 0 && (*@\label{ln109}@*)
	            /* cost of the initial fragment of the CES (stages 1 to i) */
	            DefineCost(W, i) < minCost) { (*@\label{ln111}@*)
	            /* recursive search */
	            EnumerateVariants(i + 1); }}}(*@\label{ln113}@*)
	else { /* all tree levels have been passed (i==I+1) */(*@\label{ln114}@*)
		/* check the processing time constraint (1.5) */
		if (T(W) <= Tmax) {(*@\label{ln116}@*)
			/* check the optimality criterion */
			if (DefineCost(W, I) < minCost) {(*@\label{ln118}@*)
				/* store the new optimal solution */
				Wopt = W;	minCost = DefineCost(W, I); }}} (*@\label{ln120}@*)
}
\end{lstlisting}

The traversal of the tree of variants starts with a call of the procedure \texttt{EnumerateVariants()} at the first level (line~\ref{ln101}).
This procedure is called recursively down to level \texttt{I} (line~\ref{ln104}).
The current optimal CES variant is stored in the vector \texttt{Wopt} with \texttt{I} elements, each of which holds the selected main size of the unit for the corresponding stage.
The working CES variant is stored in a similar vector \texttt{W}.

To reduce the enumeration, the algorithm uses the following scheme.
At each level $i$ of the tree, the compatibility constraint (\ref{eq_4}) for the joint operation of the CES equipment is checked (procedure \texttt{S()}, line~\ref{ln109}); however, it is checked not for the whole CES but only for its initial fragment, which consists of the levels already passed above and the current one (the CES fragment consisting of stages 1 to $i$).
If the fragment is already inoperable, there is no point in enumerating units at the lower levels.
In addition, the cost (\ref{eq_1}) of the initial fragment of the CES variant (procedure \texttt{DefineCost()}, line~\ref{ln111}) is compared with the cost \texttt{minCost} of the current optimal CES variant.
Thus, the algorithm descends to the next recursion level (a call of \texttt{EnumerateVariants()} at level \texttt{i + 1}, line~\ref{ln113}) only if the initial fragment of the CES consisting of stages 1 to $i$ is, first, operable and, second, not more expensive than the best CES variant found so far.
Otherwise, the algorithm backtracks to the previous level of the tree.
As a result, the number of enumerated variants is reduced, and the higher the level of the tree at which such a mismatch is detected, the greater the reduction.

If all levels of the tree have been passed (\texttt{i==I+1}, line~\ref{ln114}), the processing time constraint (\ref{eq_5}) is checked (procedure \texttt{T()}, line~\ref{ln116}).
If constraint (\ref{eq_5}) is satisfied, a new operable CES variant has been found, and its cost (the result of \texttt{DefineCost()}) is compared with the cost \texttt{minCost} of the current best CES variant (line~\ref{ln118}).
If a better solution has been found, it replaces the current optimal solution \texttt{Wopt}, and its cost is stored in \texttt{minCost} (line~\ref{ln120}).
The initial value of \texttt{minCost} is the sum of the maximum unit costs over all stages (the cost of a CES variant consisting of the most expensive units is the largest possible).

\section{Execution Times of Individual Parts of the Algorithm}
\label{chap2_sect2}
For the computational experiments, the algorithm presented above (see Listing~\ref{lst:sequential}) was implemented in C++.

The experiments were carried out on a computer with two Intel Westmere X5650 processors (six cores at 2.6~GHz) and 4~GB of RAM running Red Hat Linux.
The program was compiled with Intel C++ Compiler v.~11.1.
The test case was the equipment selection problem for the CES described in Section~\ref{chap1_sect2}, with $5$ standard equipment sizes available for each processing stage.

To identify the main bottlenecks of the algorithm, the execution times of its individual elements were measured.
The results are given in Table~\ref{tab:AveragedExecutionTimes}.
\begin{table}[htb]
	\centering
	\small
	\caption{Execution times of individual elements of the algorithm}
	\label{tab:AveragedExecutionTimes}
	\begin{tabular}{lc}
		\toprule
		Element of the algorithm & Execution time, $\mu$s \\
		\midrule
		Recursive call of the procedure & $0.1$ \\
		\texttt{EnumerateVariants(i)} & \\
		\texttt{S(W, i)} & $4.0$ \\
		\texttt{T(W)} & $417.0$ \\
		\texttt{DefineCost(W, i)} & $0.7$ \\
		\bottomrule
	\end{tabular}
\end{table}

The times in Table~\ref{tab:AveragedExecutionTimes} are negligible for a single evaluation.
However, since the search repeats these operations billions of times, the times accumulate and may add up to tens or hundreds of hours.
In particular, for the example considered ($16$ stages with $5$ standard sizes available for each), the total run time of the program was 27~h\,11~min.

The results show that the most expensive operation is the procedure \texttt{T(W)}, which checks the processing time constraint.
To reduce the total run time of the algorithm, the number of calls of this procedure must be minimized; to this end, the algorithm was optimized as described below.

\section{Optimization of the Sequential Algorithm}
\label{chap2_sect3}
A call of the procedure \texttt{DefineCost(W)}, which evaluates the objective function, is a relatively cheap operation (see Table~\ref{tab:AveragedExecutionTimes}).
If the optimality criterion is checked immediately after an operable CES variant has been obtained, the processing time constraint \texttt{T(W)} need not be checked for solutions that are certainly not optimal.
Thus, although the number of calls of \texttt{DefineCost(W)} increases, the number of calls of \texttt{T(W)} decreases noticeably, which ultimately reduces the total run time of the algorithm.

The changes made to the algorithm are shown in Listing~\ref{lst:sequential_mod}.
\begin{lstlisting}[label=lst:sequential_mod,%
caption={Modification of the sequential algorithm},%
float=htb,%
firstnumber=13]
...
	else { /* all tree levels have been passed (i == I + 1) */
		/* check the optimality criterion */
		if (DefineCost(W, I) < minCost) {(*@\label{ln216}@*)
			/* check the processing time constraint (1.5) */
			if (T(W) <= Tmax) {(*@\label{ln218}@*)
			/* store the new optimal solution */
				Wopt = W; minCost = DefineCost (W, I); }}}
...
\end{lstlisting}

The optimality criterion is checked (line~\ref{ln216}) before the processing time constraint (line~\ref{ln218}).
As a result, the run time was reduced approximately by half (13~h~52~min instead of 27~h~11~min).

%%%% Chapter 3
\chapter{Parallel Branch-and-Bound for Multi-Core and Cluster Systems}
\label{chapter3}
Despite the optimization of the algorithm described in Section~\ref{chap2_sect3}, the run time remains considerable.
To reduce it further, it is therefore worthwhile to develop and study a parallel version of the algorithm.
The tree-like organization of the search space in branch-and-bound is well suited for this, since different branches of the tree can be processed simultaneously~\cite{Posypkin2005,Grama2003}.
The first results on parallelizing the algorithm for the CES equipment selection problem were reported in~\cite{Borisenko2011parallel}.

The chapter is organized as follows.
Section~\ref{chap3_review} reviews methods for parallelizing search tree traversal: static and dynamic tree partitioning, dynamic load balancing methods and the strategies they use, speedup anomalies, the assigned subtrees and allocated subtrees methods, and approaches to implementing branch-and-bound on GPUs; at the end of the section, the implementations developed in this book are placed in this context.
Section~\ref{chap3_sect1} presents a parallel implementation of the algorithm for shared-memory systems based on the OpenMP standard, in which the nodes of the upper levels of the search tree are treated as independent tasks, and studies the effect of the granularity parameter on the run time.
Section~\ref{chap3_sect2} presents an implementation for distributed-memory systems based on the MPI library and the master--worker paradigm~\cite{Ershov2015,Ejova2018}.
For both implementations, the speedup and the processor efficiency are studied.

\section{Methods for Parallelizing Search Tree Traversal}
\label{chap3_review}

\subsection{Features of Parallel Search Tree Traversal}
\label{chap3_review_tree}

Many discrete optimization problems arising in planning, scheduling, computer-aided design, and robotics are formulated as a search in a graph or tree: one has to find a minimum-cost path from the initial node to one of the goal nodes~\cite{Grama2003,Grama1999}.
In general form, the problem is stated as follows: given a finite set $X$ of feasible solutions and an objective function $f(x)$ defined on it, find an element $x_{opt}$ such that $f(x_{opt}) = \min\{f(x) \mid x \in X\}$.
Constraint satisfaction problems, in which it suffices to find at least one element of a subset $S \subset X$, fit the same scheme if we set $f(x)=0$ for $x\in S$ and $f(x)=1$ for the other elements of $X$.

In practically important problems, the cardinality of $X$ is so large that exhaustive enumeration of its elements is impossible.
Branch-and-bound, dynamic programming, and heuristic search exploit the structure of the tree (graph) of variants in order to exclude entire branches from consideration and find a solution without explicitly enumerating all elements of $X$~\cite{Morrison2016,Grama2003}.
The optimal CES equipment selection problem considered in Chapter~\ref{chapter1} belongs to exactly this class: each level of the tree corresponds to a CES stage, each leaf to a complete CES variant, and checking the constraints and the cost of initial fragments makes it possible to prune subtrees that certainly do not contain the optimal solution (see Section~\ref{chap2_sect1}).

A search tree naturally decomposes into independent subtrees that can be traversed simultaneously.
This makes tree traversal methods a convenient target for parallelization~\cite{Posypkin2005,Gendron1994}.
In branch-and-bound, however, the tree is not given in advance: it is built during the computation, and its shape depends on the input data and on how early a good solution allowing branches to be pruned is found.
It is therefore unknown in advance how many nodes will be visited in each subtree, and the central problem of parallel traversal is to distribute the work among processors so as to minimize the total traversal time (or, equivalently, to maximize the efficiency of using the computing system)~\cite{Kumar1994scalable,Tim_2004}.

Efficiency losses in parallel tree traversal consist of three components:
\begin{itemize}
	\item idle time of processors that have received smaller subtrees than the others (load imbalance);
	\item communication overhead: transferring subtrees, work requests, and solutions found;
	\item redundant work, i.e., nodes that are visited by the parallel algorithm but would not be visited by the sequential one (for details, see Section~\ref{chap3_review_anomalies}).
\end{itemize}
Different parallel traversal methods trade off these components in different ways, and the choice of method depends both on the properties of the tree of the problem at hand and on the architecture of the computing system.

\subsection{Static and Dynamic Tree Partitioning}
\label{chap3_review_static}

According to how the tree is split into subtrees and how these are distributed among processes, all methods of parallel depth-first tree traversal can be divided into two groups: those with static and those with dynamic partitioning~\cite{Tim_2004,Grama1999}.

With \emph{static} partitioning, the tree is split into subtrees once, before the traversal begins, and each subtree is assigned to a specific process.
For example, each process can uniquely compute the roots of ``its'' subtrees from its rank and need not interact with the other processes afterwards.
This approach is simple to implement and requires practically no communication; however, it is efficient only when the subtrees are known to be of similar size, i.e., for trees with a regular structure known in advance.
Similarly, static decomposition of the computational domain is efficient in problems with a uniformly distributed computational load, for example, in solving the unsteady heat equation with an explicit finite-difference scheme~\cite{Borisenko2010heat}.
For branch-and-bound, this condition usually does not hold: pruning makes the tree irregular, so that some subtrees are traversed in fractions of a second while others take hours.

With \emph{dynamic} partitioning, the distribution of work changes during the traversal.
A process that has finished traversing its part of the tree becomes idle and receives part of the not-yet-traversed work of one of the busy processes.
The process that gives away work will be called the \emph{donor}, and the process that receives it, the \emph{receiver}.
Dynamic partitioning keeps the processors evenly loaded for trees of arbitrary shape at the price of additional communication and task-scheduling overhead.
This is why dynamic schemes are predominantly used for branch-and-bound~\cite{Gendron1994,Crainic2006}.

\subsection{Classification of Dynamic Load Balancing Methods}
\label{chap3_review_classes}

Dynamic load balancing methods differ in who initiates the splitting of the tree into subtrees and who initiates the transfer of subtrees to other processes.
Three groups of methods are distinguished~\cite{Tim_2004,Kumar1994scalable} (Table~\ref{tab:tree_search_methods}).

\begin{table}[htb]
	\centering
	\small
	\caption{Classification of dynamic search tree partitioning methods}
	\label{tab:tree_search_methods}
	\begin{tabular}{>{\raggedright\arraybackslash}p{0.27\textwidth}>{\raggedright\arraybackslash}p{0.33\textwidth}>{\raggedright\arraybackslash}p{0.3\textwidth}}
		\toprule
		Initiator of splitting and transfer & Typical scheme & Advantages and drawbacks \\
		\midrule
		Splitting and transfer are initiated by the donor & Generated nodes are sent to randomly chosen or neighboring processes & No requests needed; large communication volume, unbounded growth of the number of nodes held by a process \\
		\addlinespace
		Splitting is initiated by the donor, transfer by an idle process & Master processes generate a pool of subtree roots in advance; workers obtain them on request (master--worker) & Simple control, low communication volume; possible imbalance due to different subtree sizes \\
		\addlinespace
		Splitting and transfer are initiated by an idle process & An idle process requests work; a busy process hands over part of its subtree & Best balancing for irregular trees; costs of requests, rejections, and work splitting \\
		\bottomrule
	\end{tabular}
\end{table}

In the methods of the \emph{first group}, the donor sends the nodes it generates to other processes regardless of their state.
In the simplest variant, each new node is sent to a randomly chosen process~\cite{Shu1989}; on average, this balances the load, but it generates communication at every branching step and overloads the network.
Moreover, the number of nodes accumulated by an individual process is unbounded, which may lead to memory shortage.
The communication volume is reduced if nodes are sent only to neighboring processors or if nodes below a certain cutoff level may not be transferred and are processed by the process that generated them~\cite{Rao1987}.

In the methods of the \emph{second group}, work is generated in advance but handed out to idle processes only on request.
The processes are divided into masters, which generate subproblems, and workers, which solve them.
A typical representative of this group is the master--worker paradigm, in which a single master process traverses the upper part of the tree and forms a pool of subtree roots, and the worker processes receive them one at a time as they become free.
To reduce the load on the master process, schemes with several sources of subproblems are used, each serving its own group of workers; when a source runs out of subproblems, its workers switch to another source.
A similar idea underlies the fixed-packet DFS method~\cite{Reinefeld1994}: all processes first traverse the tree synchronously down to a certain level, the resulting nodes are divided into as many classes as there are processes, and then processes that have become idle request some of the root nodes from the classes of other processes.

The \emph{third group}, which is the most widely used in practice, comprises methods in which both the splitting and the transfer of work are performed at the request of an idle process~\cite{Finkel1987,Rao1987,Kumar1994scalable,Sanders1995}.
Each process traverses its assigned subtree depth-first and periodically checks whether requests from other processes have arrived.
Having finished its traversal, a process selects a potential donor and sends it a request.
The donor either rejects the request if its remaining work is too small, or splits its work into two parts and sends one of them to the receiver.
In modern terminology, this approach is called \emph{work stealing}.
For multithreaded shared-memory computations, it has been proved to provide near-optimal load balance with moderate overhead~\cite{Blumofe1999}.

\subsection{Algorithmic Aspects of Dynamic Load Balancing}
\label{chap3_review_strategies}

The methods of the third group differ in the strategies used at each step of work exchange.

\textbf{Donor selection.}
With \emph{random polling}, an idle process chooses a donor at random and repeats the choice if it is rejected~\cite{Kumar1994scalable,Sanders1995}.
This strategy scales well, but when little work remains, it leads to many rejections.
With a \emph{global round robin}, all processes access a common pointer, which is advanced to the next process after each request.
This considerably reduces the number of requests for trees of various structures; however, with many processes, contention for the shared pointer arises and it becomes a bottleneck~\cite{Kumar1994scalable}.
Finally, with \emph{targeted selection}, a master process that receives information from the workers about the traversed parts of their subtrees directs the request to the most heavily loaded process~\cite{Tim_2004}.

\textbf{Work splitting strategy.}
The donor may give the receiver a single subtree by removing one not-yet-visited node from its stack (\emph{node splitting}), or split its stack into two parts so that each part contains approximately the same number of nodes of different levels (\emph{stack splitting})~\cite{Rao1987,Kumar1994scalable}.
The second option transfers more work per exchange and therefore reduces the number of requests, but requires more data to be sent.
To transfer as large a subtree as possible, the unvisited node closest to the root is chosen.

\textbf{Initial work distribution.}
The traversal may start with a single process to which the whole tree is assigned; the other processes then obtain work through the request mechanism~\cite{Rao1987,Sanders1995}.
This is simple, but at the beginning most processes are idle waiting for work, and the number of interprocess exchanges rises sharply.
An alternative is a preliminary traversal of the upper part of the tree (usually breadth-first) until enough root nodes have been obtained, which are then distributed among the processes, or having each process compute the root of its first subtree from its own rank.

\textbf{Splitting threshold.}
Splitting very small subtrees is not worthwhile, because the communication overhead would exceed the gain from parallel traversal.
Therefore, the donor rejects a work request if, for example, the current traversal depth is close to the maximum or the subproblem contains too few nodes~\cite{Tim_2004}.
The specific threshold depends on the problem size, the cost of processing a node, and the characteristics of the communication network.

Using these strategies, parallel algorithms have been developed for the travelling salesman problem, the knapsack problem, the Boolean satisfiability problem, puzzles, and a number of other combinatorial problems.
For most of them, near-linear speedup has been achieved on multiprocessor systems~\cite{Grama1999,Gendron1994}.

\subsection{Speedup Anomalies and the Search Overhead Factor}
\label{chap3_review_anomalies}

A specific feature of parallel search is that the parallel and the sequential algorithms usually visit different sets of nodes.
If any solution is to be found (as in constraint satisfaction problems), the search stops as soon as one of the processes finds a feasible solution, and the parallel algorithm may visit either fewer or more nodes of the search tree than the sequential one.
In branch-and-bound, the same effect arises because the incumbent (the best value found so far), which is used to prune branches, is updated at different moments.
If a good solution is found by one of the processes earlier than the sequential algorithm would find it, the other processes prune more branches; if later, part of the work turns out to be redundant.

To quantify this effect, the \emph{search overhead factor} is introduced:
\begin{equation}
	\label{eq:search_overhead}
	R = \frac{W_p}{W_s},
\end{equation}
where $W_s$ is the number of nodes visited by the sequential algorithm and $W_p$ is the total number of nodes visited by all processes of the parallel algorithm~\cite{Grama2003,Rao1993}.
The speedup on $s$ processors is then bounded from above by $s/R$.
For $R>1$, the parallel algorithm performs redundant work, and the speedup is necessarily sublinear.
For $R<1$, \emph{superlinear} speedup is possible: the parallel algorithm visits fewer nodes than the sequential one.
Such situations, known as \emph{speedup anomalies}, have been studied in detail for parallel branch-and-bound~\cite{Lai1984}.
It has been shown that increasing the number of processors may either accelerate or slow down the search, and conditions have been formulated under which a slowdown is impossible.
For parallel depth-first search with backtracking, it was shown in~\cite{Rao1993} on abstract models of the search space that if solutions are distributed identically over the search regions of different processes, then the amounts of work of the sequential and parallel algorithms coincide on average.
If the solution distributions differ, the sequential algorithm visits on average more nodes than the parallel one.
A systematic superlinear speedup rather indicates that the chosen sequential traversal order is not the best one for the given problem.

\subsection{The Assigned Subtrees and Allocated Subtrees Methods}
\label{chap3_review_mnp}

For traversing the trees of combinatorial problems on distributed-memory multiprocessor systems, two methods were proposed in~\cite{Tim_2004,Tim_2009}; they represent the second and the third group of the above classification, respectively.
In both methods, the processes are divided into one master process and several workers.

The \emph{assigned subtrees method} focuses on reducing interprocess communication.
The master process traverses the tree breadth-first from the root until the number of generated but not yet processed nodes (subtree roots) reaches a given value $m$, which considerably exceeds the number of worker processes.
The root nodes are then handed out to the workers one at a time as they finish traversing the subtrees received earlier.
The workers interact with the master process only when receiving a new task and sending back a result, so the control overhead is small.
However, the method does not guarantee a uniform load: individual subtrees may be many times larger than the others, and at the end of the computation some processes will be idle.
Increasing $m$ reduces the size of individual subtrees and smooths out the imbalance, but increases the number of exchanges with the master process.
Thus, the assigned subtrees method is advisable when the partitioning yields subtrees of similar size or when there are enough of them for the differences in size to average out.

The \emph{allocated subtrees method} emphasizes uniform load.
The master process generates only as many root nodes as there are workers and hands them out one per worker.
When a worker finishes its traversal, the master process chooses, among the busy processes, the one with the largest untraversed part of its subtree (the workers periodically report the progress of their traversal to the master) and instructs it to split off a subtree for the idle process.
A subtree is split off only if a splitting condition holds, for example, if the index of the last fully traversed level is much smaller than the total number of tree levels.
The allocated subtrees method ensures a uniform load, but requires additional effort for collecting progress information and for splitting off subtrees.

\subsection{Parallelizing Branch-and-Bound on GPUs}
\label{chap3_review_gpu}

The methods discussed above were developed for multiprocessor (multi-core) systems in which each processor traverses its own subtree independently.
The GPU architecture imposes additional constraints on the parallelization of branch-and-bound.
Thousands of GPU threads are grouped into warps that execute the same instruction; the branching inevitable in the traversal of an irregular tree causes different code paths within a warp to be executed serially (divergence).
The amount of fast memory per thread is small, and dynamic memory allocation and data exchange between threads of different blocks are difficult.
Finally, host--device transfers are relatively expensive, so frequent movement of work between the CPU and the GPU is undesirable.

Existing approaches to implementing branch-and-bound on GPUs can be roughly divided into three groups.
In the first group, the search tree is managed on the CPU, while the most computationally intensive part, the simultaneous processing (branching and bounding) of a large number of nodes from the pool of active nodes, is offloaded to the GPU~\cite{Chakroun2013combining,Lalami2012}.
In the second group, the tree is partitioned in advance into many subtrees, each of which is traversed entirely by a separate GPU thread; the implementations presented in Chapter~\ref{chapter4} also belong to this group~\cite{Borisenko2015lncs,Borisenko2016}.
In the third group, both the traversal and the dynamic load balancing are implemented on the GPU; for example, the Integer-Vector-Matrix data structure makes it possible to represent subtrees compactly as intervals and to redistribute work among threads directly on the accelerator~\cite{Gmys2016}.
A comparison of branch-and-bound implementations on multi-core CPUs and on massively parallel accelerators for large problems is given in~\cite{Melab2018}.

\subsection{Methods Used in This Book}
\label{chap3_review_ours}

The above classification makes it possible to position the parallel branch-and-bound implementations presented in this chapter and in Chapter~\ref{chapter4}.

The implementation for shared-memory systems (Section~\ref{chap3_sect1}) uses the task mechanism of the OpenMP standard: each node of the upper levels of the tree, down to level $G$, is turned into an independent task, and the OpenMP runtime distributes the tasks among the threads.
This is a dynamic scheme in which the task pool serves as the work pool, and the granularity parameter $G$ determines the number and size of the generated tasks.

The implementation for distributed-memory systems (Section~\ref{chap3_sect2}) is based on the master--worker paradigm and the assigned subtrees method: the master process traverses the tree down to level $G$ and hands out initial fragments of CES variants to the workers on their requests.
In terms of the assigned subtrees method, the number of assigned subtrees $m$ equals the number of nodes at level $G$, i.e., $m=K_G=\prod_{l=1}^{G}J_l$; as in the general case, the choice of $G$ determines the trade-off between load uniformity and the amount of communication with the master process.

The GPU implementation (Chapter~\ref{chapter4}) uses static partitioning: the host traverses the tree down to level $G$, and each of the $K_G$ resulting subtrees is assigned to a separate GPU thread without subsequent redistribution of work.
This approach minimizes communication and simplifies the kernel code, but makes the run time dependent on the largest subtree, which, together with thread divergence, limits the achievable speedup (see Section~\ref{chap4_sect4}).

\section{Shared-Memory Systems (OpenMP)}
\label{chap3_sect1}
\subsection{Program Design}
In shared-memory systems, several processors have equal access to memory. Such systems are also called Symmetric Multiprocessing (SMP) systems.

In the proposed version of the algorithm, all tree nodes $N_i=\{ n_{i,1},n_{i,2},\ldots, n_{i,k}\},\,i=\overline{1,I},
k=\overline{1,K_i}, K_i=\prod_{l=1}^i J_l$ at each level $i$ are treated as independent \emph{tasks} that can be executed in parallel.
The total number of such tasks, $N_{tasks}=\sum_{i=1}^G K_i,\,1 \leq G \leq I$, may be very large.
The granularity parameter $G$ of the algorithm limits the degree of parallelism to a certain level of the tree: subtrees below this level are not split into parallel tasks but are processed sequentially~\cite{Borisenko2011}.

Tasks are an alternative to traditional threads; they provide a high level of abstraction and allow the programmer to express the logical parallelism of an application without managing parallel execution directly~\cite{Tousimojarad2015,borisenko2015}.
Tasks provide better load balancing and efficient use of the available computing resources.
The pseudocode of the proposed parallel algorithm is shown in Listing~\ref{lst:parallelOMP}.

\begin{lstlisting}[%
emph={pragma,omp,task, parallel, single, pragma_omp_task},emphstyle=\bfseries\color{red},
caption={Parallel version of the algorithm for shared-memory systems},%
float=htb,%
label=lst:parallelOMP]
FindSolution() {
    #pragma omp parallel {(*@\label{ln302}@*)
        #pragma omp single {(*@\label{ln303}@*)
            EnumerateVariants(1); }}}(*@\label{ln304}@*)
EnumerateVariants(i) {
	if (i <= I) {
		for (j=1; j <= J[i]; j++) {
			/* build the initial fragment of the CES (stages 1 to i) */
			W[i] = X[i][j];
			/* check the operability and the cost
				of the initial fragment of the CES */
			if (S(W, i) == 0 && DefineCost(W, i) < minCost) {
				if (i <= G) { /* current search depth does not exceed G */
					/* create a parallel task */
					#pragma omp task
					    EnumerateVariants(i + 1); }(*@\label{ln316}@*)
				else { /* current level is deeper than G */
					/* sequential recursive tree traversal */
					EnumerateVariants(i + 1);}}}}
	else { /* all tree levels have been passed (i == I + 1) */
		...	}
}
\end{lstlisting}
The parallel implementation is a development of the sequential program considered above (see Chapter~\ref{chapter2}, Listings~\ref{lst:sequential} and~\ref{lst:sequential_mod}); directives of the Open Multi-Processing (OpenMP) standard~\cite{openmp} were added to its source code.
Starting with version~3.0, OpenMP supports the \texttt{task} directive, which makes it possible to create separate independent tasks.

The initial call of the recursive procedure \texttt{EnumerateVariants()} (line~\ref{ln304} of Listing~\ref{lst:parallelOMP}) is made inside a \texttt{\#pragma~omp~single} directive nested in a \texttt{\#pragma~omp~parallel} directive, so that only one of the available parallel threads performs the recursive tree traversal.

Subsequent recursive calls of \texttt{EnumerateVariants()} (line~\ref{ln316}) are made inside the \texttt{\#pragma~omp~task} directive.
Thus, while recursively traversing the tree, the first thread of the program creates new independent tasks at each level $i, 1 \leq i \leq G$; these tasks are automatically placed into the task pool, from which any idle thread can take them for execution.

Note that, when these tasks are processed in parallel, each task must be given its own copy of the array \texttt{W}, the current initial fragment of the CES.

\subsection{Results of Computational Experiments}
\label{chap3_sect1_res}
For the computational experiments, a parallel version of the program was developed in C++ using OpenMP~3.0.

The test case was the equipment selection problem for the CES described in Section~\ref{chap1_sect2}, with $5$ standard equipment sizes available for each processing stage.

The experiments were carried out on a computing system with four eight-core Intel Xeon E7550 processors at 2~GHz ($32$ cores in one machine) and $128$~GB of RAM running Red Hat Linux.
The program was compiled with Intel C++ Compiler 11.1.

\begin{figure}[htb]
	\centering
	\includegraphics[width=0.97\textwidth]{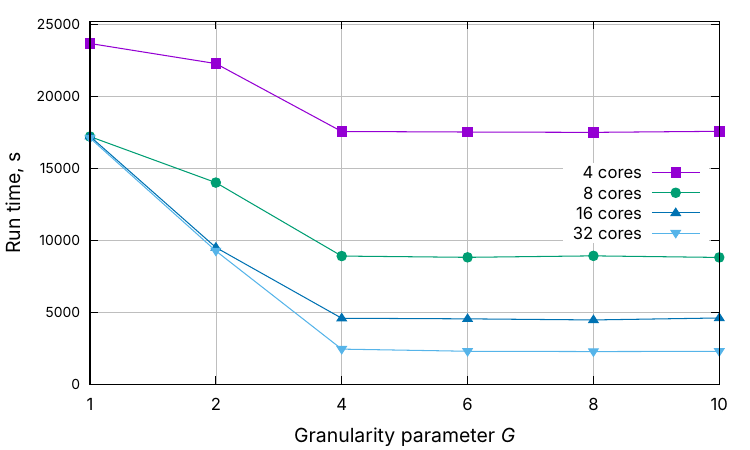}
	\caption{Run time of the parallel program versus the granularity parameter $G$}
	\label{fig:ompGranularity}
\end{figure}

Figure~\ref{fig:ompGranularity} shows how the run time of the parallel program on different numbers of processor cores depends on the granularity parameter~$G$.

For values of $G$ greater than $10$, the number of tasks became so large that there was not enough memory to store them (the program terminated abnormally).

The speedup and the efficiency of using the cores were also studied.
In general, the speedup $S_p$ is defined as the ratio of the time $T_s$ needed to execute the sequential program on one core to the time $T_p$ needed to execute the parallel program on a $p$-core machine: $S_p=T_s/T_p$.
The case $S_p = p$ corresponds to linear (ideal) speedup.
The efficiency $E_p$ is defined as the ratio of the speedup $S_p$ to the number of cores $p$: $E_p=S_p/p=T_s/(p\cdot T_p)$.

\begin{figure}[htb]
	\centering
	\includegraphics[width=0.97\textwidth]{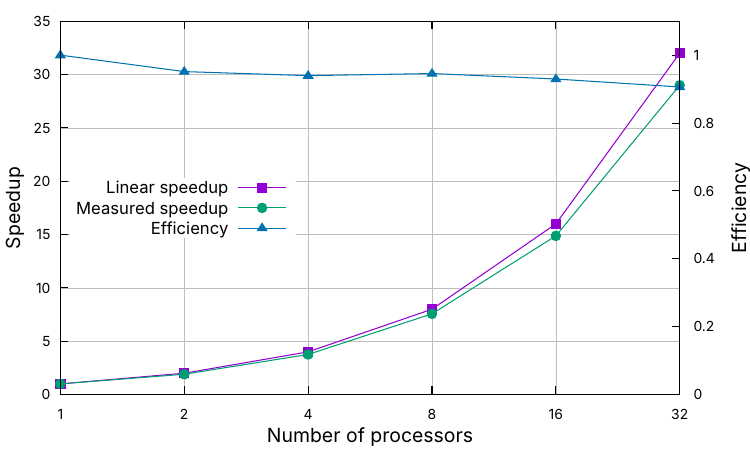}
	\caption{Speedup and efficiency of using the cores ($G = 10$)}
	\label{fig:ompSpeedup}
\end{figure}

The results are shown in Fig.~\ref{fig:ompSpeedup} (the horizontal axis is logarithmic).
The granularity parameter was $G=10$.

It is recommended to choose the granularity parameter $G$ so that the number of tasks considerably exceeds the number of physical cores of the computing device and is at least not smaller than the number of cores used.
The number of tasks is determined by the value of $G$, for which we propose the empirical estimate
$G \ge \log_{\max\{J_i\}}N_p,\ i=\overline{1,I}$, where $\max\{J_i\}$ is the maximum cardinality of the set of unit sizes available at a stage and $N_p$ is the number of cores used.
However, with a very large number of tasks (in our case, for $G>10$), memory may run out.
It is difficult to determine the required amount of memory exactly.
Nevertheless, the results in Fig.~\ref{fig:ompGranularity} show that, starting from a certain threshold value of $G$ (in our example, $G = 4$), the run time of the program remains practically unchanged despite the increasing number of tasks.
We therefore suggest first performing computations for a small problem, finding the threshold value of $G$ that yields an acceptable solution time, and then carrying out the computations for larger problems.

The study of speedup and efficiency showed that the parallel program achieves near-linear speedup and fairly high scalability.
The efficiency of using the cores is $0.9$--$0.95$.

\section{Distributed-Memory Systems (MPI)}
\label{chap3_sect2}

In distributed-memory systems, each compute node exchanges data with the other nodes only over communication links, by message passing.
The parallel implementation of the algorithm for distributed-memory systems is based on the master--worker paradigm (also known as the master--slave paradigm)~\cite{Aida2003,Bouziane2010} and on the assigned subtrees method~\cite{Tim_2009}.

The main features of the master--worker paradigm are the following:
\begin{itemize}
\item the master process (master) generates tasks and distributes them evenly among the worker processes (workers);
\item after solving the task it has received, each worker returns the results to the master and requests a new task;
\item the workers do not exchange data with one another, only with the master;
\item after receiving the results of a task, the master dynamically generates a new task for the worker that has become free or, if no tasks remain, sends it a termination signal;
\item the algorithm requires at least two processes (one master and one worker).
\end{itemize}

The pseudocode of the master and worker processes is shown in Listings~\ref{lst:parallelMpiMaster} and~\ref{lst:parallelMpiWorker}, respectively.
\subsection{Master Process}
\label{sec:MasterProcess}
Using the recursive procedure \texttt{MasterEnumerateVariants()}, the master performs a sequential depth-first traversal of the tree only down to a certain level $G$ (the granularity), $1\leq G\leq I$, forming the initial fragments of CES variants $W[i], i=\overline{1,G}$ (lines~\ref{ln423}--\ref{ln429}).

The granularity parameter $G$ is used to vary the degree of parallelism in the same way as in the version of the algorithm for shared-memory systems (see Section~\ref{chap3_sect1}): subtrees below this level are not split into parallel tasks but are processed sequentially.
Its value must be such that the number of generated subproblems is at least not smaller than the number of processing elements used.

\begin{lstlisting}[%
caption={Pseudocode of the master process},
float=htb,%
label={lst:parallelMpiMaster}]
Master() {
	/* number of workers (one of the processes is the master) */
	num_workers = NUM_PROCESSORS - 1;
	/* start the tree traversal */
	MasterEnumerateVariants(1);
	/* while some workers are still running */
	while (num_workers > 0) {
		msg = ReceiveWorkerMessage();
		/* a worker has sent a new solution */
		if     (msg.type == SOLUTION_FOUND) {
			/* check the optimality criterion */
			if (DefineCost(msg.W, I) < minCost) {
				/* store the new optimal solution */
				Wopt = msg.W; minCost = DefineCost(msg.W, I); }}
		/* a worker requests a new task */
		elseif (msg.type == REQUEST_WORK) {
			/* stop the worker */
			SendWorkerMessage(msg.workerID, QUIT);
			num_workers--; }}
}

MasterEnumerateVariants(i) {
	if (i <= G) { /* current depth does not exceed G */ (*@\label{ln423}@*)
		for (j=1; j <= J[i]; j++) {
			/* build the initial fragment of the CES (stages 1 to i) */
			W[i] = X[i][j];
			/* check the operability and the cost of the initial fragment */
			if (S(W, i) == 0 && DefineCost(W, i) < minCost) {
				MasterEnumerateVariants(i + 1); }}}/* recursive search */(*@\label{ln429}@*)
	else {/* an initial fragment of G stages has been built (i == G + 1) */
		while (true) {
			/* receive messages from workers */
			msg = ReceiveWorkerMessage();(*@\label{ln432}@*)
			if (msg.type == SOLUTION_FOUND) {/* a solution received */(*@\label{ln433}@*)
				/* check the optimality criterion */
				if (DefineCost(msg.W, I) < minCost) {(*@\label{ln434}@*)
					/* store the new optimal solution */
					Wopt = msg.W; minCost = DefineCost(msg.W, I); }} (*@\label{ln436}@*)
				elseif (msg.type == REQUEST_WORK) { /* work request */(*@\label{ln437}@*)
				/* send the initial fragment of the CES and the current criterion value */
				SendWorkerMessage(msg.workerID, DO_WORK, W, minCost);(*@\label{ln438}@*)
				break; }}}
}
\end{lstlisting}

At the lowest level of recursion, the master waits for messages from the workers (line~\ref{ln432}); these messages can be of two types: \emph{solution found} (\texttt{SOLUTION\_FOUND}) or \emph{work request} (\texttt{REQUEST\_WORK}).

If the master receives a new complete CES variant from a worker (line~\ref{ln433}), its cost is compared with the cost of the current global optimal CES variant (line~\ref{ln434}).
If a new optimal CES variant has been found, it replaces the current optimal solution \texttt{Wopt}, and its cost is stored in \texttt{minCost} (line~\ref{ln436}).

If the master receives a request for a new task from an idle worker (line~\ref{ln437}), it sends this worker a \emph{do work} message (\texttt{DO\_WORK}) containing the initial fragment of a CES variant and the cost of the current optimal CES variant (line~\ref{ln438}).

The variable \texttt{minCost} is initialized with the sum of the maximum unit costs over all stages.
\subsection{Worker Process}
\label{sec:WorkerProcess}

A worker requests a task from the master (line~\ref{ln540}) and waits for a reply.
The reply can be of two types: \emph{do work} (\texttt{DO\_WORK}) or \emph{quit} (\texttt{QUIT}).
If a message with a new task has been received from the master (line~\ref{ln541}), containing an initial fragment of a CES variant and the cost of the current optimal CES variant, the recursive procedure
\texttt{WorkerEnumerateVariants()} is called (line~\ref{ln542}). Using this procedure, the worker performs a sequential depth-first traversal of the subtrees assigned to it, from level $G+1$ down to the lowest level $I$, completing the initial fragment received from the master to a full CES variant $W[i], i=\overline{G+1,I}$ (lines~\ref{ln547}--\ref{ln549}).

\begin{lstlisting}[%
caption={Pseudocode of the worker process},
float=htb,%
label={lst:parallelMpiWorker}]
Worker() {
	while (true) {
		/* request a new task */
		SendMasterMessage(workerID, REQUEST_WORK);(*@\label{ln540}@*)
		msg = ReceiveMasterMessage();
		/* a new task has been received */
		if (msg.type == DO_WORK) {(*@\label{ln541}@*)
			minCost = msg.minCost;
			foundNewSolution = false;
			WorkerEnumerateVariants(G + 1); (*@\label{ln542}@*)
			if (foundNewSolution) {/* a new solution has been found */
				/* send the solution to the master */
				SendMasterMessage(workerID, SOLUTION_FOUND, Wopt); }} (*@\label{ln543}@*)
		/* a termination message has been received */
		elseif (msg.type == QUIT) {(*@\label{ln544}@*)
			break; }}(*@\label{ln545}@*)
}

WorkerEnumerateVariants(i) {
	if (i <= I) {(*@\label{ln547}@*)
		for (j=1; j <= J[i]; j++) {
			/* build a complete CES variant */
			W[i] = X[i][j];
			/* check the operability and the cost of the CES variant */
			if (S(W, i) == 0 && DefineCost(W, i) < minCost) {(*@\label{ln548}@*)
				/* recursive search at the next level */
				WorkerEnumerateVariants(i + 1); }}}(*@\label{ln549}@*)
	else {/* all tree levels have been passed (i == I + 1) */(*@\label{ln550}@*)
		/* check the optimality criterion */
		if (DefineCost(W, I) < minCost) {(*@\label{ln551}@*)
			/* check the processing time constraint (1.5) */
			if (T(W) <= Tmax) {(*@\label{ln552}@*)
				/* store the new optimal solution */
				Wopt = W; minCost = DefineCost(Wopt, I);(*@\label{ln554}@*)
				foundNewSolution = true; }}}
}
\end{lstlisting}

The transition to the next recursion level is made in the same way as in the sequential version of the algorithm: only if the initial fragment of the CES consisting of stages 1 to $i$ is, first, operable and, second, not more expensive than the best CES variant found so far (line~\ref{ln548}).
Otherwise, the algorithm backtracks to the previous level of the tree.
When all levels of the tree have been passed (\texttt{i==I+1}, line~\ref{ln550}), the cost of the variant is compared with the cost of the current local best CES variant (line~\ref{ln551}).
If a better solution has been found, the processing time constraint (\ref{eq_5}) is checked (procedure \texttt{T()}, line~\ref{ln552}).
If constraint (\ref{eq_5}) is satisfied, a new optimal CES variant has been found; it replaces the current local optimal solution \texttt{Wopt}, and its cost is stored in \texttt{minCost} (line~\ref{ln554}).
After the recursive procedure has finished, the worker sends the local optimal CES variant to the master (line~\ref{ln543}).
The worker then becomes free and sends the master a request for a new task.
On receiving a termination message from the master (line~\ref{ln544}), the worker terminates (line~\ref{ln545}).

\subsection{Results of Computational Experiments}
\label{chap3_sect2_res}
For the computational experiments, a program was developed in C++ using the Message Passing Interface (MPI) library~\cite{mpi}.

The test case was the equipment selection problem for the CES described in Section~\ref{chap1_sect2}, with $5$ standard equipment sizes available for each processing stage.

The program was tested on a computing cluster of $198$ nodes, each equipped with two six-core processors (Intel Westmere X5650, $2.6$~GHz, $2$ or $4$~GB of RAM) running Red Hat Linux.
The cluster nodes are connected by a high-speed, low-latency switched InfiniBand network.
The program was compiled with Intel C++ Compiler 11.1.

The parallel program requires at least two processors (one for the master and one for a worker).

\begin{figure}[htb]
	\centering
	\includegraphics[width=0.97\textwidth]{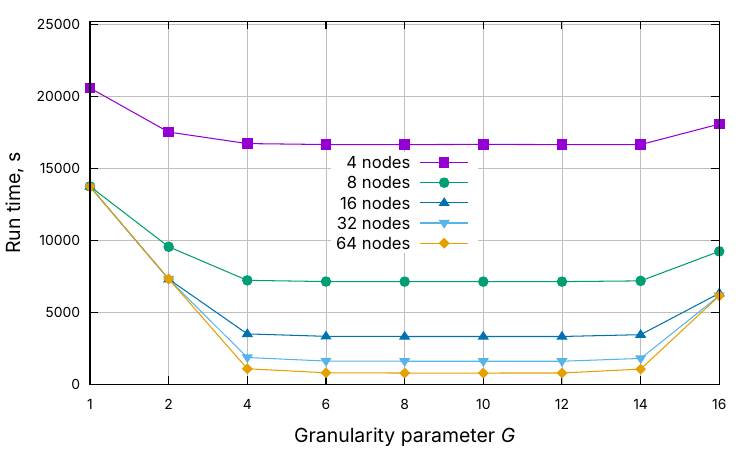}
	\caption{Run time of the program versus the granularity parameter $G$}
	\label{fig:mpiGranularity}
\end{figure}

Figure~\ref{fig:mpiGranularity} shows how the run time of the parallel program on different numbers of processors depends on the granularity parameter~$G$.

\begin{figure}[htb]
	\centering
	\includegraphics[width=0.97\textwidth]{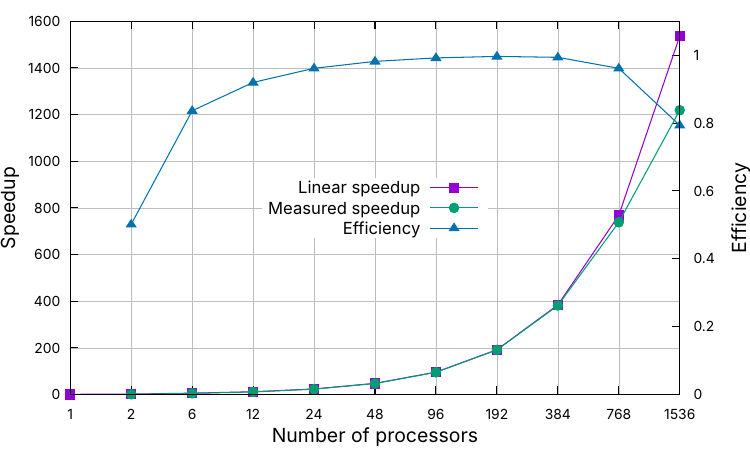}
	\caption{Speedup and efficiency of the program versus the number of processors ($G = 10$)}
	\label{fig:mpiSpeedup}
\end{figure}

The speedup and the processor efficiency were also studied.
Figure~\ref{fig:mpiSpeedup} (the horizontal axis is logarithmic) shows the speedup of the program and the processor efficiency.

The study of speedup and efficiency showed that the parallel program achieves near-linear speedup and fairly high scalability.

However, Fig.~\ref{fig:mpiSpeedup} shows that the processor efficiency and the speedup decrease when the number of processors exceeds $768$ ($64$ cluster nodes).
This happens because the master process becomes a bottleneck of the program when it serves a large number of workers, which reduces the speedup and efficiency.
This problem can be addressed by the hierarchical master--worker paradigm, in which several submasters are added to the main master process~\cite{Aida2006,Castellanos2013}.

%%%% Chapter 4
\chapter{Branch-and-Bound on Graphics Processing Units}
\label{chapter4}

Chapter~\ref{chapter3} considered the parallelization of branch-and-bound for shared-memory and distributed-memory systems.
This chapter deals with its implementation for heterogeneous computing systems equipped with graphics processing units (GPUs).

Modern high-performance computing systems have a heterogeneous architecture that includes a general-purpose multi-core Central Processing Unit (CPU) and a set of various accelerators~\cite{wen2015heterogeneous}.
The performance of heterogeneous systems is increased not only by increasing the number of processors and cores, but also by incorporating (either integrated or discrete) specialized accelerators for parallel computing, such as Graphics Processing Units (GPUs), Digital Signal Processors (DSPs), and Field-Programmable Gate Arrays (FPGAs)~\cite{Sluga2014heterogeneous}.
Such platforms can considerably improve performance in applications such as image processing, medical visualization, and data acquisition, while keeping power consumption fairly low and at a relatively small additional cost.

Today, GPUs are among the most powerful computing devices available at an affordable price~\cite{Borisenko2017}.
Architecturally, a CPU consists of a few cores optimized for sequential processing, whereas a GPU consists of thousands of smaller and more energy-efficient cores designed to handle many tasks simultaneously.

Originally, graphics accelerators were devices intended for displaying two- or three-dimensional graphics on a computer screen.
As they evolved, they acquired support for so-called shaders, small data-processing programs executed on the graphics accelerator itself.
The originally fixed algorithm for computing lighting and transforming vertex coordinates was replaced by a user-defined one.

Later, it became possible to write programs computing the color of a pixel on the screen.
For this reason, programs for GPUs came to be called vertex and pixel shaders.
Shaders enabled developers to create, for example, their own special effects, which considerably increased the complexity and realism of computer graphics and, accordingly, the attractiveness of the resulting images.
Shaders had to be programmed in the GPU assembly language, their length did not exceed $20$ instructions, branch instructions were not supported, and computations were performed in fixed-point format.
Nevertheless, the advent of the first shaders, and with them the possibility of executing program code directly on the GPU, made it possible to use graphics accelerators for problems unrelated to graphics processing, i.e., for so-called General-Purpose computing on GPUs (GPGPU)~\cite{Owens2007}.

Subsequently, GPUs became fully programmable devices.
Branching operations and loops appeared, and in 2003 support for 32-bit precision computations was added.
As GPUs grew more sophisticated, more advanced software development tools for them appeared as well.

The chapter is organized as follows.
Section~\ref{chap4_sect1} discusses the general principles and specific features of GPU programming.
Section~\ref{chap4_sect2} presents a parallel program in which the host splits the search tree into subtrees processed by independent GPU threads, as well as two ways of traversing the tree on the device: recursive and iterative.
Section~\ref{chap4_sect3} proposes ways of optimizing the program code: using shared memory and reducing branch divergence.
Section~\ref{chap4_sect4} reports the results of computational experiments and assesses the efficiency of using GPUs for the problem under consideration.

\section{General Principles and Features of GPU Programming}
\label{chap4_sect1}

One of the first programming interfaces worth mentioning is Microsoft Direct3D, which was the first to support shaders (the HLSL language).
For GPU programming, the SIMD (Single Instruction, Multiple Data) approach was proposed.
This approach assumes that a group of processors operating in parallel act on different data, but at any given moment all of them must execute the same instruction.

An important milestone in the development of general-purpose GPU computing was the introduction in November 2006 of the Compute Unified Device Architecture (CUDA), a hardware and software platform developed by NVIDIA and released together with the GeForce 8800 GTX graphics card~\cite{2022CudaProgrammingGuide}.
This technology is based on an extension of the C language and provides access to the instruction set of the graphics accelerator and control over its memory for parallel computing.
Today, CUDA is one of the most popular programming technologies for computing systems equipped with GPUs~\cite{boreskov2010}.

The operation of a GPU can be described by the stream computing model.
A stream (in the data sense) is a sequence of elements of the same type, each of which is processed independently of the others.
A stream is processed by compute kernels.
The body of a kernel contains the computations for a single element; it is invoked many times (possibly in parallel), once for each element.
The CUDA programming model is also based on the stream computing model.

An application includes code executed on the \textbf{host}, which implements only the sequential parts of the algorithm, prepares data and copies them to the \textbf{device}, sets the parameters of the \textbf{kernel}, and launches it.

A kernel is a special function (kernel function), hundreds or thousands of instances of which run on the device in parallel, each processing its own part of the input data~\cite{Kazennov2010}.
In effect, the GPU acts as a specialized coprocessor to the CPU, with its own memory and the ability to execute a large number of threads in parallel.

Threads in CUDA are organized in a two-level hierarchy.
At the upper level, a \textbf{grid} is a one-, two-, or three-dimensional array of \textbf{blocks}.
In early versions of CUDA, the third dimension of block indices always had to equal $1$.
At the lower level, a block is a one-, two-, or three-dimensional array of \textbf{threads}.
All threads run in parallel; inside a thread, information about the block and grid it belongs to is available through the built-in CUDA variables \emph{blockIdx} (the index of the block in the grid) and \emph{threadIdx} (the index of the thread in the block).
Each of these variables is a three-component integer vector.
The dimensions of the grid and of a block are available through the built-in variables \emph{gridDim} and \emph{blockDim}~\cite{boreskov2010,sanders2011}.

Threads are divided into groups of $32$ called \textbf{warps}.
In fact, only threads within one warp are executed physically simultaneously, and they execute the same instructions in parallel according to the SIMD principle.
Although this approach is very similar to the SIMD model, there are fundamental differences, and NVIDIA uses the term SIMT (Single Instruction, Multiple Threads).

The thread scheduler periodically switches execution from one warp to another, so threads of different warps may be at different points of program execution.
Warps are managed transparently by the GPU itself.

For developing GPU programs, NVIDIA released its own extensions of the C language and the NVCC compiler for building such programs, and introduced the new file extension *.cu for files containing CUDA source code.
The C language extensions comprise specifiers for functions and variables, new built-in data types, built-in variables (inside a kernel), and a directive for launching a kernel from C code.

CUDA offers two Application Programming Interfaces (APIs): the low-level driver API and the high-level runtime API. The runtime API is implemented on top of the driver API; it is less flexible but more convenient for writing programs, since it does not require a large number of initialization operations.
In general, CUDA is used according to the following scheme:
\begin{itemize}
\item allocate memory on the GPU;
\item copy data from CPU memory to the allocated GPU memory;
\item launch the kernel function (or several kernel functions in succession);
\item copy the results back to CPU memory;
\item free the allocated GPU memory.
\end{itemize}

The capabilities of a CUDA GPU are characterized by its Compute Capability (CC), whose major number corresponds to the architecture version and whose minor number corresponds to incremental changes within the architecture. For example, GPUs with CC~1.0 do not support double-precision computations, GPUs with CC below~2.0 do not support recursion in kernel functions, and so on.
Today, applications based on the CUDA architecture can be developed in various languages and APIs, including C, C++, Fortran, OpenCL, and DirectCompute.
The CUDA architecture is actively developed and supported by NVIDIA.
Its drawbacks are that it is proprietary (the technology is developed and supported by NVIDIA) and cannot be used to program graphics cards of other vendors.
Alternative tools for developing GPGPU applications include the Open Computing Language (OpenCL) and Microsoft DirectX~11 DirectCompute~\cite{Diaz2012,kirk2016programming}.

OpenCL is a framework for writing programs that perform parallel computations on various GPUs and CPUs~\cite{Inc2022}.
The OpenCL framework comprises a programming language based on the C99 standard and an application programming interface (API).
OpenCL provides both task-level and data-level parallelism and is an implementation of the GPGPU approach~\cite{Munshi2011}.
OpenCL is a fully open standard developed and maintained by the non-profit Khronos Group consortium, whose members include a number of major companies such as Apple, AMD, Intel, NVIDIA, Sun Microsystems, and Sony Computer Entertainment.

Microsoft DirectCompute is an application programming interface that is part of the Microsoft DirectX API collection and is intended for computers running Microsoft Windows operating systems~\cite{zink2016practical}.
DirectCompute is designed for general-purpose computing on GPUs and is thus an implementation of the GPGPU concept.
DirectCompute was originally released as part of DirectX~11, but later became available for DirectX~10 and DirectX~10.1 as well.

In this book, NVIDIA CUDA is used to program GPGPU applications.

\section{Development of the Parallel Program}
\label{chap4_sect2}
To split the original search tree into subtrees for parallel processing, we use the following strategy, which is largely similar to the master--worker paradigm presented in Section~\ref{chap3_sect2}.
Here the CPU plays the role of the master, while the workers run in parallel on the GPU.

The granularity parameter $G$ is used to vary the degree of parallelism: the part of the tree above this level is processed sequentially on the CPU, while the subtrees below it are turned into parallel tasks and processed by separate threads on the GPU.

\begin{figure}[htb]
	\centering
	\includegraphics[width=0.97 \textwidth]{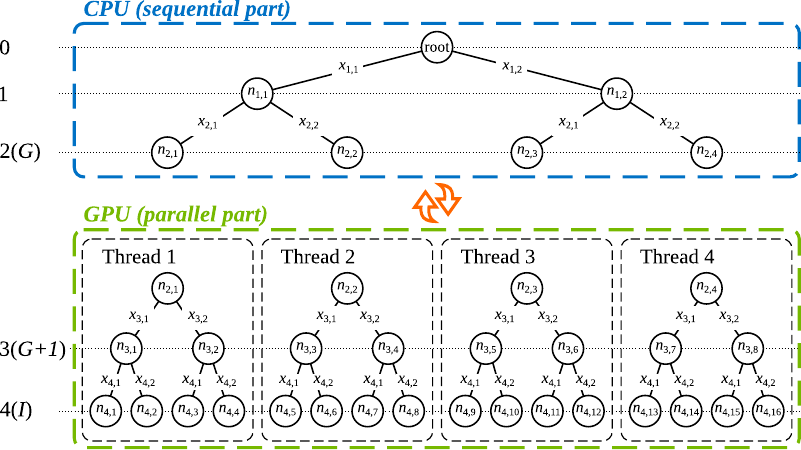}
	\caption{Splitting the search tree into subtrees for parallel processing on the GPU}
	\label{fig:SearchTreeDividing}
\end{figure}

The depth-first traversal of all tree nodes $N_i=\{ n_{i,1},n_{i,2},\ldots, n_{i,k}\},$ $i=\overline{G+1,I},\, k=\overline{1,K_i},\, K_i=\prod_{l=1}^i J_l$ is performed by independent threads on the GPU.
The total number of such threads is
$N_{threads} = K_G = \prod_{l=1}^{G} J_l, (K_{0}=1), 0\leq G \leq I-1$.
In particular, Fig.~\ref{fig:SearchTreeDividing} shows how the original search tree is split into subtrees for the simple CES example of Fig.~\ref{fig:simple_CES}.

For the granularity parameter $G=2$, levels $0$ to $2$ are processed on the CPU; the initial fragments of CES variants are then transferred to the GPU, and levels $3$ and $4$ are processed in parallel on the GPU.
The CPU does not check constraints (\ref{eq_4}) and (\ref{eq_5}), does not compute the objective function (\ref{eq_1}), and does not check the upper and lower bounds of the branch-and-bound method: all these operations are performed on the GPU.

\subsubsection*{Host (CPU)}
\label{chap4_host_code}
The host (see Listing~\ref{lst:hostCodeBandB}) starts by calling the procedure \texttt{ReadInputData()}, which reads the input data from a file (line~\ref{ln603}).

The variable \texttt{inData} (input data) is a pointer to a structure whose fields are variables and arrays of various dimensions holding all the necessary input data.
These data are the same for all threads and do not change while the problem is being solved, i.e., they are constants.

The function \texttt{ThreadsNumber()} computes, from the input data \texttt{inData} and the granularity parameter \texttt{G}, the number of parallel threads \texttt{numThreads} required to process all subtrees (line~\ref{ln606}).
This value is then used when calling the procedure \texttt{PrepOperationalData()}, which prepares the working data (line~\ref{ln608}).

The variable \texttt{oprData} (working data) is a pointer to a structure whose fields are variables and arrays of various dimensions whose values are modified independently by all threads while the problem is being solved.
Therefore, one of the array indices is the thread identifier \texttt{threadID}.
Both procedures, \texttt{ReadInputData()} and \texttt{PrepOperationalData()}, initialize the variables and perform all memory allocations needed for the arrays.

\begin{lstlisting}[%
caption={Pseudocode of the host (CPU) algorithm},%
float=htb,%
label={lst:hostCodeBandB}]
main() {	/* prepare the input data */
	ReadInputData(inData); (*@\label{ln603}@*)
	/* determine the required number of threads */
	numThreads = ThreadsNumber(inData, G); (*@\label{ln606}@*)
	/* prepare the working data */
	PrepOperationalData(oprData, inData, numThreads);(*@\label{ln608}@*)
	/* traverse the tree to split it into subtrees */
	/* and build the initial fragments of the CES */
	threadID = 0;
	EnumerateHost(1);(*@\label{ln610}@*)
	/* transfer the input and working data to the device */
	cudaMemcpyHostToDevice(inData);(*@\label{ln612}@*)
	cudaMemcpyHostToDevice(oprData);
	cudaMemcpyHostToDevice(W, G);(*@\label{ln614}@*)
	/* determine the kernel launch parameters */
	blocksPerGrid = numThreads / MAX_THREADS_PER_BLOCK + 1; (*@\label{ln616}@*)
	threadsPerBlock = MAX_THREADS_PER_BLOCK;(*@\label{ln617}@*)
	/* launch the kernel function on the device */
	FindSolution<<<blocksPerGrid, threadsPerBlock>>>(G); (*@\label{ln619}@*)
	/* synchronize with the device */
	cudaDeviceSynchronize(); (*@\label{ln621}@*)
	/* copy the search results from the device to the host */
	cudaMemcpyDevicetoHost(W);(*@\label{ln623}@*)
	cudaMemcpyDevicetoHost(Cost);
	cudaMemcpyDevicetoHost(minCost);(*@\label{ln625}@*)
	/* find the global optimal solution */
	for (n = 0; n < numThreads; n++){(*@\label{ln627}@*)
		if(Cost[n] <= minCost){
		    Wopt = W[n];
			minCost = Cost[n];}}(*@\label{ln630}@*)
}

EnumerateHost(i){(*@\label{ln633}@*)
	if (i <= G){/* current search depth does not exceed G */
		for (j=1; j <= J[i]; j++){
			Wloc[i] = X[i][j];
			EnumerateHost(i + 1);}} /* recursive search */
	else{/* an initial fragment of G stages has been built (i == G + 1) */
		/* store the initial fragment of the CES */
		W[threadID] = Wloc;
		threadID++;}(*@\label{ln643}@*)
}(*@\label{ln644}@*)
\end{lstlisting}

Next, the procedure \texttt{EnumerateHost()} is called at the first level of the search tree (line~\ref{ln610}).
Using the recursive procedure \texttt{EnumerateHost()}, the host performs a sequential depth-first traversal of the tree down to the level given by the granularity parameter \texttt{G}, forming the initial fragments of CES variants and storing them in the array \texttt{W} (lines~\ref{ln633}--\ref{ln644}).
\texttt{W} is an array of \texttt{numThreads} vectors of length \texttt{I} each.
The elements of each vector hold the selected main size of the unit for each stage of the CES.
Note that the host does not check constraints (\ref{eq_4}) and (\ref{eq_5}) and does not compute the upper and lower bounds of the objective function (\ref{eq_1}).
All these operations are performed on the GPU.

Once the initial fragments of the CES have been formed, they are transferred to the device for further computation (lines~\ref{ln612}--\ref{ln614}).
Then the required number of blocks in the grid, \texttt{blocksPerGrid} (line~\ref{ln616}), and the number of threads per block, \texttt{threadsPerBlock} (line~\ref{ln617}), are computed, and the kernel function \texttt{FindSolution()} is launched (line~\ref{ln619}).
When the required number of blocks is computed (line~\ref{ln616}), the result of the division may be non-integer and is automatically truncated to an integer; therefore, $1$ is added to the quotient to round it up.

In CUDA, a kernel function is launched asynchronously: after the kernel has been launched on the device, control returns to the host immediately (the host does not wait for the kernel to finish).
Therefore, the host must call \texttt{cudaDeviceSynchronize()} (line~\ref{ln621}) to pause and wait until all GPU threads have finished (i.e., synchronize with the device).
After the device has finished, the results are transferred to the host (lines~\ref{ln623}--\ref{ln625}).
Here \texttt{Cost} is a one-dimensional array of \texttt{numThreads} elements used to store the local costs of the CES variants found by each thread.
The smallest value in the array \texttt{Cost} of local costs is then found (lines~\ref{ln627}--\ref{ln630}) and stored in the variable \texttt{minCost}.
This is the global optimal cost of the best CES variant.
The optimal CES variant itself is stored in the array \texttt{Wopt} with \texttt{I} elements, each holding the main size of the unit for the corresponding stage.

\subsubsection*{Device (GPU)}
\label{chap4_kernel_code}
Two approaches to traversing the search tree were implemented: recursive (see Listing~\ref{lst:deviceRecursiveCode}) and iterative (see Listing~\ref{lst:deviceIterativeCode}).
The recursive approach is convenient for traversing tree structures and can be used on NVIDIA GPUs with Compute Capability $2.0$ and higher.
The iterative approach can be used on older NVIDIA devices that do not support recursion (Compute Capability below $2.0$).

\begin{lstlisting}[%
caption={Pseudocode of the recursive device (GPU) algorithm},%
float=htb,%
label={lst:deviceRecursiveCode}]
__global__ void FindSolution(G) (*@\label{ln701}@*)
{	threadID = blockDim.x * blockIdx.x + threadIdx.x; (*@\label{ln702}@*)
	/* if the identifier is less than the number of threads */
	if(threadID < numThreads) {(*@\label{ln704}@*)
		EnumerateDevice(threadID, G+1);}/* start the tree traversal */ (*@\label{ln705}@*)
}
__device__ EnumerateDevice(threadID, i){
	for (j=1; j <= J[i]; j++) {
		/* build a complete CES variant */
		Wloc[i] = X[i][j];
		if (i < I) { (*@\label{ln712}@*)
		/* check the operability and the upper bound */
			if (S(Wloc, i) == 0 && DefineCost(Wloc, i) < minCost){ (*@\label{ln714}@*)
				EnumerateDevice(threadID, i + 1); }}/* recursive search */(*@\label{ln715}@*)
		else {/* the last level of the tree has been reached */ (*@\label{ln716}@*)
			/* check the operability (constraint 1.4) and the upper bound */
			if(S(Wloc, I) == 0 && DefineCost(Wloc, I) < minCost) { (*@\label{ln717}@*)
				if (T(Wloc) <= Tmax) { /* check constraint 1.5 */ (*@\label{ln718}@*)
					/* store the new local optimal solution */
					W[threadID] = Wloc; Cost[threadID] = DefineCost(Wloc, I);(*@\label{ln720}@*)
					atomicMinDbl(minCost, Cost[threadID]); }}}}(*@\label{ln721}@*)
}
\end{lstlisting}

The kernel function \texttt{FindSolution()} (see Listing~\ref{lst:deviceRecursiveCode}, line~\ref{ln701}) determines the thread identifier \texttt{threadID} (line~\ref{ln702}).
Its value must be less than the number of threads used, \texttt{numThreads} (line~\ref{ln704}).

This check is necessary because the number of blocks in the grid is rounded up when the kernel parameters are computed (see Listing~\ref{lst:hostCodeBandB}, line~\ref{ln616}).
The number of threads launched on the GPU may exceed \texttt{numThreads}, whereas all dynamic memory allocations for the arrays in \texttt{oprData} were made for exactly \texttt{numThreads} elements.
The recursive procedure \texttt{EnumerateDevice()} is called for all threads whose identifier is less than \texttt{numThreads} (line~\ref{ln705}).

Using this procedure, each thread performs a sequential depth-first traversal of the subtrees assigned to it (from level \texttt{G+1} down to the lowest level \texttt{I}), completing the initial fragment of the CES variant received from the host to a full CES variant.

To reduce the enumeration, an algorithm based on branch-and-bound is used here as well.
The set of feasible solutions is successively partitioned into subsets~\cite{Brassard1996}.
A lower bound on the objective function is then determined for each subset.
If the lower bound is not less than the best solution found so far, the subset is discarded, and the process continues until a single solution is obtained (assuming that the optimality criterion is the minimum of the objective function).
If the current tree level \texttt{i} is not the lowest level \texttt{I} (line~\ref{ln712}), the compatibility constraint (\ref{eq_4}) for the joint operation of the CES equipment is checked at level \texttt{i} (procedure \texttt{S()}, line~\ref{ln714}) for the initial fragment of the CES consisting of the levels already passed above and the current one (the CES fragment consisting of stages \texttt{1} to \texttt{i}).
If part of the flowsheet is already inoperable, there is no point in enumerating units at the lower levels.
In addition, the cost of the initial fragment of the CES variant (procedure \texttt{DefineCost()}, line~\ref{ln714}) is compared with the cost \texttt{minCost} of the current optimal CES variant.
As a result, the number of enumerated variants is reduced.
The variable \texttt{minCost} can be read and written by all threads running on the device.
Its initial value is the sum of the maximum unit costs over all stages.

The algorithm descends to the next recursion level (a call of \texttt{EnumerateDevice()} at level \texttt{i + 1}, line~\ref{ln715}) only if the initial fragment of the CES variant consisting of stages \texttt{1} to \texttt{i} is, first, operable and, second, cheaper than the best CES variant found so far; otherwise, it backtracks to the previous level of the tree.

Thus, when the last level of the tree is reached (\texttt{i == I}, line~\ref{ln716}), a complete operable CES variant (one satisfying constraint~(\ref{eq_4})) has been formed.
Its cost (the result of \texttt{DefineCost()}) is then compared with the cost \texttt{minCost} of the current best CES variant (line~\ref{ln717}).
If a better solution has been found, the processing time constraint (\ref{eq_5}) is checked (procedure \texttt{T()}, line~\ref{ln718}).
If constraint (\ref{eq_5}) is satisfied, a new optimal CES variant has been found; it replaces the current optimal solution \texttt{W} (line~\ref{ln720}), and its cost is stored in \texttt{minCost} using the atomic function \texttt{atomicMinDbl()} (line~\ref{ln721}).
Atomicity guarantees that the operation is performed correctly when many threads attempt to execute it simultaneously.

A well-known way of traversing a tree without explicit recursion is the iterative use of a stack~\cite{graham1994}.
A stack is a data structure representing a list of elements organized according to the LIFO (Last-In, First-Out) principle.
The pseudocode of the algorithm implementing this approach is shown in Listing~\ref{lst:deviceIterativeCode}.
The kernel function \texttt{FindSolution()} is identical to the one in Listing~\ref{lst:deviceRecursiveCode} (lines~\ref{ln701}--\ref{ln705}).

\begin{lstlisting}[%
caption={Pseudocode of the iterative device (GPU) algorithm},%
float=htb,%
firstnumber=7,%
label=lst:deviceIterativeCode]
...
__device__ EnumerateDevice(threadID, i)
{	/* declare a standard Last-In, First-Out (LIFO) stack */
	stack[MAX_STACK_SIZE]; (*@\label{ln812}@*)
	stackTop = 0;
	/* define the standard stack operations */
#define push(x)	stack[stackTop++] = (x) (*@\label{ln813}@*)
#define pop()	stack[--stackTop] (*@\label{ln814}@*)
#define empty	(stackTop == 0) (*@\label{ln815}@*)
	/* push the first node of level i onto the stack */
	push(1); (*@\label{ln817}@*)
	/* repeat while the stack is not empty */
	while (!empty){ (*@\label{ln819}@*)
		j = pop();/* pop the next node from the stack */ (*@\label{ln820}@*)
		/* if there are no more nodes at this level, go one level up */
		if(j == J[i] + 1) {i--;} (*@\label{ln823}@*)
		else {
			/* otherwise push the next node of the current level */
			push(j + 1); (*@\label{ln826}@*)
			/* build a complete CES variant */
			Wloc[i] = X[i][j];(*@\label{ln830}@*)
			/* check the operability and the cost of the CES variant */
			if (S(Wloc,i) == 0 && DefineCost(Wloc, i) < minCost){(*@\label{ln828}@*)
				if (i < I) {/* the last level has not been reached */ (*@\label{ln831}@*)
					i++;/* go to the next level of the tree */(*@\label{ln832}@*)
				/* push the first node of the next level */
					push(1); }(*@\label{ln834}@*)
				else {/* the last level has been reached (i==I) */ (*@\label{ln835}@*)
					/* check the optimality criterion */
					if (DefineCost (Wloc, I) < minCost){(*@\label{ln837}@*)
				/* check the processing time constraint (1.5) */
						if (T(Wloc) <= Tmax){(*@\label{ln839}@*)
						/* store the new local optimal solution */
							W[threadID] = Wloc;(*@\label{ln841}@*)
							Cost[threadID] = DefineCost(Wloc, I);(*@\label{ln842}@*)
							atomicMinDbl(minCost, Cost[threadID]);}}}}}} (*@\label{ln843}@*)
}
\end{lstlisting}
The stack is implemented as a one-dimensional static array \texttt{stack} of length \texttt{MAX\_STACK\_SIZE} (line~\ref{ln812}).
The stack size (the value of \texttt{MAX\_STACK\_SIZE}) must be at least the number of tree levels: \texttt{MAX\_STACK\_SIZE >= I}.
Data can be added to or removed from the stack only at its top.
The variable \texttt{stackTop} points to the top of the stack.
Next, the standard elementary stack operations are defined.
The \texttt{push} operation adds a new element (line~\ref{ln813}),
the \texttt{pop} operation removes the element at the top of the stack (line~\ref{ln814}),
and the \texttt{empty} operation checks whether the stack is empty (line~\ref{ln815}).

Initially, the stack is empty, and the first tree node at level $G+1$ (the first unit variant for stage $G+1$ of the CES) is pushed onto it (line~\ref{ln817}).
Then the tree nodes are enumerated iteratively until the stack becomes empty (line~\ref{ln819}).
At each iteration of the loop, the most recently pushed node is popped from the stack (line~\ref{ln820}).
If there are no more nodes at the current level, the algorithm moves one level up (line~\ref{ln823}).
Otherwise, the next node of the current level is pushed onto the stack (line~\ref{ln826}).
If the current initial fragment of the CES consisting of stages \texttt{1} to \texttt{i} satisfies the compatibility condition (\ref{eq_4}) and its cost is lower than the cost of the best CES variant found so far (line~\ref{ln828}), the unit variant at the current level is appended to the preceding operable initial fragment of the CES consisting of stages \texttt{1} to \texttt{i-1} (line~\ref{ln830}).
Otherwise, the algorithm does not descend to deeper levels of the tree and proceeds to the next loop iteration.
If the current level is not the last one (line~\ref{ln831}), the algorithm moves to the next level (line~\ref{ln832}) and pushes its first node onto the stack (line~\ref{ln834}).
If the last level of the tree has been reached (line~\ref{ln835}), the algorithm proceeds in the same way as the recursive one (see Listing~\ref{lst:deviceRecursiveCode}).
The cost of the operable CES variant (the result of \texttt{DefineCost()}) is compared with the cost \texttt{minCost} of the current best CES variant (line~\ref{ln837}).
If a better solution has been found, the processing time constraint (\ref{eq_5}) is checked (procedure \texttt{T()}, line~\ref{ln839}).
If constraint (\ref{eq_5}) is satisfied, a new optimal CES variant has been found; it replaces the current optimal solution (line~\ref{ln841}), and its cost is stored (lines~\ref{ln842}--\ref{ln843}).

For the computational experiments, the algorithm variants presented above were implemented in the C programming language.
They use the mathematical model of CES operation (including, among other things, the expressions for checking the constraints and computing the optimality criterion) described in detail in~\cite{Malygin2005mathematical}.

The amount of input data required for the CES equipment selection problem is small (tens of kilobytes), but the data have a rather complex structure.

In particular, in~\cite{Borisenko2011,borisenko2012,borisenko2013} the model of CES operation was implemented using an object-oriented approach (the C++ programming language) and the Standard Template Library (STL), in particular the \texttt{std::vector} template whose elements were other vectors.
For example, a CES is a vector of process routes, a process route is a vector of processing stages, a processing stage is a vector of operating cycles, an operating cycle is a vector of process operations, and so on.

However, at the time of this study, CUDA had no full-fledged implementation of STL containers that could be used in code executed on the GPU.
For this reason, the program code was completely rewritten without STL.
Instead of numerous nested \texttt{std::vector} objects, arrays (about $50$ of them) of various dimensions (from $1$ to $5$) were used.

To make the arrays occupy contiguous regions of memory (flat memory), multidimensional arrays were converted into one-dimensional ones (see Listing~\ref{lst:likeVector}).
This guarantees that the data occupy a single contiguous memory region,
which is convenient when transferring data with \texttt{memcpy}-like functions.
Moreover, since the data are not fragmented in memory, the number of cache misses decreases, which considerably increases the program's performance.

On the test platform described below, the new version of the sequential CPU program (without nested vectors) turned out to be $8$--$9$ times faster than the previous STL-based version.
In what follows, all computational experiments (performance comparisons, speedup calculations, etc.) are carried out with the new version of the program.

\begin{lstlisting}[%
caption={Accessing multidimensional arrays as one-dimensional ones},%
float=htb,%
numbers=none,
label={lst:likeVector}]
	...
	typeName *A; /* A[i][j], a two-dimensional array of size I*J */
	typeName *B; /* B[i][j][k], a three-dimensional array of size I*J*K */
	/* allocate memory */
	A = malloc (I * J * sizeof (typeName));
	B = malloc (I * J * K * sizeof (typeName));
	/* initialize the arrays */
	for (i = 0; i < I; i ++){
		for (j = 0; j < J; j ++){
			A[i * J + j] = i + j;
			for (k = 0; k < K; k ++){
				B[i * J * K + j * K + k] = i * j * k ;}}}
	...
\end{lstlisting}

\section{Optimization of the Program Code}
\label{chap4_sect3}

To increase the performance of the program, two optimization approaches were applied: efficient use of shared memory and reduction of branching in the kernel function.

Physically, GPU memory can be divided into Dynamic Random-Access Memory (DRAM) and memory located directly on the chips of the streaming multiprocessors (SMs).
However, the classification of memory is not limited to its physical location.

\begin{figure}[htb]
    \centering
	\includegraphics[width=0.97 \textwidth]{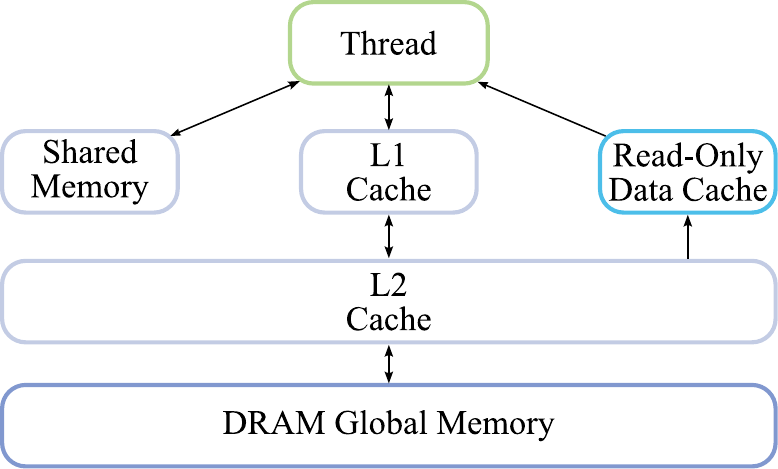}
	\caption[Memory hierarchy of NVIDIA GPUs of the Kepler and Fermi microarchitectures]%
	{Memory hierarchy of NVIDIA GPUs of the Kepler and Fermi microarchitectures
		(based on~\cite{2014Kepler,Bruel2017})}
	\label{fig:KeplerMemoryHierarchy}
\end{figure}

The memory hierarchy of NVIDIA GPUs of the Kepler and Fermi microarchitectures is shown in Fig.~\ref{fig:KeplerMemoryHierarchy}~\cite{2014Kepler,Bruel2017}.

Global memory is physically implemented as DRAM chips soldered directly onto the graphics board.
It is the largest memory on the card, measured in gigabytes.
However, since it is located outside the multiprocessors, it is the slowest type of memory~\cite{boreskov2010}.
In the Fermi and Kepler microarchitectures, global memory is cached by the L2 and L1 caches, but it nevertheless remains the slowest (with a latency of $400$ to $600$ clock cycles without caching).
Global memory can be read and written by all threads of the grid.
Optimizing global memory accesses usually has a very large effect on program performance.

In addition, each streaming multiprocessor has a small amount ($64$~KB) of on-chip memory.
In the Fermi microarchitecture, the programmer can configure this memory either as $48$~KB of shared memory and $16$~KB of L1 cache, or as $16$~KB of shared memory and $48$~KB of L1 cache.
The Kepler microarchitecture also allows this memory to be split in half: $32$~KB of shared memory and $32$~KB of L1 cache.
The lifetime of data in shared memory coincides with the lifetime of the block.
In addition to the L1 cache, the Kepler microarchitecture has a $48$~KB read-only data cache.
The L2 cache, shared by all SMs, is the main point at which data and instruction read/write requests between the streaming multiprocessors and DRAM are unified.
Constant memory physically resides in DRAM (like global memory), but is cached for reading on all devices.
Constant memory is read-only for the GPU and readable and writable for the CPU.

In the first implementation of the program, all input data \texttt{inData} were loaded into constant memory and the working data \texttt{oprData} into global memory.
Shared memory is the fastest, since it is located directly on the SM chip.
Therefore, in the first optimization (optimization \emph{Mo1}), the kernel function \texttt{FindSolution()} was modified (see Listing~\ref{lst:modificationMo1}) so as to move the local array \texttt{Wloc} (line~\ref{ln907}) and, for the iterative algorithm, also the array \texttt{stack} (line~\ref{ln909}) from global memory to shared memory using the procedure \texttt{SendDataByMemoryPtr()}~\cite{Huang2017}.

\begin{lstlisting}[%
caption={Pseudocode of the optimized algorithm: optimization \emph{Mo1}},%
float=htb,%
firstnumber=3,%
label={lst:modificationMo1}]
...
	threadID = blockDim.x * blockIdx.x + threadIdx.x; (*@\label{ln904}@*)
	__shared__ sharedMemory[];(*@\label{ln905}@*)
	/* copy data to shared memory */
	SendDataByMemoryPtr(Wloc, sharedMemory);(*@\label{ln907}@*)
#ifdef ITERATIVE
	SendDataByMemoryPtr(stack, sharedMemory);(*@\label{ln909}@*)
#endif
	/* if threadID is less than the number of threads */
	if(threadID < numThreads){(*@\label{ln912}@*)
		/* start the tree traversal */
		EnumerateDevice(threadID, G+1);}(*@\label{ln914}@*)
...
\end{lstlisting}
Both arrays, \texttt{Wloc} and \texttt{stack}, are used intensively during the traversal of the decision tree.

In the second variant (optimization \emph{Mo2}), the input data \texttt{inData}, whose size is several tens of kilobytes, were moved from constant memory to shared memory (see Listing~\ref{lst:modificationMo2}, line~\ref{ln1007}).

For this purpose, the procedure \texttt{SendDataByMemoryPtr()} is used; it runs on the GPU and transfers data to the given memory address~\texttt{sharedMemory}.

\begin{lstlisting}[%
caption={Pseudocode of the optimized algorithm: optimization \emph{Mo2}},%
float=htb,%
firstnumber=3,%
label={lst:modificationMo2}]
...
	threadID = blockDim.x * blockIdx.x + threadIdx.x; (*@\label{ln1004}@*)
	__shared__ sharedMemory[];(*@\label{ln1005}@*)
	/* copy data to shared memory */
	SendDataByMemoryPtr(inData, sharedMemory);(*@\label{ln1007}@*)
	/* if threadID is less than the number of threads */
	if(threadID < numThreads){(*@\label{ln1009}@*)
		/* start the tree traversal */
		EnumerateDevice(threadID, G+1);}(*@\label{ln1011}@*)
...
\end{lstlisting}
The amount of extremely fast shared memory \texttt{sharedMemory[]} is specified by the programmer as a kernel launch parameter, but cannot exceed $48$~KB because of GPU hardware limitations.
The threads, grouped into blocks by the programmer, are executed on the GPU.
For each SM separately, the warp scheduler divides the blocks into warps, and all threads of a warp always belong to the same block.
It is within a warp that threads are executed physically simultaneously.
In the CUDA implementation used, a warp consists of $32$ threads.
An SM executes warps one after another on its streaming processors, and switching between warps incurs minimal overhead.

All threads execute the same instruction but process different data.
Depending on the program logic and on the data of each thread, different threads may be in different states at the same moment.
If the control flow of different GPU threads follows different execution paths, so-called branch divergence occurs.

A warp executes all branches of the kernel code serially, disabling the threads whose code is not in the current branch~\cite{Han2011reducing,Chakroun2013reducing}.
The fewer strongly divergent branches (i.e., branches executed differently depending on the arguments), the higher the performance.
The most common cause of divergence is the use of an if-then-else statement in the kernel code.
If the condition evaluates to true for some threads of a warp and to false for the others, the execution path of the control flow branches.

The problem is that the \emph{then} and \emph{else} branches are executed serially rather than in parallel.
First, all instructions of the \emph{then} branch are executed, while all threads for which the condition is false are suspended.
Then the instructions of the \emph{else} branch are executed, and all threads for which the condition is true are suspended in turn.
This causes a considerable loss of performance, which cannot be compensated for by a higher degree of parallelism~\cite{Han2011reducing,Lin2018}.

Besides conditional statements, this problem can be caused by irregular loops, i.e., loops whose number of iterations is not known in advance, such as loops with a precondition or a postcondition (\emph{while}, \emph{do-while})~\cite{Fukuhara2020}.
The number of iterations of such loops may differ from thread to thread, so that some threads stay idle waiting for the loop to finish in other threads~\cite{Zhang2011}.
The conditions that cause the control flow to branch often depend on the data and cannot be determined at compile time.

In this book, we attempt to reduce the effect of branch divergence by removing from the kernel function the expressions that check the operability of the CES variant and the upper bound, i.e., the comparison of the cost of the current CES variant with the cost of the best variant found so far (see Listing~\ref{lst:deviceRecursiveCode}, line~\ref{ln714}, and Listing~\ref{lst:deviceIterativeCode}, line~\ref{ln828}).
All conditions are now checked only at the lowest level of the tree (\texttt{i==I}), so the execution paths of the control flow branch less.
Removing the checks at each level of the tree traversal effectively turns the branch-and-bound scheme into exhaustive search~\cite{Nievergelt2000}.
The names of the algorithm versions optimized for divergence reduction carry the suffix ``dr'' (\emph{Mo1dr} and \emph{Mo2dr}).

\section{Results of Computational Experiments}
\label{chap4_sect4}
For the computational experiments, a program was developed in C using CUDA~\cite{2022CudaProgrammingGuide}.
The experiments were carried out on a computing system with an Intel Xeon processor (E5-1620~v2, $4$ cores with Hyper-Threading, $10$~MB Smart Cache, $3.7$~GHz) and $16$~GB of RAM, equipped with an NVIDIA Tesla K20c GPU.
GPU specifications: $13$ streaming multiprocessors (SMs) with $192$ CUDA cores each ($2496$ CUDA cores in total), $5$~GB of global memory, and $48$~KB of shared memory per SM.
The software comprised Ubuntu 14.04.1, NVIDIA driver 340.29, CUDA 6.5, and the GNU C++ compiler 4.8.2.

The test case was the CES equipment selection problem described in Section~\ref{chap1_sect2} (see Fig.~\ref{fig:test_CES}), with $3$ standard equipment sizes available for each of the $16$ processing stages ($3^{16}$ = $43\,046\,721$ CES variants in total).

The program was run with values of the granularity parameter $G$ from $0$ to $10$.
The value $G = 0$ corresponds to the case where the host forms no initial fragments of CES variants and the whole search tree is traversed by a single GPU thread ($N_{threads} = K_0 = 1$), i.e., to sequential execution of the algorithm on the GPU. This value serves as the baseline for assessing the effect of parallelization.

For $G > 10$, too many threads are created on the GPU, and the working data \texttt{oprData} require more memory than the available DRAM (an out-of-memory error occurs).

In optimization \emph{Mo1}, shared memory is used to store the array \texttt{Wloc} in the recursive approach, and to store \texttt{Wloc} and \texttt{stack} in the iterative approach.
For granularity values greater than $6$ (recursive approach) and greater than $5$ (iterative approach), a shared memory shortage error occurs.

\begin{figure}[htb]
    \centering
	\includegraphics[width=0.97 \textwidth]{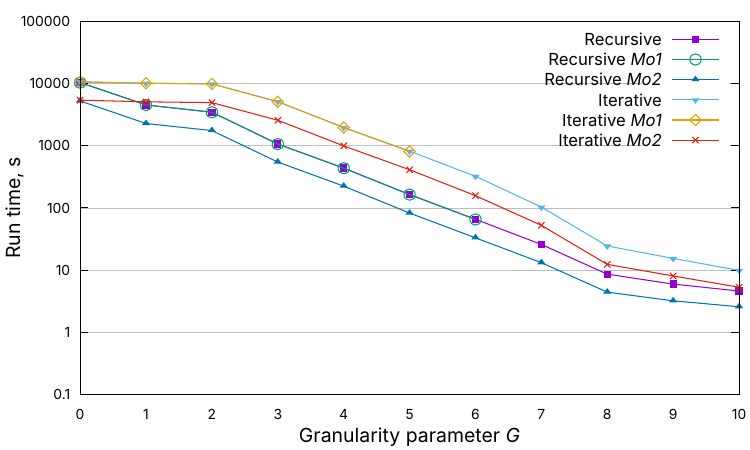}
	\caption{Shared memory optimization: run time of the parallel program versus granularity}
	\label{fig:MemoryOptimization}
\end{figure}

The results of the \emph{Mo1} and \emph{Mo2} optimizations are shown in Fig.~\ref{fig:MemoryOptimization} (the vertical axis is logarithmic).

The total run time of a program on a heterogeneous computing system consists of the run time of the sequential part on the CPU and the run time of the parallel part on the GPU.
The experiments showed that the sequential part on the CPU accounts for a negligible fraction of the total run time (no more than $0.1\%$).
For $G > 10$, global memory runs out for all versions of the program (for $G > 10$, at least $3^{11}=177\,147$ threads are required).
For $G > 6$ for the ``Recursive~\emph{Mo1}'' version and for $G > 5$ for the ``Iterative~\emph{Mo1}'' version, shared memory runs out.

Optimization \emph{Mo1} had no significant effect on the total run time (the average change in run time was less than $1\%$).
Optimization \emph{Mo2} reduced the run time by $48\%$ on average.
It should also be noted that the recursive approach turned out to be faster than the iterative one by $120\%$ on average, i.e., other things being equal, the recursive approach is on average $2.2$ times faster than the iterative one, which uses a stack.

\begin{figure}[htb]
    \centering
	\includegraphics[width=0.97 \textwidth]{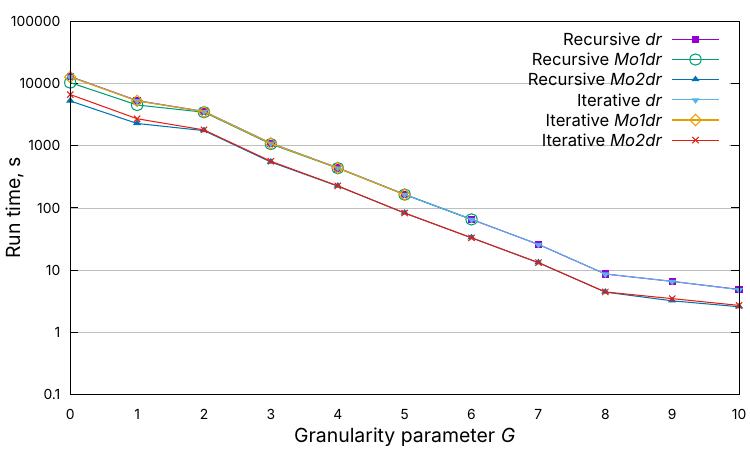}
	\caption{Branching reduction: run time of the parallel program versus granularity}
	\label{fig:BranchOptimization}
\end{figure}

The results of reducing the amount of branching are shown in Fig.~\ref{fig:BranchOptimization} (the vertical axis is logarithmic).

In this case, the branching function was removed as described above, and a series of computational experiments similar to that for the memory optimization was carried out.
In general, the pattern of the results remained the same as before: optimization \emph{Mo2dr} turned out to be the most effective.

At the same time, removing the branching function from the algorithm had markedly different effects on the recursive and iterative approaches.
For the recursive algorithm, the run time increased slightly (by no more than $5$--$6\%$) compared with the original version.
For the iterative approach, the run time decreased by $55$--$58\%$ on average.

The experimental data suggest that, in terms of run time reduction, optimization \emph{Mo2} is the most effective for the recursive approach, and \emph{Mo2dr} for the iterative approach.
For the test case used, the fastest recursive version, \emph{Mo2}, is on average $5\%$ faster than the fastest iterative version, \emph{Mo2dr}.

\begin{figure}[htb]
    \centering
	\includegraphics[width=0.97 \textwidth]{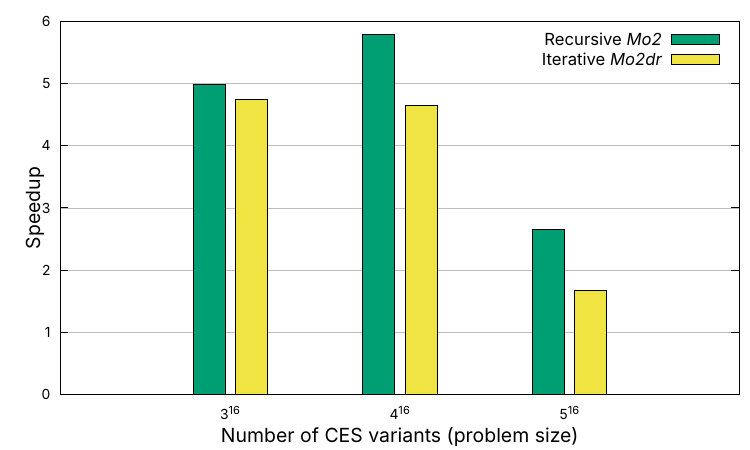}
	\caption{Speedup for problems of different sizes}
	\label{fig:GPUSpeedup}
\end{figure}

A number of tests for larger problems were also carried out to measure the speedup relative to a fully sequential CPU program.
The experiments used the same test CES, with $3$ ($3^{16}$ = $43\,046\,721$ CES variants in total), $4$ ($4^{16} = 4\,294\,967\,296$ variants), and $5$ ($5^{16} = 152\,587\,890\,625$ CES variants) standard equipment sizes available for each stage.
The results for the recursive and iterative approaches are shown in Fig.~\ref{fig:GPUSpeedup}.

For the recursive implementation, the speedup ranges from $2.66$ to $5.79$, and for the iterative implementation, from $1.67$ to $4.75$.
In all computational experiments, the recursive approach was faster than the iterative one.
Although the program as a whole accelerates the solution of the problem considered in this book, a direct implementation of branch-and-bound on the GPU does not yield a large speedup (by hundreds or thousands of times).
The reason is that the traversal of the search tree involves a great deal of branching in the program code, which considerably reduces the efficiency of the GPU.
Similar problems are described, for example, in~\cite{Rocki2010parallel}, where the authors report the results of implementing minimax search using CUDA.

%%%% Chapter 5
\chapter{Parallel Hybrid Metaheuristic Algorithm on GPUs}
\label{chapter5}

As shown in Chapters~\ref{chapter2}--\ref{chapter4}, branch-and-bound is guaranteed to find the globally optimal solution, but its run time grows rapidly with the problem size.
Despite the optimization of the sequential algorithm and its parallelization for shared-memory and distributed-memory systems, solving large problems still takes tens of hours, and a direct implementation of branch-and-bound on the GPU does not yield a substantial speedup (Chapter~\ref{chapter4}).
In this chapter, we therefore propose to solve the problem with metaheuristic algorithms, which do not guarantee finding the global optimum but produce near-optimal solutions in acceptable time.

The algorithm proposed in this book combines two metaheuristics: Ant Colony Optimization (ACO) and Simulated Annealing (SA)~\cite{Lee2012}.
In the proposed approach, ACO is used to find a feasible initial solution for SA~\cite{Solnon2013}.
This division of roles is due to the specific features of the problem.
For classical combinatorial optimization problems such as the travelling salesman problem, a randomly generated variant can serve as the initial solution.
For the CES equipment selection problem, this approach is not applicable: the initial solution must satisfy both the compatibility constraint~(\ref{eq_4}) and the processing time constraint~(\ref{eq_5}), and the probability that a randomly chosen variant is feasible is low~\cite{Rossi2006}.
Therefore, ACO is used for a targeted search for a feasible CES variant, and SA is used to improve it subsequently with respect to the capital cost of the equipment.

Both algorithms are stochastic and well suited to the massively parallel architecture of the GPU: each thread can search for a solution independently, and the results of all threads are then compared.
This makes it possible to explore a much larger part of the search space in the same time than with sequential execution.

The chapter is organized as follows.
Section~\ref{chap5_sect1} gives a brief overview of metaheuristic algorithms for combinatorial optimization problems.
Sections~\ref{chap5_sect2} and~\ref{chap5_sect3} describe Simulated Annealing and Ant Colony Optimization, together with the ways of constructing a candidate solution and defining the heuristic information and the fitness function proposed for the problem under consideration.
Section~\ref{chap5_sect4} presents the implementation of the hybrid algorithm (ACO+SA) on the GPU, and Section~\ref{chap5_sect5} reports the results of computational experiments and a comparison with branch-and-bound.

\section{A Brief Overview of Metaheuristic Algorithms for Combinatorial Optimization}
\label{chap5_sect1}

Metaheuristic algorithms combine basic heuristic methods within higher-level algorithmic frameworks aimed at an efficient exploration of the search space.
The main difficulty lies in adapting general metaheuristic frameworks to specific problems.
The fundamental properties of metaheuristics include the following~\cite{Sherbina2014}:
\begin{itemize}
    \item metaheuristics are strategies that guide the search process;
    \item the goal of a metaheuristic is to explore the search space efficiently in order to find near-optimal solutions;
    \item metaheuristic algorithms are approximate and, as a rule, nondeterministic;
    \item metaheuristic algorithms may include mechanisms for avoiding being trapped in a confined region of the search space;
    \item metaheuristics are not necessarily problem-specific and can be described at an abstract level;
    \item metaheuristics use domain-specific knowledge in the form of heuristics controlled by a higher-level strategy;
    \item modern metaheuristics use search experience stored in memory to guide the search.
\end{itemize}

Each metaheuristic algorithm has its own domain-dependent features; nevertheless, all of them share a number of invariable basic components, which can be represented at an abstract level as a generic scheme.
This generic scheme is as follows~\cite{Sherbina2014,Shabelnikov2015}.
\begin{enumerate}
    \item Initialization: finding an initial solution.
    \item Forming one or more local neighborhoods of the initial solution.
    \item Selecting the most promising neighborhood.
    \item Selecting a candidate in the chosen neighborhood using an evaluation function (fitness function) that depends on such parameters of two solutions as the objective function value, penalties for violating certain constraints, etc.
    The result of the selection is the best solution with respect to this criterion.
    \item Stopping according to one of the criteria, which may be the computation time, the number of iterations, or the rate of improvement of the solutions.
\end{enumerate}

Any classification of metaheuristic methods is somewhat arbitrary, since typical groups of methods are based on similar ideas and there is a persistent trend towards hybrid algorithms~\cite{Anghinolfi2008,Hibler2012}.
Four groups of metaheuristic algorithms are distinguished in~\cite{Panteleev2013}: evolution-based algorithms, swarm intelligence-based algorithms, physics-based algorithms, and multi-start algorithms.

\subsubsection*{Evolution-based algorithms}
Genetic algorithms (with binary and real-valued encoding); methods that imitate the immune systems of organisms (artificial immune systems); scatter search; the covariance matrix adaptation evolution strategy; variable mesh optimization; differential evolution; weed colonization; and cuckoo search.

\subsubsection*{Swarm intelligence-based algorithms}
Particle swarm optimization; ant colony optimization; bacterial foraging optimization; bee colony algorithms, including the artificial bee colony algorithm and the bees algorithm; fish school search; the bat-inspired algorithm; the firefly algorithm; glowworm swarm optimization; and the shuffled frog-leaping algorithm.

\subsubsection*{Physics-based algorithms}
Central force optimization; simulated annealing; adaptive simulated annealing; deterministic annealing; harmony search; and the electromagnetism-like mechanism.

\subsubsection*{Multi-start algorithms}
The greedy randomized adaptive search procedure (GRASP) and tabu search.

\section{Simulated Annealing (SA)}
\label{subsubsec:simulated_annealing}
\label{chap5_sect2}
Simulated Annealing (SA)~\cite{Kirkpatrick1983}, an optimization technique that uses a controlled random search to avoid getting trapped in local optima, was chosen as the algorithm for minimizing the objective function (the equipment cost).
The idea of the method arose from observing the processes that take place in a metal as it cools after being heated to its melting point~\cite{Tihomirov2009,Valadi2014}.

Simulated annealing is a stochastic search method in which, at each step, the current solution is replaced by another solution randomly chosen from its neighborhood.
A solution that improves the objective function is always accepted, while a worsening solution is accepted with a probability that depends on a control parameter called the temperature.
The temperature is gradually lowered according to a cooling schedule, so that worsening solutions are often accepted at the beginning of the search and are practically never accepted towards its end.

Simulated annealing can be used to search for solutions of complex linear and nonlinear problems and is able to escape from local minima.
Its advantages include the ability to search for the global optimum, its generality, and its relative ease of implementation.
The search strategy allows the algorithm to jump over local minima of the objective function and reach the basin of attraction of the global minimum~\cite{Nikolaev2010,Aarts2014}.
This basin is reached only with a certain probability; therefore, simulated annealing does not guarantee that the global minimum will be found.
However, with a proper choice of the ``temperature'' trajectory (the parameter of the method), the algorithm not only improves the initial approximation but also comes substantially closer to the global optimum.

The general scheme of simulated annealing is shown in Listing~\ref{lst:SimulatedAnnealing}.
\begin{lstlisting}[%
float=htb,%
caption = {Simulated Annealing (SA)},%
label=lst:SimulatedAnnealing]
SimulatedAnnealing(){
    /* initialize the temperature and the initial solution */
    t = Tinit; W = Winit; (*@\label{lstsa:initialization}@*)
    while(t > Tfinal){/* until the final temperature is reached */ (*@\label{lstsa:while_t}@*)
        for(l = 0; l < Lmax; l++){ /* repeat Lmax times */  (*@\label{lstsa:for_l}@*)
            Wcand = ConstructSolution(W); /* construct a new candidate solution */ (*@\label{lstsa:perturb}@*)
                deltaCost = DefineCost(Wcand) - DefineCost(W); (*@\label{lstsa:delta}@*)
                if (deltaCost < 0){ /* if the candidate is better */(*@\label{lstsa:calc_delta}@*)
                    W = Wcand; } /* accept the candidate solution */ (*@\label{lstsa:accept_solution}@*)
                else { /* otherwise */
                    p = exp (-deltaCost / t); /* compute the probability */(*@\label{lstsa:compute_probability}@*)
                    r = rand(0, 1); /* generate a random number */ (*@\label{lstsa:generate_random_number}@*)
				    if (p > r){ /* if the probability exceeds the random number */ (*@\label{lstsa:compare_probability}@*)
					    W = Wcand; }}} /* accept the candidate solution */ (*@\label{lstsa:accept_non_better_solution}@*)
        t = sigma * t; } /* lower the temperature */ (*@\label{lstsa:cooling_rule}@*)
}
\end{lstlisting}

At the start of the algorithm, the temperature \texttt{t} is initialized with the initial temperature \texttt{Tinit}, and the current solution \texttt{W} with a feasible initial solution \texttt{Winit} (line~\ref{lstsa:initialization}).
The feasible initial solution is the result of the ant colony algorithm.
A solution \texttt{W} is a vector whose length equals the number of CES stages.
Each element of the vector holds the index of one of the units available for installation at the corresponding stage.

The cooling schedule is the law according to which the temperature changes during cooling.
It is defined by the initial temperature \texttt{Tinit} (line~\ref{lstsa:initialization}), the final temperature \texttt{Tfinal} (line~\ref{lstsa:while_t}), and the cooling rule, which changes the temperature at each cooling iteration (line~\ref{lstsa:cooling_rule}).
In practice, a good cooling schedule is designed by trial and error, by trying different parameter values and observing the results.
Nevertheless, certain recommendations are given in the literature~\cite{Aarts2014}.

The initial temperature \texttt{Tinit} is defined as the largest possible difference in cost between two feasible solutions~\cite{Aarts2014}.
For the problem under consideration, the maximum difference between solutions was computed as follows: first, the most expensive unit is installed at each CES stage, and the cost of the resulting solution is calculated.
This is the cost of the most expensive solution.
The initial temperature is thus the difference between the potentially most expensive and the cheapest solution.

The final temperature \texttt{Tfinal} is defined as the smallest possible difference in cost between two neighboring solutions~\cite{Aarts2014}.
For the problem under consideration, the final temperature was taken to be the minimum unit cost over all CES stages.
This value is of the same order as the minimum nonzero difference between the costs of two units of the same stage.

The algorithm contains two nested loops.
The outer loop (line~\ref{lstsa:while_t}) gradually lowers the temperature \texttt{t}, which reduces the probability of accepting solutions that do not improve the objective function (\ref{eq_1}).

At each particular temperature, \texttt{Lmax} iterations of the algorithm must be performed; this is done by the inner loop (line~\ref{lstsa:for_l}).

The number of iterations \texttt{Lmax} affects the speed and convergence of the algorithm: if it is too large, the run time increases; if it is too small, the solutions obtained are far from the global minimum.
In general, simulated annealing generates a finite inhomogeneous Markov chain. A Markov chain is a sequence of trials in which the probability of the outcome of each trial depends only on the outcome of the preceding trial.
In simulated annealing, a trial consists in choosing the next solution at each step.
In practice, an infinite chain cannot be executed, so its length must be limited~\cite{Aarts2014}.

There is no single generally accepted approach to determining $L_{max}$; different researchers recommend different choices.
For example, Skiena~\cite{Skiena2011} recommends taking the number of iterations before each temperature reduction between $100$ and $1000$.

As described below, a new solution in the neighborhood of the current feasible solution is found by randomly changing one randomly chosen stage of this solution.
For each CES stage, a set of available equipment is given.
We propose the following approach to determining $L_{max}$: the number of repetitions at the current temperature equals the total number of units available for the whole CES.

For example, consider a CES of $5$ stages in which $3$ standard unit sizes are available for each stage.
The total number of units is $3+3+3+3+3=15$, i.e., $L_{max}=15$.
This gives a theoretical chance of trying every available unit at every stage of the current solution.
This is not an enumeration of combinations, which has exponential complexity, because at each iteration we randomly change only one randomly chosen stage of the solution, keeping the others fixed.
In general, $L_{max}=\sum_{i=1}^IJ_i$.
An advantage of this approach is that $L_{max}$ adapts to the problem size: the more units per stage and the more stages, the more iterations are performed before the temperature is lowered.

At each iteration of the inner loop (line~\ref{lstsa:for_l}), a candidate solution \texttt{Wcand} is constructed in the neighborhood of the current feasible solution (line~\ref{lstsa:perturb}).
For this purpose, the procedure \texttt{ConstructSolution()} is used.

Then the change in the objective function, \texttt{deltaCost}, is determined (line~\ref{lstsa:delta}).
If a better solution has been obtained (line~\ref{lstsa:calc_delta}), it is accepted as the current solution
(line~\ref{lstsa:accept_solution}).
Otherwise, the probability \texttt{p} of accepting the worse candidate solution as the current one is computed (line~\ref{lstsa:compute_probability}).

Historically, the first annealing scheme was the so-called \emph{Boltzmann annealing}~\cite{Kirkpatrick1983}.
The probability is computed as $p=\exp(-\Delta \mathcal{E} /(k_{\mathrm{B}}\cdot t ))$, where $k_{\mathrm{B}}$ is the Boltzmann constant and $\Delta \mathcal{E}$ is the change in the energy level.
The probability formula has a certain thermodynamic meaning~\cite{Skiena2011}.
The probability of a transition from a high-energy state to a low-energy state is very high.
Nevertheless, there is a nonzero probability of a transition to a higher-energy state, and small transitions are more likely than large ones.
Moreover, the higher the temperature, the higher the probability of energy transitions.
For optimization problems, this means that at the beginning of the search, when the temperature is high, non-optimal solutions are accepted as current more often, and as the system cools, the probability of accepting non-optimal solutions decreases with the temperature. This is a valuable property of simulated annealing, since it allows the search to escape from local optima and, at least theoretically, to find the global optimum.

In practical problems, the change in the energy level is usually taken to be the change in the objective function, and $k_{\mathrm{B}}~=~1$~\cite{yang2010}.
Thus, the transition probability formula takes the form \texttt{p = exp (-deltaCost / t)} used in the algorithm (line~\ref{lstsa:compute_probability}).

After the probability \texttt{p} has been computed, a random number \texttt{r} between $0$ and $1$ is generated (line~\ref{lstsa:generate_random_number}).
If \texttt{p} exceeds \texttt{r}, the candidate solution is accepted as the current solution
(line~\ref{lstsa:accept_non_better_solution}).

The temperature \texttt{t} is then lowered according to the chosen cooling rule (line~\ref{lstsa:cooling_rule}).
There are various ways of lowering the temperature, for example, linear, $t = t - \Delta t$, or exponential, $t=\sigma \cdot t$.
Cooling is faster with the exponential rule, so we use it, following the recommendations of~\cite{Valadi2014}.
The constant is usually chosen in the range $0.8 \le \sigma \le 0.99$~\cite{Aarts2014}; for the problem under consideration, we take $\sigma=0.9$.

Note that, at the initial step, the current solution must be initialized with a feasible initial solution (a feasible CES variant \texttt{W} at the temperature \texttt{Tinit}).
For typical problems such as the travelling salesman problem, the initial solution can be chosen at random.
For the problem under consideration, a random solution amounts to choosing, at random, a unit for each CES stage from the set of units available for that stage.
A CES variant can be generated in this way, but it may be either inoperable (violating the compatibility constraint (\ref{eq_4})) or unable to produce the required products within the given time (violating the processing time constraint (\ref{eq_5})).
The time needed to obtain an operable CES variant in this way is completely unpredictable, so this approach is not applicable in practice.
Instead, a Constraint Satisfaction Problem (CSP) is solved, in which the solution cost (\ref{eq_1}) is ignored and only constraints (\ref{eq_4}) and (\ref{eq_5}) must be satisfied~\cite{Rossi2006,Sherbina2011}.
The constraint satisfaction problem can be formulated as follows: find an operable CES variant that satisfies the compatibility constraint (\ref{eq_4}) and the processing time constraint (\ref{eq_5}).

Problems of this type can be solved by Ant Colony Optimization (ACO), presented in Section~\ref{subsubsec:ant_colony_optimization}.
The result is a CES variant that is operable and able to produce the required products within the given time, although its cost may be far from optimal.
This CES variant is used as the starting point for simulated annealing, which in turn searches for a cost-optimal CES variant.

\section{Ant Colony Optimization (ACO)}
\label{subsubsec:ant_colony_optimization}
\label{chap5_sect3}
The main idea of this approach is to model the behavior of ants searching for a path from their colony to a food source~\cite{dorigo2019ant}.
Ants are able to find the shortest path from the nest to a food source quickly and to adapt to changing conditions~\cite{Shtovba2003,Zaichenko2012}.
An ant colony can be regarded as a multi-agent system in which the agents (ants) interact to achieve a common goal.
The ants interact indirectly through a chemical substance, the \emph{pheromone}, which they deposit along their path (trail).
The pheromone concentration on a trail determines how attractive it is (when choosing a path, an ant is more likely to choose the one with a higher pheromone concentration).
The pheromone evaporates over time, which allows the ants to adapt their behavior to changes in the environment.

\begin{lstlisting}[%
caption={Ant Colony Optimization (ACO)},%
float=htb,%
label={lst:AntColonyOptimization}]
AntColonyOptimization() {
	isFound = false; iterCounter = 0; (*@\label{lstaco:initialize_is_found}@*)
	Initialize(); /* initialize the pheromone concentration */ (*@\label{lstaco:initialize_pheromone}@*)
	/* until a solution has been found */
	while(!isFound && iterCounter < maxIterationNumber){   (*@\label{lstaco:while_not_solution_is_found}@*)
		foreach(ant in swarm){/* for each ant in the colony */
			ConstructSolution(alpha, beta);} /* construct a new solution */ (*@\label{lstaco:construct_solution}@*)
		if(isFound) return; /* stop if a solution has been found */(*@\label{lstaco:solution_is_found}@*)
		PheromoneUpdate(); /* update the pheromone concentration */(*@\label{lstaco:update_pheromone}@*)
		EvaporatePheromone(rho); /* pheromone evaporation */(*@\label{lstaco:pheromone_evaporation}@*)
		iterCounter ++; /* increment the iteration counter */
}
\end{lstlisting}

The pseudocode of the ant colony algorithm is shown in Listing~\ref{lst:AntColonyOptimization}.

The flag \texttt{isFound}, which controls the search, is initialized to \texttt{false} (line~\ref{lstaco:initialize_is_found}).
The search continues until a feasible solution is found or the number of attempts exceeds the maximum number \texttt{maxIterationNumber} (line~\ref{lstaco:while_not_solution_is_found}).

The local iteration counter \texttt{iterCounter}, initialized to 0 (line~\ref{lstaco:initialize_is_found}), protects the algorithm from looping forever if no ant of the colony manages to find a feasible solution (which is quite possible for a stochastic algorithm).
In this case, \texttt{iterCounter} exceeds \texttt{maxIterationNumber}, and the algorithm terminates without a result; it is then restarted with new random initial values of the pheromone concentration.

Before the algorithm starts, the pheromone concentration is initialized (line~\ref{lstaco:initialize_pheromone}).

While constructing a solution of the problem (line~\ref{lstaco:construct_solution}), each ant of the colony (line~\ref{lstaco:while_not_solution_is_found}) can move in the search tree (see Fig.~\ref{fig:search_tree}) only from top to bottom.
Suppose that an ant is at node $r=n_{i,j}$ at level $i$ of the tree.
It can then move only to a child node $s=n_{i+1,j'}$, $s \in C_{r}$, at level $i+1$, where $C_{r}$ is the set of child nodes of node $r$.
The complete path of an ant from level $1$ to level $I$ corresponds to a potential solution of the problem.

The probability $p_{rs}$ that an ant moves from node $r$ to node $s$ is determined from two types of information, the pheromone level $\tau_{rs}$ and the heuristic information $\eta_{rs}$, according to the random proportional rule $p_{rs} = {\tau_{rs}^\alpha \cdot \eta_{rs}^\beta } / {\sum_{k\in C_{r}} {(\tau_{rk}^\alpha \cdot \eta_{rk}^\beta)}}$, where $C_{r}$ is the set of child nodes of node $r$, $k$ is the index of a child node, and $\alpha$ and $\beta$ are empirical parameters of the algorithm that determine the relative influence of the pheromone concentration and of the heuristic information, respectively, on the probability of choosing a node~\cite{dorigo2019ant,Solnon2013,Valadi2014}.
As a rule, the values of $\alpha$ and $\beta$ are chosen from the interval $(0, 2)$~\cite{Agarwal2012}.
For $\alpha=0$, only the heuristic information is taken into account.
For $\beta=0$, only the pheromone concentration is taken into account, so that the search degenerates: all ants follow a single suboptimal route (the path with the highest pheromone concentration, which they thereby increase further).
The first version of the algorithm used the values $\alpha=0.4$ and $\beta=0.6$, obtained empirically: the sequential algorithm was run just a few times with different values of $\alpha$ and $\beta$, and the values yielding the highest speed were chosen.
Looking ahead, we note that these values were not the best ones.
The problem of finding the best values of the algorithm parameters is considered in detail in Chapter~\ref{chapter6}.
The node to move to is selected by the ``roulette wheel'' principle: each child node has its own sector of the wheel with an area proportional to its probability.
To select a node, a random number is generated, and the sector into which it falls is determined;
the node corresponding to this sector is chosen.

For the problem of finding a feasible CES variant, we propose the following way of defining the heuristic information, based on the following considerations.
First, it is known that a CES whose stages are equipped with larger units is more expensive, but has a larger batch size and, as a rule, produces the required amount of product faster, since fewer batches are needed.
For the same required production amount, a small batch size means producing many batches over a long time, whereas a large batch size means producing few batches over a short time.
For example, suppose that $1000$~kg of product must be produced.
If the batch size (the amount of product produced in one CES cycle) is $10$~kg, then $1000/10=100$ batches must be produced ($100$ CES cycles), whereas with a batch size of $100$~kg only $1000/100=10$ batches are needed.
The fewer the cycles (i.e., the larger the units and hence the batch size), the more likely the processing time constraint (\ref{eq_5}) is to be satisfied.
Thus, when a unit is chosen for the next stage, preference should be given to larger units.

Second, the compatibility condition (\ref{eq_4}) must be satisfied for all initial fragments of the CES consisting of stages $1$--$2$, $1$--$2$--$3$, \ldots, $1$--$2$--$3$--$\ldots$--$I$.
Suppose that, while constructing a solution, an ant is at stage $i$ of the CES.
Then the operability of the initial fragment of the CES formed at the preceding tree levels is checked.
If the corresponding CES fragment is inoperable, its weight is zero; otherwise, it is one.

Thus, the heuristic $\eta_{rs}$ for the move from node $r$ to node $s$ is the sum of the main size of the unit on the tree edge $(r,s)$ and a value ($0$ or $1$) characterizing the operability of the initial fragment of the CES. The second term is the same for all child nodes of node $r$.

After an ant of the colony has reached the last level $I$ of the search tree, the pheromone level on all edges of the ant's path must be updated.
The amount of pheromone deposited by an ant depends on the quality of the solution obtained (how close the solution is to the goal).

Each ant deposits pheromone on the edges of its own path (line~\ref{lstaco:update_pheromone}): for every edge $(r,s)$ traversed by the $m$th ant, $\tau_{rs} = \tau_{rs} + Q/L_m$, $m=\overline{1,M}$,
where $Q$ is a constant, $L_m$ is the fitness value of the solution constructed by the $m$th ant, and $M$ is the number of ants in the colony.
This rule has the form of the pheromone deposit rule of classical ant colony algorithms for the travelling salesman problem, where $L_m$ is the length of the tour of the $m$th ant and $Q$ is chosen to be of the same order as the length of the optimal tour~\cite{Shtovba2003,dorigo2019ant}.
Here $L_m$ is the value of the fitness function, a measure of how well the solution obtained satisfies the requirements of the problem~\cite{Khan2009}.

For the problem of finding a feasible CES variant, we propose the following algorithm for computing the fitness function.
For a CES to be able to produce the required amount of product within the given time, it must be operable, i.e., satisfy the compatibility constraint (\ref{eq_4}) and the processing time constraint (\ref{eq_5}).
As the ant moves down the decision tree, the number of satisfied compatibility constraints (\ref{eq_4}) for the initial fragments of the CES consisting of stages $1$ to $i$ is counted.
For example, at tree level $i=1$, the initial fragment consists of stage $1$; at level $i=2$, of stages $1,2$; at level $i=3$, of stages $1,2,3$; at level $i=4$, of stages $1,2,3,4$; and so on.

Operability of the CES is a necessary but not sufficient condition, since the product range must also be produced within the given time, i.e., the processing time constraint (\ref{eq_5}) must be satisfied.
Therefore, if constraint (\ref{eq_5}) is satisfied, we add one to the number of satisfied compatibility constraints.
Thus, the ideal value of the fitness function (the goal) for a feasible CES variant equals the total number of stages plus one, and we use this value as the constant $Q=I+1$.

The fitness function $L_m$ makes it possible to assess how close the solution found by a particular ant is to the goal.
For the example shown in Fig.~\ref{fig:search_tree}, $Q=5$, and the fitness function may range from $1$ to $5$: an initial fragment consisting of the first stage alone is always operable, so $L_m\geq 1$, and $L_m=5$ means that all constraints are satisfied and the goal $Q$ has been reached.

The fitness values make it possible to compare the paths of different ants.
For example, if $L_1=4$ for the first ant and $L_2=2$ for the second, then the first ant has found a solution closer to the goal.
Note that the amount of pheromone deposited, $Q/L_m$, equals $1$ for a feasible solution ($L_m=Q$) and grows as the fitness value decreases (in this example, $Q/L_1=1.25$ and $Q/L_2=2.5$).
Thus, in the implementation used in all the experiments reported in this book, the deposit is inversely proportional to the fitness value, so the pheromone does not selectively reinforce the paths that are closest to the goal.
A deposit that grows with the fitness value (for example, $L_m/Q$) is a natural alternative; comparing the two rules is a subject for further study.

Note that different modifications of ACO may differ in the formula used to update the pheromone concentrations.
For example, elitist ants, which have found the best solutions, may be introduced, and only their contributions to the pheromone concentration are taken into account, while all other ants are ignored~\cite{White2003}.
In the proposed modification, the pheromone deposited by all ants is taken into account.

Over time, the pheromone deposited at previous iterations evaporates, i.e., its concentration decreases (line~\ref{lstaco:pheromone_evaporation}).
In the proposed version of the algorithm, the pheromone concentration is updated with evaporation taken into account~\cite{stuetzle2011parameter} after the solution of each ant has been evaluated: $\tau_{rs} = \tau_{rs} \cdot \rho + \Delta \tau_{rs}$, where $\Delta\tau_{rs} = Q/L_m$ for the edges of the path of the $m$th ant and $\Delta \tau_{rs}=0$ for the other edges, and $\rho \in [0,1]$ is the evaporation coefficient.
Since this update is applied after each ant, evaporation acts $M$ times per iteration of the colony, and the pheromone values reflect mainly the paths of the most recently evaluated ants.
The first version of the solution of the problem used the value $\rho=0.9$.

\section{GPU Implementation of the Hybrid Algorithm (ACO+SA)}
\label{chap5_sect4}

\begin{figure}[htb]
    \centering
	\includegraphics[width=0.97\textwidth]{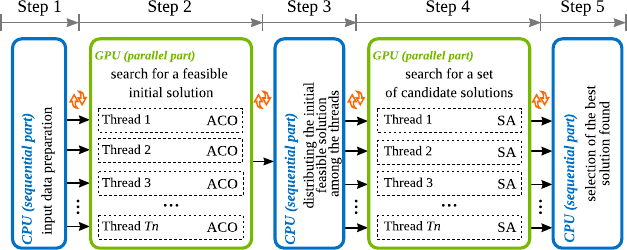}
	\caption{Structure of the hybrid algorithm (ACO + SA)}
	\label{fig:hybrid_ACO_SA}
\end{figure}

Figure~\ref{fig:hybrid_ACO_SA} illustrates the five-step implementation of the hybrid algorithm (ACO~+~SA) on computing systems with GPUs.

\begin{enumerate}
\item The CPU reads the input data, initializes the parameters of ACO and SA, transfers all necessary data to the GPU, and then launches the kernel function for ant colony optimization (the ACO kernel).
\item The ACO kernel performs a parallel search for a feasible CES variant as described in Section~\ref{subsubsec:ant_colony_optimization}.
It uses the Multiple Ant Colonies (MAC) method~\cite{Delevacq2013}: all colonies run as parallel threads and solve the problem independently of one another.
As soon as any colony finds a solution, the computation stops.
Not all parallel threads need to find a feasible solution; one is enough.
Thus, as the number of parallel threads increases, the probability of finding a feasible solution increases and, accordingly, the search time decreases.
\item The CPU receives the feasible initial solution found, distributes it among the threads, and launches the kernel function for simulated annealing (the SA kernel).
\item The SA kernel searches for the optimal CES variant by simulated annealing, using the feasible CES variant obtained at the previous step as the starting point.
It uses Multiple Independent Runs (MIR)~\cite{Syberfeldt2017}, in which every parallel thread must find its own feasible CES variant.
The threads exchange only the cost \texttt{minCost} of the best solution found, which is used in the acceptance criterion for candidate solutions.
The run time is practically independent of the number of threads, but as the number of threads increases, so does the probability that some thread will find the global optimum or a solution close to it.
The kernel function finishes when every thread has completed its search.
\item The CPU receives the solutions obtained by all threads at the previous step and selects the best of them.
This best solution is the final solution of the problem.
\end{enumerate}

The pseudocode of the algorithm implementing the proposed approach for the host (CPU) and the device (GPU) is shown in Listings~\ref{lst:ACO_SA_host}, \ref{lst:ACO_SA_device_ACO}~(ACO), and~\ref{lst:ACO_SA_device_SA}~(SA), respectively.
Most variable names coincide with those used in the problem statement.

\subsubsection*{Host (CPU)}
\label{chap5_host_code}

The host (CPU) starts by calling the procedure \texttt{ReadInputData()}, which reads the input data from a file (line~\ref{ln1102}).
The variable \texttt{inData} is a pointer to a structure whose fields are variables and arrays of various dimensions holding all the necessary \emph{input data}.
These data are the same for all threads and do not change while the problem is being solved, i.e., they are constants.

\begin{lstlisting}[%
caption={Pseudocode of the host (CPU) algorithm},%
float=htb,%
label={lst:ACO_SA_host}]
main(){	/* prepare the input data */
	ReadInputData(inData); (*@\label{ln1102}@*)
	/* the number of threads is a command-line parameter */
	numThreads = argv[1];(*@\label{ln1104}@*)
	/* prepare the working data */
	PrepOperationalData(oprData, inData, numThreads); (*@\label{ln1106}@*)
	/* transfer the necessary data to the device */
	cudaMemcpyHtoD(inData, oprData);(*@\label{ln1108}@*)
	/* determine the kernel launch parameters */
	blocksPerGrid = numThreads / MAX_THREADS_PER_BLOCK + 1;(*@\label{ln1110}@*)
	threadsPerBlock = MAX_THREADS_PER_BLOCK;(*@\label{ln1111}@*)
	/* initialize the array threadIdx with '-1' */
	for (n = 0; n < numThreads; n++){ threadIdx[n] = -1;} (*@\label{ln1113}@*)
	isFound = 0;/* initialize the global thread control flag for ACO */ (*@\label{ln1114}@*)
	cudaMemcpyHtoD(isFound, threadIdx); (*@\label{ln1115}@*)
	do {(*@\label{ln1114d}@*) /* launch the kernel to find a feasible initial solution */
		ACO<<<blocksPerGrid, threadsPerBlock>>>();(*@\label{ln1117}@*)
		cudaDeviceSynchronize(); /* synchronize with the device */(*@\label{ln1118}@*)
		cudaMemcpyDtoH(isFound);}
	while (isFound == 0); (*@\label{ln1114w}@*)
	/* copy the feasible initial solution from the device to the host */
		cudaMemcpyDtoH(Wfirst, threadIdx); (*@\label{ln1119}@*)
	/* find the ID of the thread that has found a feasible solution */
	for (n = 0; n < numThreads; n++){ (*@\label{ln1122}@*)
		if(threadIdx[n] != -1){
			tid = threadIdx[n]; break;}}(*@\label{ln1124}@*)
	/* the feasible solution is the initial approximation for all threads */
	for (n = 0; n < numThreads; n++){(*@\label{ln1126}@*)
		W[n] = Wfirst[tid]; }(*@\label{ln1127}@*)
	/* copy the initial approximation to the device */
	cudaMemcpyHtoD(W);(*@\label{ln1129}@*)
	/* launch the kernel to find the optimal solution */
	SA<<<blocksPerGrid, threadsPerBlock>>>();(*@\label{ln1131}@*)
	cudaDeviceSynchronize(); /* synchronize with the device */ (*@\label{ln1132}@*)
	/* copy the search results from the device to the host */
	cudaMemcpyDtoH(W, Cost, minCost);	(*@\label{ln1134}@*)
    /* find the best solution among all solutions found */
	for (n = 0; n < numThreads; n++){(*@\label{ln1136}@*)
		if(Cost[n] <= minCost){
			Wopt = W[n];
			minCost = Cost[n]; }}(*@\label{ln1139}@*)
}
\end{lstlisting}

The number of threads used for the computation, \texttt{numThreads}, is a parameter of the program and is passed to it as a command-line argument (line~\ref{ln1104}).
This value is then used when calling the procedure \texttt{PrepOperationalData()}, which prepares the \emph{working data} (line~\ref{ln1106}).
The variable \texttt{oprData} is a pointer to a structure whose fields are variables and arrays of various dimensions whose values are modified independently by all threads while the problem is being solved.
Therefore, one of the array indices is the thread identifier \texttt{threadID}.
The procedures \texttt{ReadInputData()} and \texttt{PrepOperationalData()} perform all memory allocations needed for the arrays and initialize the variables, including the pseudorandom number generators of all threads.

The data are then transferred to the device for use in subsequent kernel calls (line~\ref{ln1108}).
Next, the kernel launch parameters are determined, i.e., the required number of blocks in the grid, \texttt{blocksPerGrid} (line~\ref{ln1110}), and the number of threads per block, \texttt{threadsPerBlock} (line~\ref{ln1111}), are computed.
When the required number of blocks is computed (line~\ref{ln1110}), the result of the division may be non-integer and is automatically truncated to an integer; therefore, one is added to the quotient to round it up.

Next, the host initializes the elements of the array \texttt{threadIdx} with \texttt{'-1'} (line~\ref{ln1113}).
The length of \texttt{threadIdx} equals the total number of threads \texttt{numThreads}.
If a thread with index \texttt{tid} finds a feasible solution, the index of this thread is written to the array: \texttt{threadIdx[tid] = tid}.

Knowing the index of the thread that has found a solution, one can then retrieve the feasible solution found from the array of solutions, \texttt{Wfirst[tid]}; this solution is in turn used as the feasible initial solution for simulated annealing.

The thread control flag \texttt{isFound} is used to control the threads on the GPU. The flag is initialized to zero (line~\ref{ln1114}).
If \texttt{isFound==0}, a thread continues working.
Otherwise, the first feasible solution has been found, \texttt{isFound!=0}, and all threads in the grid stop.
The host then transfers the data to the device (line~\ref{ln1115}) and launches the kernel function \texttt{ACO()} (line~\ref{ln1117}).

The launch is repeated until at least one thread has found a feasible solution, i.e., as long as the value of \texttt{isFound}, copied from the device after each launch, is zero (lines~\ref{ln1114d}--\ref{ln1114w}).

In CUDA, a kernel function is launched asynchronously:
control returns to the CPU immediately after the kernel has been launched, without waiting for it to finish on the GPU.
By calling \texttt{cudaDeviceSynchronize()} (line~\ref{ln1118}), the host pauses and waits until all threads on the GPU have finished (i.e., synchronizes with the device).

After the GPU has finished, the results of the search for a feasible solution are transferred to the host (line~\ref{ln1119}).
Here \texttt{Wfirst} is a two-dimensional array of \texttt{numThreads} vectors of length \texttt{I} each.
The elements of each vector hold the selected standard size of the unit for each stage.
Having received the solution, the host determines the identifier \texttt{tid} of the thread that has found it by searching for the first element of \texttt{threadIdx} not equal to \texttt{'-1'} (lines~\ref{ln1122}--\ref{ln1124}).
The solution \texttt{Wfirst[tid]} with this identifier is then used as the feasible initial solution for simulated annealing.

The host then initializes the initial approximations of all threads with the feasible solution found (lines~\ref{ln1126}--\ref{ln1127}).
Here \texttt{W} is a two-dimensional array of \texttt{numThreads} vectors of length \texttt{I} each.
The elements of each vector hold the selected standard size of the unit for each stage.

Next, the host transfers the data to the device (line~\ref{ln1129}) and launches the kernel function for simulated annealing, \texttt{SA()} (line~\ref{ln1131}).

After synchronization (line~\ref{ln1132}), the results are copied from the device to the host (line~\ref{ln1134}).
Here \texttt{Cost} is a one-dimensional array of \texttt{numThreads} elements used to store the local costs of the CES variants found by each thread.
The host then finds the minimum cost and stores it in the variable \texttt{minCost} (lines~\ref{ln1136}--\ref{ln1139}).
The optimal CES variant is stored in the array \texttt{Wopt} with \texttt{I} elements, each holding the standard size of the unit for the corresponding stage.

\subsubsection*{Device (GPU)}
\label{chap5_device_code}
In the proposed parallel implementation, each thread simulates the search of one ant colony (swarm) consisting of $M$ ants~\cite{nosheen2013,Gupta2020}.
In particular, $M=100$ was used in the computational experiments.
There is no parallelization over individual ants:
each colony searches for its own solution independently.
Pseudorandom numbers are generated with NVIDIA's specialized CUDA Random Number Generation library, cuRAND~\cite{2022NvidiaCudaRandom}.
Its function \texttt{curand\_uniform()} returns a uniformly distributed random value between $0.0$ and $1.0$, excluding the lower bound ($0.0$) and including the upper bound ($1.0$).
The cuRAND library is specifically optimized for NVIDIA GPUs and is widely used in implementations of stochastic algorithms on GPUs~\cite{Leonenko2015,Askar2021}.

\begin{lstlisting}[%
caption={Pseudocode of the ant colony optimization (ACO) kernel},%
float=htb,%
label={lst:ACO_SA_device_ACO}]
__global__ void ACO()(*@\label{ln1201}@*)
{	/* get the thread identifier */
	threadID = blockDim.x * blockIdx.x + threadIdx.x; (*@\label{ln1203}@*)
	/* if threadID is less than the number of threads */
	if(threadID < numThreads) {(*@\label{ln1205}@*)
		/* initialize the pheromone matrix */
		for (i = 1; i <= I; i++) {(*@\label{ln1207}@*)
			for (j = 1; j <= J[i]; j++) {tau[i][j] = curand_uniform();}}(*@\label{ln1208}@*)
		iterCounter = 0; /* initialize the local iteration counter */(*@\label{ln1209}@*)
		bestAntId = -1;
		/* until a solution has been found */
		while (isFound == 0 && iterCounter < maxIterationNumber){(*@\label{ln1211}@*)
			/* construct a path for each ant m of the colony of size M */
			for(m = 1; m <= M; m++){ (*@\label{ln1213}@*)
				for (i = 1; i <= I; i++){ sum = 0.0;
					for (j = 1; j <= J[i]; j++){
						eta[i][j] = (S(Want[m], i) == 0 ? 1:0) + X[i][j];(*@\label{ln1216}@*)
						sum += pow(tau[i][j], alpha) * pow(eta[i][j], beta);}
				    	r = curand_uniform(); sump = 0.0;(*@\label{ln1218}@*)
					for (j = 1; j <= J[i]; j++){
						p = pow(tau[i][j],alpha) * pow(eta[i][j],beta) / sum;(*@\label{ln1220}@*)
						sump += p;
						if(sump > r) {Want[m][i] = j; break;}}}} (*@\label{ln1222}@*)
			/* compute the change in pheromone concentration */
			for(m = 1; m <= M; m++){(*@\label{ln1224}@*)
				L[m] = NumberS(Want[m],I) + (T(Want[m]) <= Tmax ? 1:0);(*@\label{ln1225}@*)
				if (L[m] == Q) {atomicAdd(isFound, 1); bestAntId = m;}(*@\label{ln1226}@*)
				for (i = 1; i <= I; i++) {(*@\label{ln1227}@*)
					for (j = 1; j <= J[i]; j++) {dtau[i][j] = 0.0;}}
				for (i = 1; i <= I; i++) {
					idx = Want[m][i]; dtau[i][idx] += Q / L[m]; }(*@\label{ln1230}@*)
			/* update the pheromone concentration with evaporation */
			for (i = 1; i <= I; i++) {(*@\label{ln1232}@*)
				for (j = 1; j <= J[i]; j++) {
					tau[i][j] = tau[i][j] * rho + dtau[i][j];}}}(*@\label{ln1234}@*)
					iterCounter++;} /* increment the iteration counter */(*@\label{ln1235}@*)
			/* store the first feasible solution and the ID of the thread that found it */
			if(bestAntId != -1){(*@\label{ln1237}@*)
			Wfirst[threadID] = Want[bestAntId]; (*@\label{ln1238}@*)
			threadIdx[threadID] = threadID;}} (*@\label{ln1239}@*)
}
__device__ int NumberS(W, level){(*@\label{ln1241}@*)
	count = 0;
	for (i = 1; i <= level; i++){
		if(PartS(W, i) == 0) count++; }
	return count;
}(*@\label{ln1246}@*)
\end{lstlisting}

The kernel function~\texttt{ACO()} starts by determining the thread identifier~\texttt{threadID} (line~\ref{ln1203}).
Its value must be less than the required number of threads \texttt{numThreads} (line~\ref{ln1205}).
This check is necessary because the number of blocks in the grid is rounded up when the kernel parameters are computed (see Listing~\ref{lst:ACO_SA_host}, line~\ref{ln1110}).
The number of threads actually launched on the GPU may exceed \texttt{numThreads}, whereas all dynamic memory allocations for the arrays in \texttt{oprData} were made for exactly \texttt{numThreads} elements.

For all threads whose identifier is less than \texttt{numThreads}, the matrix of pheromone concentrations is initialized with pseudorandom values between $0$ and $1$ (lines~\ref{ln1207}--\ref{ln1208}),
and the local iteration counter \texttt{iterCounter} is reset to zero (line~\ref{ln1209}).

The global flag \texttt{isFound} and the local iteration counter \texttt{iterCounter} are used to control the threads: each thread works only while \texttt{isFound==0} and \texttt{iterCounter<maxIterationNumber} (line~\ref{ln1211}).
The value of the global flag \texttt{isFound} can be changed by any thread that has found a feasible solution, using the atomic function \texttt{atomicAdd()}~\cite{Hussain2016}.
Atomicity guarantees that the operation is performed correctly when several threads attempt to execute it simultaneously~\cite{2022CudaProgrammingGuide}.

The local iteration counter \texttt{iterCounter} protects the algorithm from looping forever if no feasible solution can be found.
If no ant of the colony manages to find a feasible solution within \texttt{maxIterationNumber} iterations (which is quite possible for a stochastic algorithm), the thread terminates.

After initialization, a path from the first to the last level of the tree is built for each ant $m$ of the colony of size $M$ (i.e., solutions of the problem are constructed) (lines~\ref{ln1213}--\ref{ln1222}).
Here \texttt{Want} is a local two-dimensional array of \texttt{M} elements, each of which is a vector of length \texttt{I}.
The elements of each vector hold the selected main size of the unit for each stage.

The heuristic information \texttt{eta[i][j]} is computed as the sum of the operability indicator of the initial fragment of the CES variant consisting of stages $1$ to $i$ (line~\ref{ln1216}) and the main size of the unit \texttt{X[i][j]}.
The pheromone values \texttt{tau[i][j]} and the heuristic values \texttt{eta[i][j]} are stored per stage \texttt{i} and unit \texttt{j}, i.e., they are shared by all edges of level \texttt{i} of the search tree that correspond to the same unit.

The probability is then computed (line~\ref{ln1220}), and the node to move to is selected (lines~\ref{ln1218}--\ref{ln1222}) as described in Section~\ref{subsubsec:ant_colony_optimization}.

The pheromone concentrations are changed according to the value of the fitness function, which measures the quality of the solution found by the \texttt{m}th ant.
The fitness value \texttt{L[m]} (line~\ref{ln1225}) of each ant \texttt{m} (line~\ref{ln1224}) equals the number of stages in the solution for which constraint (\ref{eq_4}) is satisfied, plus $1$ if constraint (\ref{eq_5}) is satisfied for this solution.

The number of stages in the solution for which constraint (\ref{eq_4}) is satisfied is determined by the function \texttt{NumberS()} (lines~\ref{ln1241}--\ref{ln1246}).
If \texttt{L[m]==Q} (a solution has been found, line~\ref{ln1226}), the global flag \texttt{isFound} is changed and the index of the ant that has found the solution is stored in \texttt{bestAntId} (line~\ref{ln1226}); the kernel \texttt{ACO()} then stops at the next iteration of the algorithm.

Next, the changes in pheromone concentration \texttt{dtau[i][j]} are computed (lines~\ref{ln1227}--\ref{ln1230}).
At the next step, the pheromone concentrations \texttt{tau[i][j]} are updated with evaporation taken into account (lines~\ref{ln1232}--\ref{ln1234}), after which the iteration counter \texttt{iterCounter} is incremented (line~\ref{ln1235}).
If a feasible solution has been found (line~\ref{ln1237}), it is stored in \texttt{Wfirst} (line~\ref{ln1238}), and the thread index \texttt{threadID} is stored in the array \texttt{threadIdx} (line~\ref{ln1239}).

The kernel function for simulated annealing, \texttt{SA()} (Listing~\ref{lst:ACO_SA_device_SA}), starts by determining the thread identifier \texttt{threadID} (line~\ref{ln1303}), in the same way as the ant colony kernel (Listing~\ref{lst:ACO_SA_device_ACO}).
Its value must be less than the required number of threads \texttt{numThreads} (line~\ref{ln1305}).

\begin{lstlisting}[%
caption={Pseudocode of the simulated annealing (SA) kernel},%
float=htb,%
label={lst:ACO_SA_device_SA}]
__global__ void SA()(*@\label{ln1301}@*)
{	/* get the thread identifier */
	threadID = blockDim.x * blockIdx.x + threadIdx.x; (*@\label{ln1303}@*)
	/* if threadID is less than the number of threads */
	if(threadID < numThreads) {(*@\label{ln1305}@*)
	/* initialize the current solution with the feasible initial solution */
	Wloc = W[threadID]; t = Tinit; (*@\label{ln1307}@*)
	while(t > Tfinal){/* until the final temperature is reached */(*@\label{ln1308}@*)
		for(l = 0; l < Lmax; l++) {(*@\label{ln1309}@*)
		/* construct a new candidate solution */
			Wcand = Perturb(Wloc);(*@\label{ln1311}@*)
		/* check the compatibility and processing time constraints */
			if (S(Wcand) == 0 && T(Wcand) <= Tmax){(*@\label{ln1313}@*)
			/* compute the change in the objective function */
				deltaCost = DefineCost(Wcand) - minCost; (*@\label{ln1315}@*)
				/* compute the probability */
				p = deltaCost < 0 ? 1.0 : exp (-deltaCost / t);(*@\label{ln1317}@*)
				/* if the probability exceeds a random number */
				if (p > curand_uniform()) { (*@\label{ln1319}@*)
					/* accept the new solution */
					Wloc = Wcand; (*@\label{ln1321}@*)
					/* if the solution found is better */
					if(DefineCost(Wloc) < minCost){ (*@\label{ln1323}@*)
						W[threadID] = Wloc;	(*@\label{ln1324}@*)
						Cost[threadID] = DefineCost(Wloc); (*@\label{ln1325}@*)
						atomicMinDbl(minCost, Cost[threadID]);(*@\label{ln1326}@*)
					}}}}(*@\label{ln1327}@*)
		t *= sigma; }} /* update the temperature */(*@\label{ln1328}@*)
}
__device__ int * Perturb(W){(*@\label{ln1330}@*)
	Wn = W;
	stage = (int) curand(1, I); /* choose a random stage */ (*@\label{ln1331}@*)
	/* choose a random standard unit size for this stage */
	Wn[stage] = (int) curand(1, J[stage]);(*@\label{ln1333}@*)
	return Wn;
}(*@\label{ln1335}@*)
\end{lstlisting}
For all threads whose identifier is less than \texttt{numThreads}, the current local solution \texttt{Wloc} is initialized with the feasible solution \texttt{W[threadID]} found by ACO (line~\ref{ln1307}),
and the temperature \texttt{t} is initialized with the initial value \texttt{Tinit} (line~\ref{ln1307}).
Here \texttt{Wloc} and \texttt{Wcand} are one-dimensional arrays of \texttt{I} elements, each of which stores the standard size of the unit for the corresponding stage.

The search continues until the temperature \texttt{t} drops below the final temperature \texttt{Tfinal} (line~\ref{ln1308}).
Before the current temperature \texttt{t} is lowered, the process is repeated \texttt{Lmax} times (line~\ref{ln1309}).

At each iteration, a candidate solution \texttt{Wcand} is constructed in the neighborhood of the current solution \texttt{Wloc} by the procedure \texttt{Perturb()} (line~\ref{ln1311}).
To do so, \texttt{Perturb()} (lines~\ref{ln1330}--\ref{ln1335}) randomly chooses a stage \texttt{stage} in the current feasible solution (line~\ref{ln1331}) and then, for this stage only, randomly chooses a unit from the set \texttt{J[stage]} of units admissible for it (line~\ref{ln1333}).
That is, at each iteration we replace one unit at one stage of the current feasible solution.

The candidate solution \texttt{Wcand} is checked for feasibility, i.e., for satisfying the compatibility constraint (\ref{eq_4}) and the processing time constraint (\ref{eq_5}) (line~\ref{ln1313}).

If the candidate is feasible, the difference \texttt{deltaCost} between its cost and the cost \texttt{minCost} of the best solution found so far by all threads is computed (line~\ref{ln1315}).
If \texttt{deltaCost} is negative, the candidate is accepted with probability \texttt{p = 1}; otherwise, it is accepted with probability \texttt{p = exp(-deltaCost/t)},
which depends on \texttt{deltaCost} and on the temperature \texttt{t} (line~\ref{ln1317}).
If \texttt{p} exceeds a random number between $0$ and $1$ (line~\ref{ln1319}), the candidate \texttt{Wcand} becomes the current solution \texttt{Wloc} (line~\ref{ln1321}) and is used for the further search.

If the cost of the new current solution \texttt{Wloc} is lower than the cost \texttt{minCost} of the current global optimal solution (line~\ref{ln1323}), the new solution replaces the current global optimal solution \texttt{W} (line~\ref{ln1324}), and its cost is stored in the array \texttt{Cost} (line~\ref{ln1325}) and in the global variable \texttt{minCost} using the atomic function \texttt{atomicMinDbl()} (line~\ref{ln1326}).

The temperature \texttt{t} is then lowered (line~\ref{ln1328}).
If no feasible candidate solution is obtained during the whole run, as the temperature decreases from its maximum to its minimum value, the result is the feasible initial solution used as the starting point (line~\ref{ln1307}).

\section{Results of Computational Experiments}
\label{chap5_sect5}
For the computational experiments, a program was developed in C using CUDA~\cite{2022CudaProgrammingGuide}.
The experiments were carried out on a computing system with an Intel Xeon processor (E5-1620~v2, $4$ cores with Hyper-Threading, $10$~MB Smart Cache, $3.7$~GHz) and $16$~GB of RAM, equipped with an NVIDIA Tesla K20c GPU. GPU specifications: $13$ streaming multiprocessors (SMs) with $192$ CUDA cores each ($2496$ CUDA cores in total), $5$~GB of global memory, and $48$~KB of shared memory per SM. The software comprised Ubuntu 16.04.1, NVIDIA driver 367.57, CUDA 8.0, and the GNU C++ compiler 5.4.0.

The test case was the CES equipment selection problem described in Section~\ref{chap1_sect2} (see Fig.~\ref{fig:test_CES}),
with $2$ to $12$ standard equipment sizes available for each of the $16$ processing stages ($2^{16}$ to $12^{16}$ CES variants in total).

In Chapters~\ref{chapter2}--\ref{chapter4}, the optimal CES equipment selection problem was solved by branch-and-bound, which yields globally optimal solutions~\cite{Borisenko2011,Borisenko2015lncs,Borisenko2016}.
Here, on the same test case and the same test platform, we compare the solutions obtained by the hybrid metaheuristic algorithm (ACO + SA) with those obtained by branch-and-bound.
Since both ant colony optimization and simulated annealing are stochastic (probabilistic) algorithms, the results for the same problem may differ even for identical input data.
Therefore, for each example, the algorithm was run $100$ times and the results were averaged.

\begin{figure}[htb]
    \centering
	\includegraphics[width=0.97\textwidth]{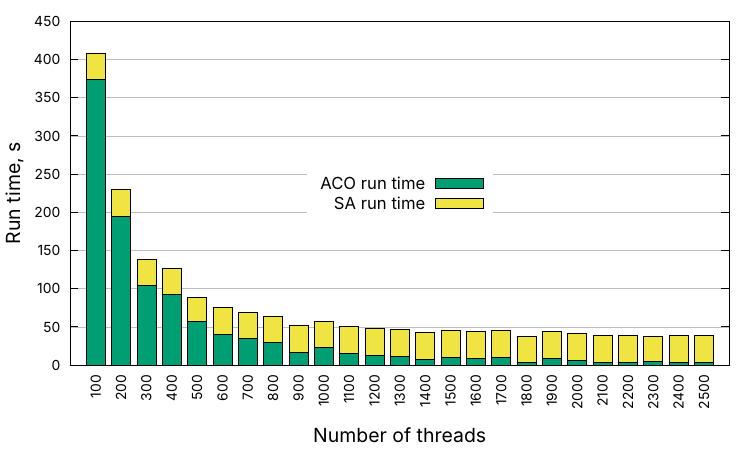}
	\caption{Run time of the parallel program versus the number of threads}
	\label{fig:ACO_SA_GPUspeedup}
\end{figure}

The first experiment studied the effect of the number of threads on the program's performance (see Fig.~\ref{fig:ACO_SA_GPUspeedup}).

The experiments were carried out for the test CES with $10$ standard equipment sizes available at each stage ($10^{16}$ possible CES variants in total).
The program was run with $100$ to $2500$ threads in steps of $100$. The run time of the algorithm decreases as the number of threads increases.
With $100$ threads, the ant colony algorithm (ACO) accounts for $91\%$ of the total run time; this share then decreases and, beyond $2100$ threads, remains practically constant at about $11\%$ of the total run time.

\begin{figure}[htb]
    \centering
	\includegraphics[width=0.97\textwidth]{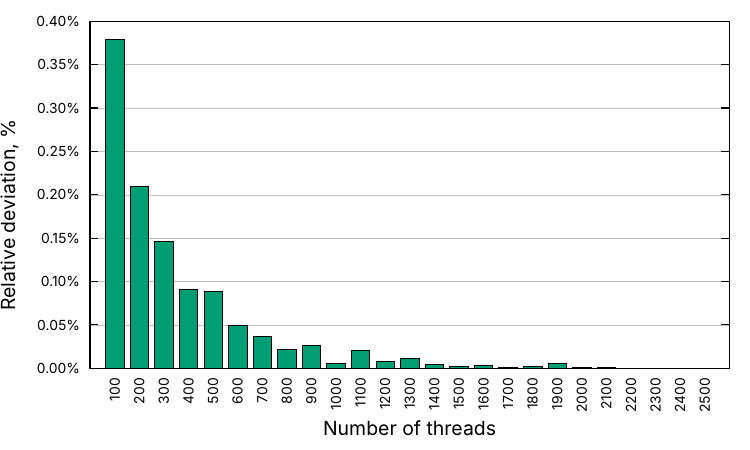}
	\caption{Relative deviation of the objective function value versus the number of threads}
	\label{fig:ACO_SA_GPU_deviation}
\end{figure}

Figure~\ref{fig:ACO_SA_GPU_deviation} shows the relative deviation of the objective function value (the cost of the CES variant) versus the number of threads used.

The relative deviation was computed as
$$\text{deviation}=\frac{|\text{observed}-\text{expected}|}{\text{expected}} \cdot 100\%,$$
where the observed value is the cost of the CES variant obtained by the metaheuristic algorithm and the expected value is the globally optimal cost of the CES variant obtained by branch-and-bound.
With $100$ threads, the relative deviation is about $0.38\%$; starting from about $2000$ threads, it decreases practically to $0\%$.

\begin{figure}[htb]
    \centering
	\includegraphics[width=0.97\textwidth]{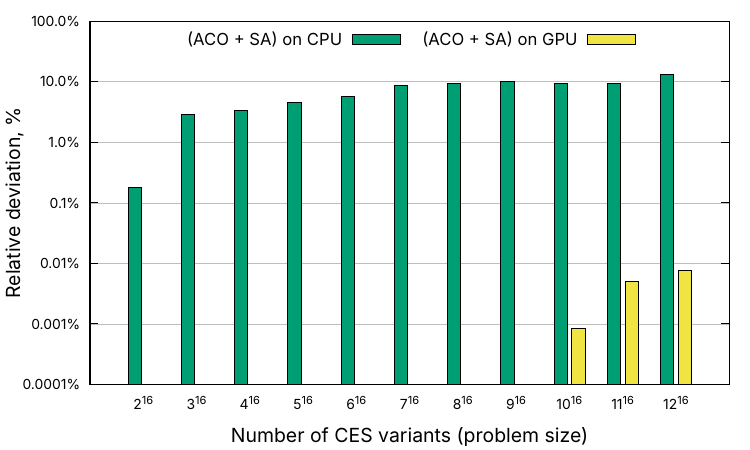}
	\caption{Relative deviation of the objective function value versus the problem size}
	\label{fig:ACO_SA_GPU_deviationup}
\end{figure}

\begin{figure}[htb]
    \centering
	\includegraphics[width=0.97\textwidth]{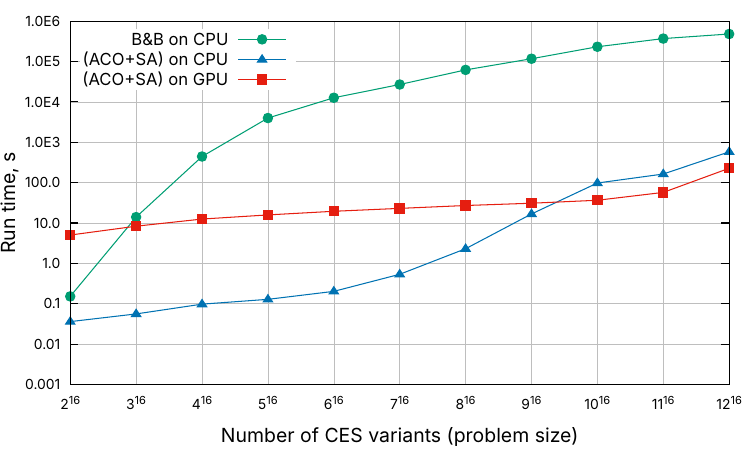}
	\caption{Run time of the program versus the problem size}
	\label{fig:ACO_SA_GPU_time}
\end{figure}

A number of tests for problems of different sizes were carried out to measure the objective function values and run times of the proposed hybrid metaheuristic algorithm (sequential CPU and parallel GPU versions) and of the sequential branch-and-bound algorithm (B\&B).
The results for the different approaches are shown in Figs.~\ref{fig:ACO_SA_GPU_deviationup} and~\ref{fig:ACO_SA_GPU_time} (the vertical axis is logarithmic).

Figure~\ref{fig:ACO_SA_GPU_deviationup} shows the relative deviation of the objective function value (the cost of the CES variant) versus the problem size.
The program was run on the GPU with $2500$ parallel threads.

The parallel version ``(ACO+SA) on GPU'' obtains near-optimal solutions (for problems of size $2^{16}$--$9^{16}$, the relative deviation is $0\%$; for problems of size $10^{16}$--$12^{16}$, it is less than $0.01\%$).
The relative deviation of the sequential version ``(ACO+SA) on CPU'' is below $1\%$ for a small problem ($2^{16}$), but then increases to $14\%$ for the largest problem ($12^{16}$).
This is because a single run of simulated annealing has a low probability of finding a good solution for large problems.

Figure~\ref{fig:ACO_SA_GPU_time} shows the run time of the program versus the problem size (the vertical axis is logarithmic).

The run time of branch-and-bound grows exponentially with the problem size.
For the smallest problem ($2^{16}$ possible CES variants in total), the run time is less than $1$~second.
For a large problem ($12^{16}$), however, it is about $134$~hours.

The run time of the sequential hybrid algorithm is less than $0.1$~second for a small problem ($2^{16}$) and increases to $581$~seconds for the largest problem ($12^{16}$).

The run time of the parallel hybrid algorithm (ACO~+~SA) increases smoothly from $5$ to $227$~seconds.
The parallel hybrid algorithm is slower than the sequential one for problems with $2^{16}$ to $9^{16}$ variants.
For the largest problems, with $10^{16}$ to $12^{16}$ variants, the parallel implementation is approximately $2.6$--$2.9$ times faster than the sequential one.

At first glance, a speedup of $2.6$--$2.9$ over the sequential version is modest.
However, the quality of the solutions obtained by the parallel implementation ``(ACO+SA) on GPU'' is higher than that of the sequential ``(ACO+SA) on CPU'': the relative deviation of the solution obtained from the globally optimal solution found by branch-and-bound is less than $0.01\%$, compared with $9$--$14\%$ for the sequential program.

This is achieved because the parallel version performs many simultaneous independent searches for the optimal solution, and the more parallel threads there are, the more likely it is that, in practically the same run time, some thread will find the globally optimal solution.
Running the parallel algorithm with $2500$ threads is equivalent to running the sequential algorithm $2500$ times and choosing the best of the solutions found.

The experiments confirm that the parallel implementation of the hybrid metaheuristic algorithm (ACO+SA) yields high-quality solutions for the test problem that coincide with the globally optimal values obtained by the deterministic branch-and-bound algorithm, but much faster: in a few tens of seconds instead of a few tens of hours.
%%%% Chapter 6
\chapter{Tuning the Parameters of the Ant Colony Algorithm}
\label{chapter6}

The hybrid algorithm presented in Chapter~\ref{chapter5} works in two stages: (1) finding a feasible initial solution (a fully operable CES variant that produces the given product range within the given time) by ant colony optimization (ACO), and (2) improving this solution by simulated annealing (SA).
The first stage is thus the basis for the second, and without it the hybrid algorithm cannot work.

In Chapter~\ref{chapter5}, a feasible initial solution was successfully found by ant colony optimization for problems with up to $12^{16}$ ($\approx10^{17}$) possible CES variants~\cite{Borisenko2017lncs,Borisenko2019jos}.
However, when the problem size exceeds $11^{16}$, the performance of the algorithm deteriorates considerably despite the use of high-performance GPUs.
Problems larger than $12^{16}$ were not solved, which motivated us to turn to tuning the parameters of the algorithm.

The chapter is organized as follows.
Section~\ref{chap6_sect1} reviews existing approaches to tuning the parameters of metaheuristic algorithms.
Section~\ref{chap6_sect2} discusses the effect of the parameters $\alpha$, $\beta$, and $\rho$ on the ant colony algorithm.
Section~\ref{chap6_sect3} proposes a method for tuning these parameters based on statistical analysis of the results of multiple runs of the parallel program, together with three ways of applying it: random, constant, and multi-constant.
Section~\ref{chap6_sect4} reports the results of computational experiments and compares the approaches in terms of the time needed to find a feasible solution and the time spent on the tuning study.
Section~\ref{chap6_sect5} presents the results of running the program on GPUs of three generations (Kepler, Pascal, Volta) and on compute nodes with several GPUs.

\section{Approaches to Tuning the Parameters of Metaheuristic Algorithms}
\label{chap6_sect1}

It is well known that the efficiency of metaheuristic algorithms depends to a large extent on their parameters~\cite{Pellegrini2012,Joshi2020}.
Identifying these parameters (so-called tuning of metaheuristics) is a nontrivial problem to which a considerable body of research has been devoted.
As a rule, there are no universally optimal parameter values for metaheuristic algorithms~\cite{Talbi2009}.

There are two strategies for setting the parameters of an algorithm: offline parameter initialization and online parameter initialization~\cite{Talbi2009,Eryoldas2022}.
With offline initialization, the parameter values are searched for before the algorithm is actually applied.
The best values are determined once at the design stage and are not changed thereafter when the algorithm is used~\cite{fallahi2014}.

Traditionally, offline tuning has mainly been performed in a trial-and-error fashion~\cite{stuetzle2011parameter}.
The drawback of this approach is that trial-and-error tuning depends heavily on the intuition and experience of the algorithm designer, which are usually difficult or impossible to document and, therefore, to reproduce.
Recently, much attention has been paid to search-based offline tuning methods, which allow the parameters of algorithms to be tuned (semi-)automatically with minimal human involvement.
Nevertheless, these methods yield specific parameter values before the algorithm is applied, and these values do not change while a particular problem is being solved.

An alternative to offline tuning is online tuning~\cite{Sun2014}.
Typically, online tuning makes it possible to modify the parameters of the algorithm while a particular problem is being solved. The potential advantage of online tuning is that algorithms can adapt to the characteristics of a particular problem and solve it more efficiently.
When the problems to be solved are relatively heterogeneous, parameter values found in advance, which generally lead to good performance on average, may lead to considerably worse results in some individual cases.
Adapting the parameters during the run may also be useful for achieving the best performance depending on the stage of the search.
The operation of an algorithm can often be divided into explorative and exploitative search phases, and the optimal parameter values in these phases may differ considerably.
Furthermore, if algorithms are applied in situations very different from those for which they were designed or tuned offline, adapting the parameters during the run increases the robustness of the algorithm.
There are various ways of changing the parameters of an algorithm during its run~\cite{fallahi2014}.
One of the simplest is to specify a rule for changing the parameters before the algorithm is actually launched.
In this approach, the static parameters are replaced offline by (deterministic or stochastic) functions of the computation time or of the number of iterations.
Some authors use the term deterministic parameter control~\cite{Diao2010}.
Others, however, consider this term inaccurate and propose the term pre-scheduled parameter variation instead, since the adjective ``deterministic'' does not quite characterize this method of parameter control, given that the parameters may be chosen at random~\cite{Maur2010}.
Although pre-scheduled parameter variation is an online tuning method, it does not eliminate the problem of offline tuning, because in this case, too, the parameters must be set appropriately before the algorithm is launched.

An alternative approach is to use adaptive parameter settings, in which the scheme for changing the parameters is defined as a function of some statistics of the algorithm's behavior~\cite{Mezura-Montes2009}.
The indicators used for online adaptation can be roughly grouped according to whether they are based on absolute or relative information.
In the first case, the adaptation strategy monitors the occurrence of certain events during the run, for example, reaching some fixed threshold distance between the solutions found.
Exceeding this threshold triggers a set of predefined rules for changing the parameters of the algorithm.
In the second case, the adaptation strategy takes into account the relative difference between the performance achieved with different parameter values and adapts the parameter values towards the most successful ones.
Parameter adaptation requires certain design decisions that are not directly related to the implementation of the algorithm itself.
In particular, the expressions describing the parameter updates must be defined a priori, and one must be confident that the mechanisms used are very robust with respect to their particular implementation.

Another object of study may be tuning parameters that are modified by the algorithm itself during its run.
For example, the tuning parameters of the search algorithm are added to the search space of the main problem being solved,
and the optimization problem is then solved in the new, extended search space. This approach is called \emph{self-adaptation}~\cite{DeMendoncaMesquita2020,Cicirello2021,Santos2022}.
Some authors extend the notion of self-adaptation to \emph{search-based adaptation}, in which the parameters of the algorithm are adapted using approaches different from the one used to solve the main problem~\cite{Anghinolfi2008performance,Pellegrini2012}.

Moreover, all existing approaches to studying the relationship between parameter settings, problem characteristics, and algorithm performance can be divided into three categories: (1) analytical approaches, (2) automated approaches, and (3) empirical approaches~\cite{dobslaw2010,Swan2022}.

The analytical approach to understanding algorithm parameters and recommending their values is based on mathematical proofs.
The automated approach consists in applying another algorithm or heuristic to tune the parameters of the main algorithm.
If such an automated approach uses the algorithm or heuristic in an introspective sense (for self-analysis during the run), the term \emph{automated self-tuning} is often used.
The empirical approach collects experimental data on the behavior of the algorithm and analyzes them to draw conclusions by trial and error or by the One-Factor-At-a-Time (OFAT) method, and to give recommendations on choosing parameter values.
Design of Experiments (DOE) methods are typically used for this purpose~\cite{dobslaw2010parameter,Arin2011}.

Among automated offline tuning methods, the most widely used are so-called algorithm configurators, general-purpose software tools that treat the algorithm being tuned as a black box and search for its best configuration on a training set of problem instances within a given computational budget~\cite{Huang2020,Eryoldas2022}.
Configurators differ in their search strategy.
Racing methods, in particular F-Race~\cite{Birattari2002} and its iterated version \emph{irace}~\cite{LopezIbanez2016}, generate a set of candidate configurations, evaluate them sequentially on more and more problem instances, and use statistical tests (e.g., the Friedman test) to discard clearly inferior candidates early.
The distributions from which the candidates of the next iteration are generated gradually concentrate around the best (``elite'') configurations.
ParamILS~\cite{Hutter2009} performs an iterated local search in the space of discretized parameter values, while SMAC~\cite{Hutter2011} builds a regression model of algorithm performance as a function of the parameter values and instance features and uses it to select the most promising configurations. What these methods have in common is that they are designed for expensive sequential runs of the algorithm being tuned: their main goal is to reduce the number of runs needed to find a good configuration.

\section{Parameters of the Ant Colony Algorithm}
\label{chap6_sect2}

Before the ant colony algorithm is started, a number of parameters that affect its operation must be set~\cite{Trindade2019}.
These parameters have an extremely strong influence on the quality of the solutions obtained and on the run time of the algorithm~\cite{Li2016}.

The parameters $\alpha$ (the information elicitation factor) and $\beta$ (the expected heuristic factor) control the balance between the influence on the search of the pheromone concentration (which reflects the history of successful solutions) and of the heuristic information (which expresses the desirability of a particular choice based on the features of the problem), respectively.
For $\alpha=0$, the pheromone information is not used, and the previous experience of the ant colony is ignored.
This reduces the convergence rate of ACO but increases its capacity for global search.
If $\alpha$ is large, an ant will predominantly choose the same path as the previous ants~\cite{Li2016}.
The convergence rate of ACO then increases, but so does the probability of getting trapped in a local optimum, while the capacity for global search decreases.
Conversely, if $\alpha$ is too small, the convergence of ACO slows down, but the global search capability of the algorithm increases.

If $\beta=0$, the heuristic information is not taken into account, and an ant chooses its path solely on the basis of the pheromone concentration, which leads to suboptimal solutions.
If $\beta$ is large, the heuristic information dominates the choice of path, and the algorithm approaches a greedy algorithm~\cite{Temlyakov2011,Nuriyev2018}.
For example, in the travelling salesman problem, the city nearest to the current one will be chosen~\cite{Kizilates2013}; in the search for a feasible initial CES variant, the unit of the largest size will be chosen.

The pheromone evaporation rate $\rho\in(0,\,1]$ controls the degree to which the pheromone concentration decreases.
In the literature, the expression $(1-\rho)$ is also used as the factor by which the current pheromone concentration is multiplied upon evaporation~\cite{dorigo2019ant}.
This can be interpreted as follows: as the evaporation rate $\rho$ increases, the factor $(1-\rho)$ decreases, and the pheromone evaporates faster: a value close to one means strong evaporation, and a value close to zero means weak evaporation.
For example, for $\rho=0.9$ (strong evaporation), the factor is $1-\rho=0.1$, and multiplying by $0.1$ reduces the concentration tenfold.
Conversely, for $\rho=0.1$ (weak evaporation), multiplying by $1-\rho=0.9$ reduces the concentration by $10\%$.
In our implementation, the value $\rho$ itself is used as the factor, so the logic is reversed: a value of $\rho$ close to one means weak evaporation, and a value close to zero means strong evaporation~\cite{stuetzle2011parameter}.

As noted in Section~\ref{chap4_sect3}, differences in the control flow of individual threads in a grid degrade overall performance because of warp divergence.
Different numbers of ants in the colonies would lead to different numbers of loop iterations in the GPU threads, which would degrade the computational performance of the GPU.
For this reason, the number of ants is not tuned in this book; it is the same for all colonies and equal to $100$.

\section{Proposed Tuning Method}
\label{chap6_sect3}
The goal of tuning is to find values of the parameters $\alpha$, $\beta$, and $\rho$ for which ant colony optimization finds a feasible initial solution as fast as possible.
Here a constraint satisfaction problem is solved, i.e., the goal is to find a feasible solution rather than a solution that is optimal in terms of capital cost.
The parameters are tuned by statistical analysis of the results of multiple experimental runs of the program.
We analyze how frequently solutions are found for particular intervals of the parameters $\alpha$, $\beta$, and $\rho$.
Checking how often successful results occur within a range of parameter values does not require any specific and non-obvious information, such as the shape of membership functions in fuzzy logic~\cite{olivas2015dynamic,Castillo2015} or automated tuning methods~\cite{Barbosa2017,Skakov2018}.
Owing to the parallelization of the algorithm and the use of modern GPUs, a large number of computational experiments can be performed to accumulate statistics in acceptable time.

The idea of the proposed method is close to iterated racing~\cite{LopezIbanez2016}: in both cases, the search region for the parameter values is successively narrowed around the most successful values. However, there are several differences.

First, the evaluation of configurations is built into the parallel algorithm itself.
Each GPU thread is an independent ant colony with its own triple of values $\alpha$, $\beta$, and $\rho$; therefore, a single run of the program tests thousands of configurations simultaneously, and no external control software is needed for tuning.

Second, the success criterion for a configuration is not averaged performance but the very fact that a feasible solution has been found.
For a thread that has found a feasible solution, its triple of values is stored, and the intervals are narrowed according to the frequency distribution of such triples.
This matches the constraint satisfaction problem being solved, in which the result of each run is effectively binary: a feasible solution is either found or not.

Third, the multi-constant approach produces not a single ``best'' configuration but a parallel portfolio of configurations~\cite{Lindauer2017} executed simultaneously by different threads, which is natural for the massively parallel GPU architecture.

The price paid for the simplicity of the method is the absence of statistical tests for comparing configurations, and the fact that tuning was performed on a single family of problems (one CES with different problem sizes), whereas configurators are usually trained on many heterogeneous instances.
The applicability of the parameter values found to other CES requires further verification.

We propose three variants of tuning the algorithm: the random approach, the constant approach, and the multi-constant approach~\cite{Borisenko2019aco,Borisenko2021}.

\begin{figure}[htb]
    \centering
	\includegraphics[width=0.97\textwidth]{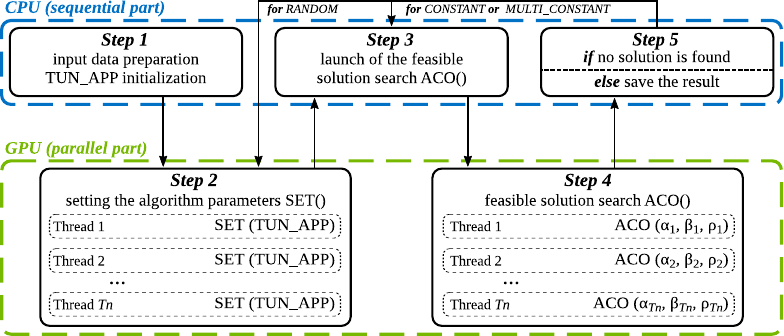}
	\caption{General method for tuning the ACO parameters}
	\label{fig:ACO_tuning}
\end{figure}

Figure~\ref{fig:ACO_tuning} shows the general approach to tuning the parameters of the algorithm:
\begin{enumerate}
    \item The host (CPU) reads the input data and initializes the parameters on the device (GPU).
    \item The device initializes the tuning parameters in one of the ways shown in Listing~\ref{lst:device_SET}. \label{set_tuneapp}
    \item The host launches the kernel function \texttt{ACO()} shown in Listing~\ref{lst:ACO_SA_device_ACO}. \label{start_aco_kernel}
    \item The kernel function \texttt{ACO()} searches for a feasible solution. \label{run_aco_kernel}
    \item The host receives the search results for further processing; if no feasible solution has been found, step~\ref{set_tuneapp} or step~\ref{start_aco_kernel} is repeated, depending on the tuning approach adopted. Otherwise, the feasible solution found is stored.
\end{enumerate}

\begin{lstlisting}[%
caption={Ways of setting the algorithm parameters},%
float=htb,%
label={lst:device_SET}]
__global__ void SET(TUN_APP) {
	...	/* get the thread identifier */
	threadID = blockDim.x * blockIdx.x + threadIdx.x;
	switch(TUN_APP){
		case RANDOM: /* parameters are initialized with random values */
			alpha[threadID] = rand(alpha_a, alpha_b); (*@\label{set:random_begin}@*)
			beta[threadID] = rand(beta_a, beta_b);
			rho[threadID] = rand(rho_a, rho_b);	(*@\label{set:random_end}@*)
		break;
		case CONSTANT: /* the same parameter values for all threads */
			alpha[threadID] = constant_alpha; (*@\label{set:const_begin}@*)
			beta[threadID] = constant_beta;
			rho[threadID] = constant_rho; (*@\label{set:const_end}@*)
		break;
		case MULTI_CONSTANT: /* different parameter values for all threads */
			alpha[threadID] = constant_alpha[threadID]; (*@\label{set:multiconst_begin}@*)
			beta[threadID] = constant_beta[threadID];
			rho[threadID] = constant_rho[threadID]; (*@\label{set:multiconst_end}@*)
		break;
	}	...}
\end{lstlisting}

\subsubsection*{Random approach}
The parameters $\alpha$, $\beta$, and $\rho$ are initialized with uniformly distributed random values from the given intervals $[\alpha_a, \alpha_b]$, $[\beta_a, \beta_b]$, and $[\rho_a, \rho_b]$, respectively (see Listing~\ref{lst:device_SET}, lines~\ref{set:random_begin}--\ref{set:random_end}).
The interval bounds are taken according to the recommendations of~\cite{Khan2009}.
To generate random numbers in a given interval $(a, b]$ other than $(0.0,\, 1.0]$, the expression \texttt{rand(a,b) = curand\_uniform() * (b-a) + a} is used; because of the specifics of the function \texttt{curand\_uniform()}, the lower bound is excluded and the upper bound is included.
By running this algorithm many times, we obtain a set of triples of the ACO parameters $\alpha$, $\beta$, and $\rho$ for which feasible solutions were found.
The stored parameter values are then processed statistically.
If no feasible solution has been found at step~\ref{run_aco_kernel} of the method after \texttt{maxIterationNumber} iterations, the method returns to step~\ref{set_tuneapp} (see Fig.~\ref{fig:ACO_tuning}).

\subsubsection*{Constant approach}
This approach differs from the random one in that the values of $\alpha$, $\beta$, and $\rho$ are the same for all threads; they are initialized once before the algorithm is started and do not change during its execution (see Listing~\ref{lst:device_SET}, lines~\ref{set:const_begin}--\ref{set:const_end}).
These initial values are obtained by statistical analysis of the set of parameter triples produced by the random approach, although they could also be taken, for example, from recommendations in the literature~\cite{Khan2009}.
If no feasible solution has been found at step~\ref{run_aco_kernel} of the method after \texttt{maxIterationNumber} iterations, the method returns to step~\ref{start_aco_kernel} (see Fig.~\ref{fig:ACO_tuning}).

\subsubsection*{Multi-constant approach}
This approach differs from the constant one in that the values of $\alpha$, $\beta$, and $\rho$ differ from thread to thread; they are initialized once before the algorithm is started and do not change during its execution (see Listing~\ref{lst:device_SET}, lines~\ref{set:multiconst_begin}--\ref{set:multiconst_end}).
These values are taken from the set of stored triples found with the random approach.
In this case, the random approach in effect trains the parallel ACO algorithm, since the random approach stores only those triples of $\alpha$, $\beta$, and $\rho$ for which a solution was found.
If no feasible solution has been found at step~\ref{run_aco_kernel} of the method after \texttt{maxIterationNumber} iterations, the method returns to step~\ref{start_aco_kernel} (see Fig.~\ref{fig:ACO_tuning}).

\section{Results of Computational Experiments}
\label{chap6_sect4}
For the computational experiments, a program was developed in C using CUDA~\cite{2022CudaProgrammingGuide}.
The experiments were carried out on a computing system with an Intel Xeon Gold $5118$ processor ($12$ cores with Hyper-Threading, $2.3$~GHz) and $192$~GB of RAM, equipped with an NVIDIA Tesla V100-SXM2 GPU.
GPU specifications: $80$ streaming multiprocessors (SMs) with $64$ CUDA cores each ($5120$ CUDA cores in total), maximum clock rate $1.53$~GHz, and $16$~GB of global memory.
The software comprised CentOS Linux 7.5.1804, NVIDIA driver 410.72, CUDA 10.0, and the GNU C++ compiler 6.4.0.

The test case was the CES equipment selection problem described in Section~\ref{chap1_sect2} (see Fig.~\ref{fig:test_CES}), with $11$ to $20$ standard equipment sizes available for each of the $16$ processing stages ($11^{16}$ to $20^{16}$ CES variants in total).

Since both ant colony optimization and simulated annealing are stochastic (probabilistic) algorithms, the results for the same problem may differ even for identical input data.
Therefore, for each problem size from $11^{16}$ to $20^{16}$ ($10$ series in total), the algorithm was run $100$ times; the time needed to find a feasible solution was measured, the average time was computed, and the values of $\alpha$, $\beta$, and $\rho$ were stored.

The program was run with $5120$ threads, which equals the number of CUDA cores of the GPU used.

\subsubsection*{Random approach}
The initial search bounds for the first iteration were chosen following the recommendations of~\cite{Khan2009}: $\alpha\in(0,2]$, $\beta\in(0,2]$, and $\rho\in(0,1]$.
After a complete series of experiments, $10\cdot100=1000$ triples of the parameter values $\alpha$, $\beta$, $\rho$ were obtained for problems of various sizes.
The total GPU time spent on the first series of experiments with the random approach was $51\,054$~s $\approx14$~h.

\begin{figure}[htb]
    \centering
	\includegraphics[width=0.97\textwidth]{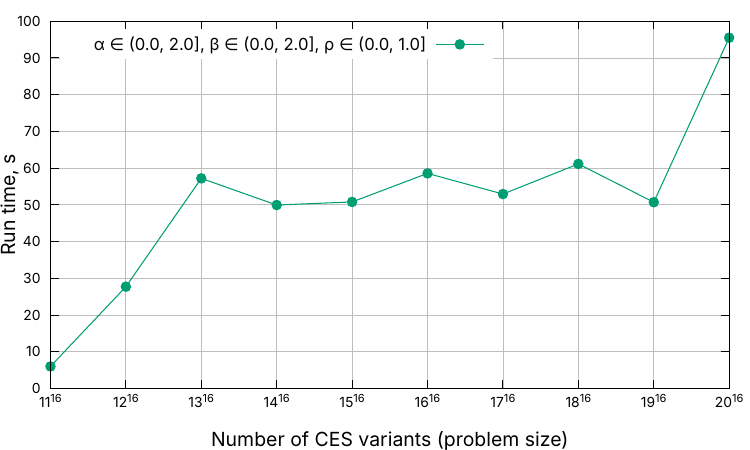}
	\caption{Run time of the parallel program versus the problem size at the first tuning iteration}
	\label{fig:random221_runtime}
\end{figure}

Figure~\ref{fig:random221_runtime} shows the run time of the program versus the problem size.
Note that the program does not enumerate all possible CES variants; the search stops at the first feasible solution.

\begin{figure}[htb]
    \centering
	\includegraphics[width=0.97\textwidth]{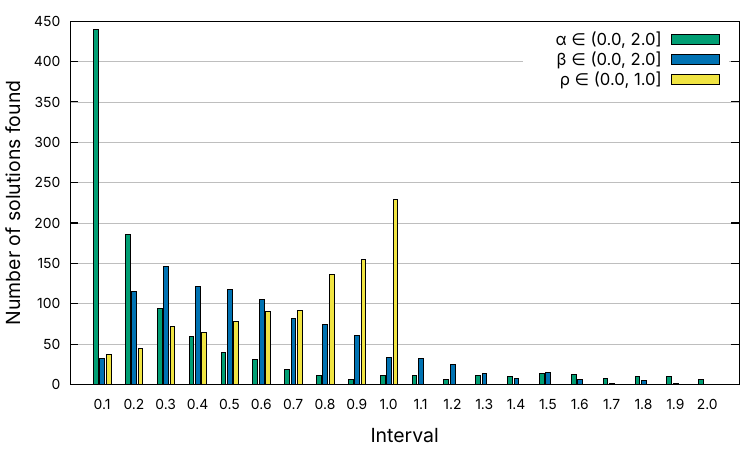}
	\caption{Frequency distribution of the feasible solutions found after the first iteration}
	\label{fig:random221_frequency}
\end{figure}

Figure~\ref{fig:random221_frequency} shows the frequency histogram of the feasible solutions found.

The histogram is constructed as follows.
The whole interval $(0.0,\,2.0]$ of the algorithm parameters is divided into equal subintervals (bins) of width $0.1$: $(0.0,\,0.1]$, $(0.1,\,0.2]$, $(0.2,\,0.3]$, \ldots, $(1.9,\,2.0]$.
These bins are plotted along the horizontal axis, and above each of them a bar is drawn for each of the algorithm parameters $\alpha$, $\beta$, and $\rho$.
The height of a bar is proportional to the number of feasible solutions found with the corresponding parameter value falling into this bin.

Analysis of the frequency distribution showed that $\approx90\%$ of the solutions found were obtained with the ACO parameters in the ranges $\alpha\in(0.0,\,1.2]$, $\beta\in(0.1,\,1.2]$, $\rho\in(0.2,\,1.0]$.
The idea of the proposed method for tuning the ACO parameters is to move step by step towards those ranges of the parameters $\alpha$, $\beta$, and $\rho$ in which feasible solutions occur more often (i.e., are found with higher frequency).

The search for $1000$ new solutions was then repeated with the adjusted parameter ranges, and the frequency of the solutions found was again analyzed for each parameter.

This process (finding 1000 solutions, analyzing the frequencies, and adjusting the parameter intervals to those containing $90\%$ of the solutions) was repeated $7$ times.
The total GPU time after seven tuning iterations was $86\,509$\,s $\approx24$\,h.

\begin{figure}[htb]
    \centering
	\includegraphics[width=0.97\textwidth]{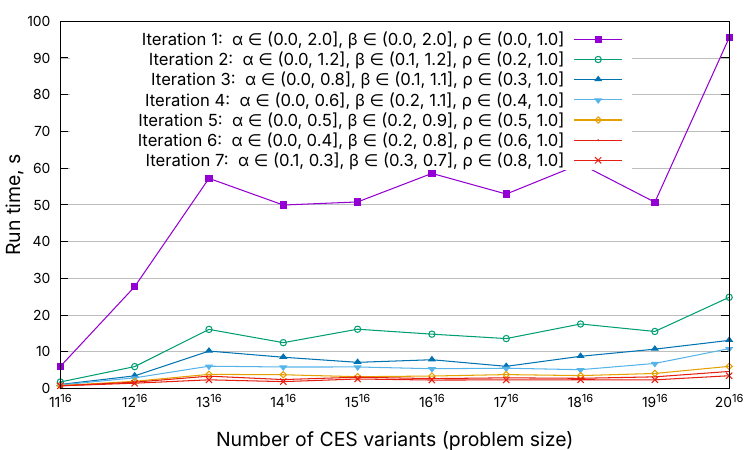}
	\caption{Run time of the parallel program versus the problem size over seven tuning iterations}
	\label{fig:random_runtime}
\end{figure}

The change in the average run time of the algorithm over the seven iterations for problems of size $11^{16}$--$20^{16}$ is shown in Fig.~\ref{fig:random_runtime}.

\begin{figure}[htb]
    \centering
	\includegraphics[width=0.97\textwidth]{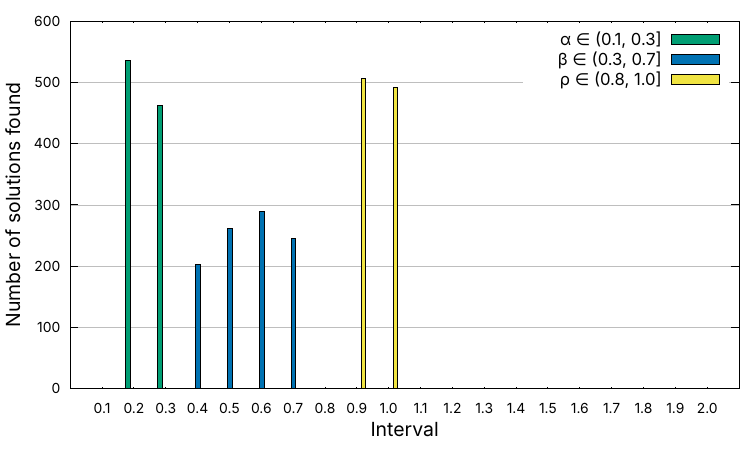}
	\caption{Frequency distribution of the feasible solutions found after the seventh iteration}
	\label{fig:random0007_frequency}
\end{figure}

Figure~\ref{fig:random0007_frequency} shows the frequency distribution of the feasible solutions found at the last, seventh, iteration (parameter ranges $\alpha\in(0.1,\,0.3]$, $\beta\in(0.3,\,0.7]$, $\rho\in(0.8,\,1.0]$).

After the last, seventh, iteration, the parameter intervals contracted to the point $(\alpha=0.2, \, \beta=0.5, \, \rho=0.9)$.
Thus, the statistical analysis of the random approach yielded parameter values that were subsequently used in the study of the constant approach.

\subsubsection*{Constant approach}

In the constant approach, every thread running on the GPU uses the same tuple of constants $\alpha$, $\beta$, $\rho$.

\begin{figure}[htb]
    \centering
	\includegraphics[width=0.97\textwidth]{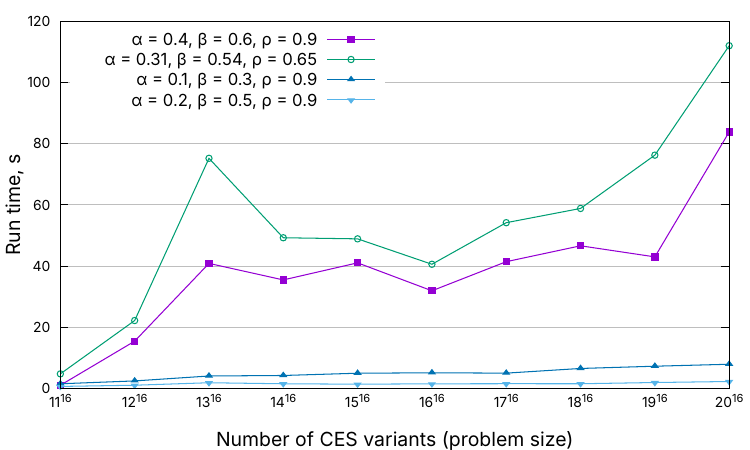}
	\caption{Run time of the program for problems of various sizes, constant approach}
	\label{fig:time_const}
\end{figure}

Figure~\ref{fig:time_const} shows the run time of the program versus the problem size for the constant approach.

Table~\ref{tab:AveragedExecTimesTuning} gives the average run times of the algorithm over the whole range of problem sizes (from $11^{16}$ to $20^{16}$).

\begin{table}[htb]
	\centering
	\small
	\caption{Average run times of the algorithm}
	\label{tab:AveragedExecTimesTuning}
	\begin{tabular}{cccc}
		\toprule
		$\alpha$ & $\beta$ & $\rho$ & Run time, s \\
		\midrule
		$0.4$ & $0.6$ & $0.9$ & $38.1$ \\
		$0.31$ & $0.54$ & $0.65$ & $54.2$ \\
		$0.1$ & $0.3$ & $0.9$ & $4.9$ \\
		$0.2$ & $0.5$ & $0.9$ & $1.5$ \\
		\bottomrule
	\end{tabular}
\end{table}

The first set of constants, $\alpha=0.4$, $\beta=0.6$, $\rho=0.9$, was chosen empirically and used in the earlier experiments presented in Section~\ref{chap5_sect5}~\cite{Borisenko2017lncs,Borisenko2019jos}.
The performance of the algorithm with the new parameter values is compared below with its performance with these values.

The second set of constants, $\alpha=0.31$, $\beta=0.54$, $\rho=0.65$, was taken as the mean values of the parameters obtained after the first iteration of the random approach, which produced the first $1000$ triples of $\alpha$, $\beta$, and $\rho$.
The intuitive assumption was that averaging known good parameter values would yield good candidate parameters.
However, using the averaged parameter values increased the average run time by a factor of $\approx1.4$ compared with the parameters used in Chapter~\ref{chapter5}
(from $\approx38.1$\,s for $\alpha = 0.4$, $\beta = 0.6$, $\rho = 0.9$ to $\approx54$~s for $\alpha = 0.31$, $\beta = 0.54$, $\rho = 0.65$).

The third set of constants, $\alpha=0.1$, $\beta=0.3$, $\rho=0.9$, was chosen on the basis of the frequency analysis after the first iteration of the random approach (see Fig.~\ref{fig:random221_frequency}) as the parameter values approximately corresponding to the highest frequency of feasible solutions.
This reduced the average run time of the algorithm by a factor of $\approx7.7$ (from $\approx38.1$\,s for $\alpha = 0.4$, $\beta = 0.6$, $\rho = 0.9$ to $\approx4.9$\,s for $\alpha = 0.1$, $\beta = 0.3$, $\rho = 0.9$).

After the seventh iteration, the parameter intervals contracted to the point $\alpha=0.2$, $\beta=0.5$, $\rho=0.9$, which was used as the fourth set of constants.
This reduced the average run time of the algorithm by a factor of $\approx25$ (from $\approx38.1$~s for $\alpha = 0.4$, $\beta = 0.6$, $\rho = 0.9$ to $\approx1.5$~s for $\alpha = 0.2$, $\beta = 0.5$, $\rho = 0.9$).

Since the constants were determined from the data obtained after seven iterations of the random approach, the tuning time can be considered the same as for the random approach, $86\,509$\,s $\approx24$\,h.

\subsubsection*{Multi-constant approach}

In the multi-constant approach, each thread running on the GPU uses its own tuple of constants $\alpha$, $\beta$, $\rho$.

For this purpose, we used the set of constants obtained in the first series of experiments with the random approach, for $\alpha\in(0.0,\,2.0]$, $\beta\in(0.0,\,2.0]$, $\rho\in(0.0,\, 1.0]$.

After $100$ experiments for the problem size $11^{16}$, we have a set of $100$ parameter triples.
After the next $100$ experiments for the problem size $12^{16}$, we have $200$ triples ($100$ for $11^{16}$ + $100$ for $12^{16}$ = $200$); after the next $100$ experiments for the problem size $13^{16}$, $300$ triples ($100$ for $11^{16}$ + $100$ for $12^{16}$ + $100$ for $13^{16}$ = $300$); and so on, up to $1\,000$ triples after the experiments for the problem size $20^{16}$.

\begin{figure}[htb]
    \centering
	\includegraphics[width=0.97\textwidth]{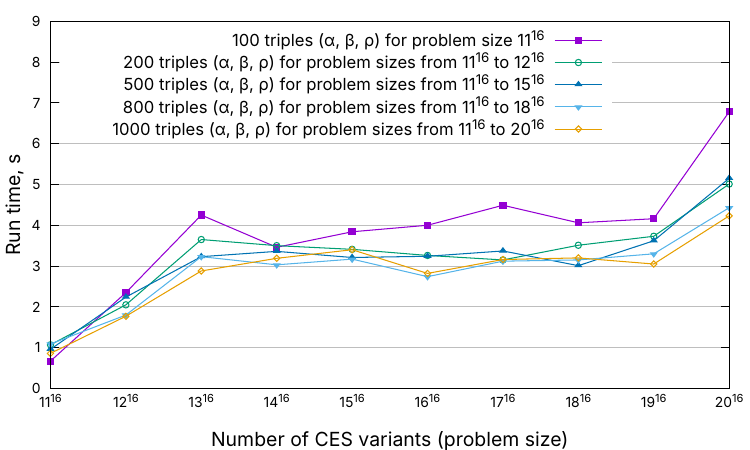}
	\caption{Run time of the program for problems of various sizes}
	\label{fig:time_multi_const}
\end{figure}

When the program is run on the GPU (in this case, $5\,120$ threads were used on the Tesla V100, corresponding to the total number of CUDA cores of this accelerator), the algorithm parameters are initialized cyclically, as described below.
Suppose there are $100$ unique parameter triples.
These $100$ unique triples must be used to initialize $5\,120$ threads;
therefore, the parameter values repeat every $100$ threads.
For $200$ triples, they repeat every $200$ threads, and so on;
for $1\,000$ unique triples, the constants repeat every $1\,000$ threads.
The run time of the program versus the problem size is shown in Fig.~\ref{fig:time_multi_const}.

In general, the time to solve the problem does not exceed $\approx7$\,s.
The worst run time (especially for the largest problem, $20^{16}$) corresponds to $100$ unique triples, and the best one to $1000$ triples.
Starting from $200$ unique triples, the average run time of the algorithm differs from that for the maximum of $1\,000$ triples by a factor of only $1.13$ ($3.23$~s versus $2.86$~s).

At the same time, the time needed to find $1000$ unique triples for problem sizes from $11^{16}$ to $20^{16}$ equals the time of the first iteration of the random approach, $51\,054$\,s $\approx14$\,h, while the time needed to find a set of $200$ triples for problem sizes from $11^{16}$ to $12^{16}$ is $3\,373$\,s $\approx56$\,min, i.e., $15.1$ times less.

\subsubsection*{Comparison of the approaches}

\begin{figure}[htb]
    \centering
	\includegraphics[width=0.97\textwidth]{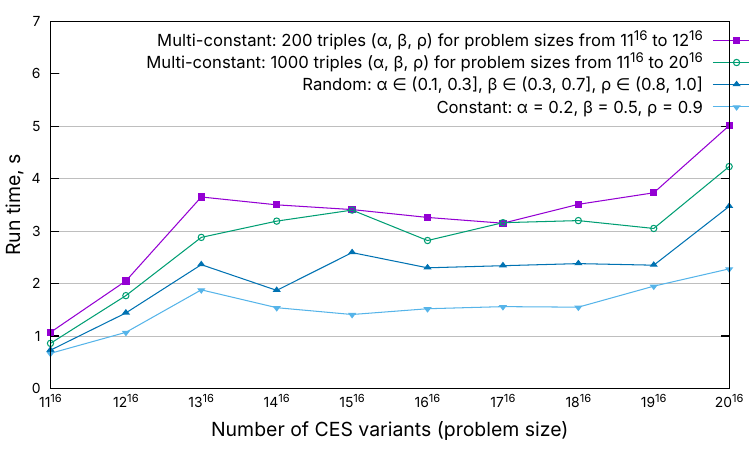}
	\caption{Run time of the program for problems of various sizes: comparison of the three tuning approaches}
	\label{fig:time_best}
\end{figure}

Figure~\ref{fig:time_best} shows the best results obtained with each approach.
For the multi-constant approach, the variant with $200$ triples is also shown, since it gave acceptable results with the minimum time spent on finding the parameter values.
The fastest algorithm, tuned with the constant approach, $\alpha = 0.2$, $\beta = 0.5$, $\rho = 0.9$ (average run time $\approx 1.5$~s), is $\approx2.1$ times faster than the slowest one, tuned with the multi-constant approach using $200$ unique triples (average run time $\approx3.2$\,s).

\begin{table}[htb]
	\centering
	\small
	\caption{Tuning time for the different approaches}
	\label{tab:ResearchTime}
	\begin{tabular}{lc}
		\toprule
		Approach & Tuning time \\
		\midrule
		Random & $86\,509$\,s $(\approx24$\,h) \\
		Constant & $86\,509$\,s $(\approx24$\,h) \\
		Multi-constant ($1\,000$) & $51\,054$\,s $(\approx14$\,h) \\
		Multi-constant ($200$) & $3\,373$\,s ($\approx56$\,min) \\
		\bottomrule
	\end{tabular}
\end{table}

However, the time spent on tuning (see Table~\ref{tab:ResearchTime}) is $86\,509$\,s $\approx24$\,h in the first case and $3\,373$\,s $\approx 56$\,min in the second, i.e., $\approx26$ times less.
\section{Experiments on GPUs of Different Architectures and on Multiple GPUs}
\label{chap6_sect5}

The results presented in Section~\ref{chap6_sect4} were obtained on a single Tesla~V100 GPU.
To assess to what extent the conclusions depend on the hardware used, and to explore ways of further reducing the time needed to find a feasible solution, the program was additionally studied on GPUs of three generations and in configurations with several GPUs within a single compute node.

In all experiments, as in Section~\ref{chap6_sect4}, the problem of finding a feasible CES variant was solved for the test case (Section~\ref{chap1_sect2}) with problem sizes from $11^{16}$ to $20^{16}$ variants; for each problem size, the program was run $100$ times, and the run time of the ant colony algorithm until the first feasible solution was found was measured.
The number of threads launched on one GPU was always equal to the number of CUDA cores of the corresponding accelerator.

\subsection{Organization of Computations on Multiple GPUs}
\label{chap6_multigpu_impl}

In the parallel implementation of the ant colony algorithm (Section~\ref{chap5_sect4}), each GPU thread simulates an independent ant colony, and the search stops as soon as one of the colonies finds a feasible solution.
The colonies do not exchange information; therefore, extending this scheme to several accelerators requires no changes to the kernel function \texttt{ACO()}: it suffices to launch it simultaneously on all GPUs of the node and to make sure that the search stops on all devices once a solution has been found on one of them.
Thus, the total number of simultaneously running colonies increases in proportion to the number of accelerators.

The pseudocode of the corresponding part of the host program is shown in Listing~\ref{lst:multiGPU_host}.

\begin{lstlisting}[%
	caption={Pseudocode of the search for a feasible solution on multiple GPUs (host)},%
	float=htb,%
	label={lst:multiGPU_host}]
	FindFirstSolutionMultiGPU(){
		for (d = 1; d <= numDevices; d++){ /* prepare each device */(*@\label{lnmg02}@*)
			cudaSetDevice(d);
			cudaMemcpyHtoD(inData, oprData);(*@\label{lnmg04}@*)
			initRandomStates<<<blocksPerGrid, threadsPerBlock>>>(seed * d);(*@\label{lnmg05}@*)
			SET<<<blocksPerGrid, threadsPerBlock>>>(TUN_APP);}(*@\label{lnmg06}@*)
		do {
			for (d = 1; d <= numDevices; d++){ /* asynchronous ACO launch */(*@\label{lnmg08}@*)
				cudaSetDevice(d);
				ACO<<<blocksPerGrid, threadsPerBlock, 0, streamKernel[d]>>>();
				cudaEventRecord(kernelDone[d], streamKernel[d]);}(*@\label{lnmg11}@*)
			finished = 0; bestDevice = -1;
			while (finished < numDevices){ /* round-robin polling of the devices */(*@\label{lnmg13}@*)
				for (d = 1; d <= numDevices; d++){
					if (!isFinished[d] && cudaEventQuery(kernelDone[d]) == cudaSuccess){(*@\label{lnmg15}@*)
						isFinished[d] = true; finished++;
						cudaMemcpyDtoH(isFound, d);
						if (bestDevice == -1 && isFound != 0){ (*@\label{lnmg18}@*)
							bestDevice = d;
							/* stop the search on all devices */
							for (k = 1; k <= numDevices; k++)
							cudaMemcpyHtoDAsync(isFound, k, streamStop[k]);}}}}(*@\label{lnmg22}@*)
			cudaDeviceSynchronizeAll(); (*@\label{lnmg23}@*)
		} while (bestDevice == -1);(*@\label{lnmg24}@*)
		cudaSetDevice(bestDevice);
		cudaMemcpyDtoH(Wfirst, threadIdx);(*@\label{lnmg26}@*)
	}
\end{lstlisting}

In the preparatory phase (lines~\ref{lnmg02}--\ref{lnmg06}), the input and working data are copied to each device, after which the pseudorandom number generators and the parameters of the ant colony algorithm are initialized.
The generator seed is multiplied by the device number (line~\ref{lnmg05}) so that the random number sequences, and hence the behavior of the colonies, differ between GPUs.
The parameters $\alpha$, $\beta$, and $\rho$ are set in one of the ways described in Section~\ref{chap6_sect3} (line~\ref{lnmg06}; see Listing~\ref{lst:device_SET}).
With the random approach, each device generates its own parameter values from the given intervals.
With the multi-constant approach, each device is assigned its own contiguous block of triples from a common file, so that all threads of all devices receive different triples.

The host then launches the kernel function \texttt{ACO()} on all devices (lines~\ref{lnmg08}--\ref{lnmg11}).
The launches are asynchronous, each in its own CUDA stream, and after each launch an event (CUDA event) marking the completion of the kernel is placed into the same stream.
The host then polls the state of the events in a loop without blocking on any single device (lines~\ref{lnmg13}--\ref{lnmg22}).
As soon as the kernel on some device has finished, the host reads the flag \texttt{isFound} from it.
If a solution has been found (line~\ref{lnmg18}), the flag value is copied asynchronously to all devices through separate streams.
Since the continuation condition of the main loop of the kernel includes a check of this flag (see Listing~\ref{lst:ACO_SA_device_ACO}, line~\ref{ln1211}), the colonies on the other devices stop at their next iteration.
After all devices have been synchronized (line~\ref{lnmg23}), the solution found is copied to the host from the device on which it was obtained (line~\ref{lnmg26}).
If no device has found a solution within the allotted number of iterations, the attempt is repeated (line~\ref{lnmg24}), as in the single-GPU version (see Listing~\ref{lst:ACO_SA_host}, lines~\ref{ln1114d}--\ref{ln1114w}).

In terms of the classification given in Section~\ref{chap3_review}, this scheme is static: from the very beginning, each device receives its own fixed set of independent colonies, and no work is redistributed between devices.
For the problem under consideration, this causes no losses due to load imbalance, since the search terminates not after all work has been done but at the first success of any colony.

\subsection{Experimental Platforms}
\label{chap6_platforms}

The specifications of the GPUs used are given in Table~\ref{tab:gpu_platforms}.
The peak single-precision performance was computed as twice the product of the number of CUDA cores and the clock rate.

\begin{table}[htb]
	\centering
	\small
	\caption{Specifications of the GPUs used in the experiments}
	\label{tab:gpu_platforms}
	\begin{tabular}{>{\raggedright\arraybackslash}p{0.4\textwidth}ccc}
		\toprule
		Specification & Tesla K20m & TITAN Xp & Tesla V100-SXM2 \\
		\midrule
		Microarchitecture & Kepler & Pascal & Volta \\
		Compute Capability & 3.5 & 6.1 & 7.0 \\
		Number of streaming multiprocessors & 13 & 30 & 80 \\
		CUDA cores per multiprocessor & 192 & 128 & 64 \\
		Total CUDA cores & 2496 & 3840 & 5120 \\
		Clock rate, MHz & 705 & 1582 & 1530 \\
		Global memory, GB & 5 & 12 & 16 \\
		Peak performance (FP32), TFLOPS & 3.5 & 12.1 & 15.7 \\
		\midrule
		Number of GPUs in the compute node & 3 & 8 & 4 \\
		Total number of colonies using all GPUs & 7488 & 30720 & 20480 \\
		\bottomrule
	\end{tabular}
\end{table}

\subsection{Comparison of GPU Architectures}
\label{chap6_arch}

Table~\ref{tab:gpu_arch_compare} gives the time to find a feasible solution on a single accelerator, averaged over all problem sizes (from $11^{16}$ to $20^{16}$), for the different approaches to parameter tuning.
For the TITAN Xp, results are available for only two parameter settings.

\begin{table}[htb]
	\centering
	\small
	\caption{Average time to find a feasible solution on a single GPU, s}
	\label{tab:gpu_arch_compare}
	\begin{tabular}{>{\raggedright\arraybackslash}p{0.42\textwidth}cccc}
		\toprule
		Approach and parameter values & K20m & TITAN Xp & V100 & K20m / V100 \\
		\midrule
		Random: $\alpha\in(0,2]$, $\beta\in(0,2]$, $\rho\in(0,1]$ & 339.5 & 125.9 & 51.1 & 6.6 \\
		\addlinespace[3pt]
		Random: $\alpha\in(0,1.2]$, $\beta\in(0.1,1.2]$, $\rho\in(0.2,1]$ & 108.9 & -- & 13.9 & 7.9 \\
		\addlinespace[3pt]
		Random: $\alpha\in(0,0.8]$, $\beta\in(0.1,1.1]$, $\rho\in(0.3,1]$ & 57.6 & -- & 7.7 & 7.5 \\
		\addlinespace[3pt]
		Random: $\alpha\in(0,0.6]$, $\beta\in(0.2,1.1]$, $\rho\in(0.4,1]$ & 44.7 & -- & 5.5 & 8.1 \\
		\addlinespace[3pt]
		Random: $\alpha\in(0,0.5]$, $\beta\in(0.2,0.9]$, $\rho\in(0.5,1]$ & 28.7 & -- & 3.4 & 8.4 \\
		\addlinespace[3pt]
		Random: $\alpha\in(0,0.4]$, $\beta\in(0.2,0.8]$, $\rho\in(0.6,1]$ & 21.3 & -- & 2.7 & 7.7 \\
		\addlinespace[3pt]
		Constant: $\alpha=0.4$, $\beta=0.6$, $\rho=0.9$ & 347.5 & 111.2 & 38.1 & 9.1 \\
		\addlinespace[3pt]
		Constant: $\alpha=0.2$, $\beta=0.5$, $\rho=0.9$ & 10.1 & -- & 1.5 & 6.5 \\
		\addlinespace[3pt]
		Multi-constant: 500 triples for problem sizes from $11^{16}$ to $15^{16}$ & 18.3 & -- & 3.1 & 5.8 \\
		\bottomrule
	\end{tabular}
\end{table}

The search time on the Tesla V100 was $5.8$--$9.1$ times shorter than on the Tesla K20m and $2.5$--$2.9$ times shorter than on the TITAN Xp.
These ratios exceed the ratios of the peak performance of the accelerators ($4.5$ and $1.3$, respectively; see Table~\ref{tab:gpu_platforms}).
The reason is that a more powerful accelerator not only performs the computations of each colony faster but also allows more colonies to run simultaneously: $5120$ on the V100 versus $3840$ on the TITAN Xp and $2496$ on the K20m.
As the number of colonies grows, the probability that at least one of them finds a feasible solution before the allotted number of iterations is exhausted increases, and the number of repeated attempts decreases: for example, for the intervals $\alpha\in(0,2]$, $\beta\in(0,2]$, $\rho\in(0,1]$, the average number of attempts was $3.7$ on the K20m, $2.7$ on the TITAN Xp, and $2.3$ on the V100.

At the same time, the relative efficiency of the tuning approaches does not depend on the architecture.
On the K20m, as on the V100, narrowing the parameter intervals in the random approach successively reduces the search time (from $339.5$ to $21.3$~s, i.e., by a factor of $16$), and the best result is achieved by the constant approach with the values $\alpha=0.2$, $\beta=0.5$, $\rho=0.9$ found in Section~\ref{chap6_sect4}: compared with the original values $\alpha=0.4$, $\beta=0.6$, $\rho=0.9$, the search time on the K20m decreased by a factor of $34$ (from $347.5$ to $10.1$~s).
Thus, parameter values obtained by tuning on one architecture remain effective on another.

\subsection{Scaling the Parallel Program to Multiple GPUs}
\label{chap6_scaling}

\begin{figure}[htb]
	\centering
	\includegraphics[width=0.97\textwidth]{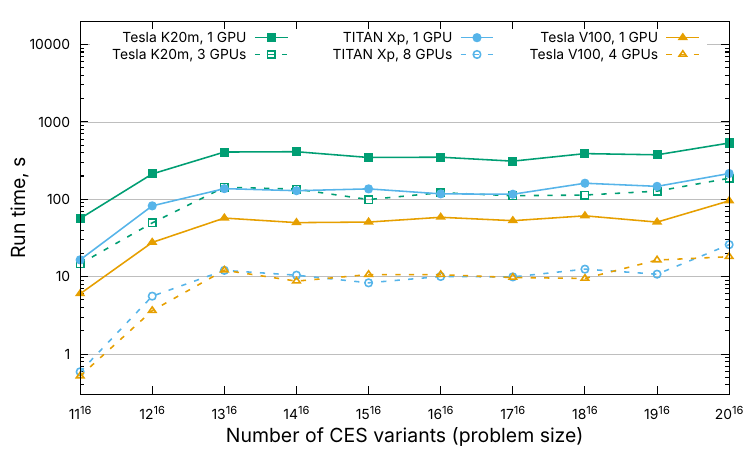}
	\caption{Time to find a feasible solution on one and on several GPUs}
	\label{fig:multiGPU}
\end{figure}

\begin{figure}[htb]
	\centering
	\includegraphics[width=0.97\textwidth]{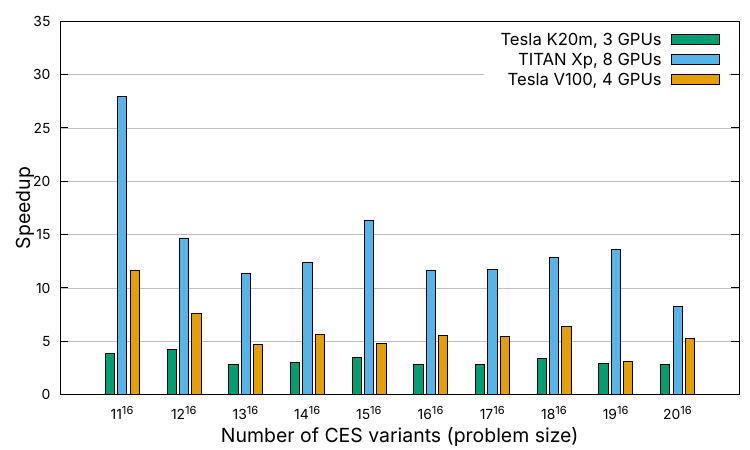}
	\caption{Speedup of the search for a feasible solution on several GPUs}
	\label{fig:multiGPUspeedup}
\end{figure}

To assess scalability, we used the random approach with the original intervals $\alpha\in(0,2]$, $\beta\in(0,2]$, $\rho\in(0,1]$, for which results are available on all three platforms, as well as the constant approach with $\alpha=0.4$, $\beta=0.6$, $\rho=0.9$ on the Tesla K20m platform.
The results are shown in Figs.~\ref{fig:multiGPU} and~\ref{fig:multiGPUspeedup} and in Table~\ref{tab:multiGPU}, where $T_1$ and $T_N$ are the average search times on one and on $N$ accelerators, $S=T_1/T_N$ is the speedup, and $E=S/N$ is the efficiency.

Figures~\ref{fig:multiGPU} and~\ref{fig:multiGPUspeedup} refer to the random approach with the original parameter intervals; because of the large spread of values, the search time in Fig.~\ref{fig:multiGPU} is shown on a logarithmic scale.

\begin{table}[htb]
	\centering
	\small
	\caption{Speedup and efficiency with several GPUs}
	\label{tab:multiGPU}
	\begin{tabular}{lcccccc}
		\toprule
		GPU, parameters & $N$ & $T_1$, s & $T_N$, s & $S$ & $E$ & Attempts ($1$ / $N$ GPUs) \\
		\midrule
		K20m, random & 3 & 339.5 & 110.3 & 3.1 & 1.03 & 3.7 / 1.7 \\
		K20m, constant & 3 & 347.5 & 76.7 & 4.5 & 1.51 & 4.3 / 1.7 \\
		TITAN Xp, random & 8 & 125.9 & 10.6 & 11.9 & 1.48 & 2.7 / 1.0 \\
		V100, random & 4 & 51.1 & 10.0 & 5.1 & 1.28 & 2.3 / 1.1 \\
		\bottomrule
	\end{tabular}
\end{table}

In all cases, the speedup was at least linear and, in most cases, superlinear: the efficiency ranged from $1.03$ to $1.51$.
The largest speedup was obtained on the node with eight TITAN Xp accelerators: the average search time decreased from $125.9$ to $10.6$~s, i.e., by a factor of $11.9$.
For individual problem sizes, the speedup was even higher: for example, for the problem size $11^{16}$, it was $28$ on eight TITAN Xp and $11.6$ on four V100 (Fig.~\ref{fig:multiGPUspeedup}).

The superlinear speedup is not a measurement artifact but follows from the nature of the problem.
As noted in Section~\ref{chap3_review_anomalies}, if the search stops when the first solution is found, the parallel algorithm may perform less work than the sequential one, which corresponds to a search overhead factor $R<1$ (expression~(\ref{eq:search_overhead})).
In this case, the run time is determined not by the total amount of computation but by the time to the first success among all simultaneously running colonies, i.e., by the minimum of random variables whose number is proportional to the number of accelerators.
If the time a single colony needs to find a solution varies strongly from run to run (such distributions are called heavy-tailed), then, alongside long runs, there is a noticeable fraction of very short ones, and increasing the number of independent colonies reduces the expected search time faster than linearly~\cite{Gomes2000}.
A similar effect is well known for parallel evolutionary and other population-based algorithms~\cite{Alba2002}.
In practice, it manifests itself as a reduction in the number of repeated attempts: when all accelerators of the node are used, a feasible solution is usually found at the very first kernel launch (Table~\ref{tab:multiGPU}), whereas on a single GPU an average of $2.3$ to $4.3$ attempts were needed in the cases considered.

Note also that the node with eight TITAN Xp accelerators ($30\,720$ colonies) and the node with four V100 accelerators ($20\,480$ colonies) achieved practically the same search time ($10.6$ and $10.0$~s), even though on a single GPU the V100 is $2.5$ times faster than the TITAN Xp.
Hence, for the problem under consideration, the lower performance of an individual accelerator can be largely compensated for by increasing the total number of simultaneously running colonies.

Combining multiple GPUs with parameter tuning gives an additional gain.
On the node with eight TITAN Xp, using the narrowed intervals of the random approach, the average search time ranged from $1.3$ to $2.7$~s: for example, $1.3$~s for the intervals $\alpha\in(0,0.5]$, $\beta\in(0.2,0.9]$, $\rho\in(0.5,1]$ versus $3.4$~s on a single V100, and $2.7$~s for the intervals $\alpha\in(0,1.2]$, $\beta\in(0.1,1.2]$, $\rho\in(0.2,1]$ versus $13.9$~s.
With the multi-constant approach using $500$ parameter triples, the search time on eight TITAN Xp was $1.9$~s versus $3.1$~s on a single V100.

The study on GPUs of three generations showed that the parameter values found remain effective regardless of the GPU architecture, and that using several accelerators within a single compute node provides at least linear speedup of the search for a feasible solution (up to $11.9$ times on eight TITAN~Xp accelerators) without changing the compute kernel.

These results show that the proposed parallel algorithm for finding a feasible solution scales well to multiple GPUs without any changes to the compute kernel.
Since the interaction between devices reduces to sending a flag when a solution is found and copying the solution to the host, the communication overhead is negligible, and increasing the number of accelerators in a node is a simple and effective way of reducing the search time for large problems.

%%%% Conclusion
\chapter*{Conclusion}
\phantomsection
\addcontentsline{toc}{chapter}{Conclusion}
\label{conclusion}
The problem of optimal equipment selection for chemical-engineering systems of multiproduct batch plants has been formulated (Chapter~\ref{chapter1}).

A sequential algorithm based on the branch-and-bound method has been developed for the optimal CES equipment selection problem (Chapter~\ref{chapter2}).

The algorithm has been implemented in C++.
The performance of individual elements of the algorithm has been studied, and on this basis the algorithm has been optimized to increase its speed (Chapter~\ref{chapter2}).

The algorithm has been parallelized.
For shared-memory systems, a program has been developed in C++ using the OpenMP standard (Section~\ref{chap3_sect1}).
The speedup and processor efficiency of this program have been studied.
The study showed that the parallel program achieves near-linear speedup and fairly high scalability,
with a core efficiency of $0.9$--$0.95$.

For distributed-memory systems, a program has been developed in C++ using the MPI library, implementing the algorithm according to the master--worker paradigm (Section~\ref{chap3_sect2}).
The study of speedup and efficiency showed that the parallel program achieves near-linear speedup and fairly high scalability.
However, the master process becomes a bottleneck of the program when it serves a large number of workers, which reduces the speedup and efficiency of the program.

For heterogeneous computing systems equipped with GPUs, a parallel program has been developed in C using CUDA, taking into account the technical features of GPUs (Chapter~\ref{chapter4}).
Two ways of traversing tree structures have been implemented: recursive and iterative.
Two ways of optimizing the algorithm have also been proposed: efficient use of high-performance shared memory and reduction of branch divergence.
The efficiency of using GPUs for the problem under consideration has been evaluated.
Although the program accelerates the solution of the problem considered in this book, a direct implementation of branch-and-bound on the GPU does not yield a large speedup (by hundreds or thousands of times).

A hybrid metaheuristic algorithm (ACO+SA) combining two metaheuristics, Ant Colony Optimization (ACO) and Simulated Annealing (SA), has been developed (Chapter~\ref{chapter5}).
The hybrid algorithm works in two stages: first, a feasible initial solution (a fully operable CES variant that produces the given product range within the given time) is found by ACO, and then this solution is improved by SA.

In developing the algorithm, an original way of computing the fitness function for ant colony optimization, on the basis of which the pheromone concentration is updated, has been proposed, together with a way of defining the heuristic information.
For simulated annealing, an original way of searching for a candidate solution in the neighborhood of the current feasible solution has been proposed, as well as a way of determining the maximum and minimum temperatures of the cooling schedule.
The algorithm has been implemented in C using the CUDA programming model.
The program includes code executed on the CPU, which implements only the sequential parts of the algorithm, prepares data and copies them to the GPU, sets the kernel parameters, and launches the kernels.
The efficiency of using GPUs for solving the problem with the hybrid algorithm has been evaluated.
The parallel implementation of the hybrid metaheuristic algorithm (ACO+SA) yields high-quality solutions for the test problem that coincide with the globally optimal values obtained by the deterministic branch-and-bound algorithm, but much faster: in a few tens of seconds instead of a few tens of hours.
However, when the problem size exceeds $11^{16}$, the performance drops considerably.
This is because, as the problem size grows, finding a feasible initial solution with ACO takes more and more time.

Problems larger than $12^{16}$ were not solved, which motivated the transition to tuning the parameters of the algorithm.

Before ACO is started, a number of parameters that affect its operation must be set (tuned) (Chapter~\ref{chapter6}).
A family of methods for tuning the parameters $\alpha$, $\beta$, and $\rho$ of the algorithm has been developed: random, constant, and multi-constant.
Owing to modern heterogeneous CPU--GPU computing systems, a large number of computational experiments could be performed in relatively short time, and their results were then processed statistically.
By analyzing how often solutions fall into the ranges of the algorithm parameters, it was possible to set the bounds of these ranges to values that yield the largest number of solutions and to accelerate the algorithm considerably.
Several successive series of experiments with the random approach yielded values of the constants for the constant approach that provide the highest performance.
To initialize the algorithm parameters in the multi-constant approach, the values obtained in the first series of experiments with the random approach were used.
The study of the multi-constant approach showed that parameters tuned for small problems can be used to solve large problems.
Although this approach is not the best one in terms of algorithm speed, it considerably reduces the time spent on searching for the best algorithm parameters.
The parameter values found for the metaheuristic algorithm reduced the average time needed to find a feasible initial solution for problems of size $11^{16}$ to $20^{16}$ by more than an order of magnitude: from $\approx38.1$~s for $\alpha = 0.4$, $\beta = 0.6$, $\rho = 0.9$ to $\approx1.5$~s for $\alpha = 0.2$, $\beta = 0.5$, $\rho = 0.9$.
The multi-constant approach provides a search time of $\approx2.9$--$3.2$~s while reducing the time spent on tuning the parameters by a factor of $\approx 26$.

The program has additionally been studied on GPUs of three generations (Tesla K20m, TITAN~Xp, and Tesla V100) and on compute nodes with several GPUs (Section~\ref{chap6_sect5}).
The parameter values found by tuning on one architecture remain effective on the others: on the Tesla K20m, they reduce the average time needed to find a feasible solution by a factor of $34$ (from $347.5$ to $10.1$~s).
Since the ant colonies are independent, the algorithm extends to several GPUs without any changes to the kernel function.
The speedup obtained is at least linear and, in most cases, superlinear: up to $11.9$ on eight TITAN~Xp accelerators (from $125.9$ to $10.6$~s).
The reason is that the search time is determined by the first success among all simultaneously running colonies, and the number of repeated attempts decreases as the number of colonies grows.

\bibliographystyle{unsrtnat}
\bibliography{refs_en}
\end{document}